\documentclass[fleqn,usenatbib]{mnras}

\usepackage{newtxtext,newtxmath}

\usepackage[T1]{fontenc}
\usepackage{ae,aecompl}

\usepackage{subfig}
\usepackage{tikz}
\usepackage{standalone}

\usepackage{graphicx}	
\usepackage{amsmath}	
\usepackage{booktabs}

\def\beq{\begin{eqnarray}}
\def\eeq{\end{eqnarray}}

\newcommand{\Mpch}{h^{-1}\mathrm{Mpc}}
\newcommand{\hMpc}{h\,\mathrm{Mpc}^{-1}}

\newcommand{\av}[1]{\langle{#1\rangle}} 
\let\vec\mathbf

\newcommand{\nd}[1]{\hat{n}_D(\vec x_{#1})} 
\newcommand{\np}[1]{n_D(\vec x_{#1})} 
\newcommand{\tj}[6]{ \begin{pmatrix}
   #1 & #2 & #3 \\
   #4 & #5 & #6 
\end{pmatrix}}

\defcitealias{2020MNRAS.492.1214P}{PE20}

\newcommand{\resub}[1]{#1}

\numberwithin{equation}{section}

\title[Efficient Power Spectrum and Bispectrum Estimators]{A Faster Fourier Transform?\\
\Large Computing Small-Scale Power Spectra and Bispectra for Cosmological Simulations in $\mathcal{O}(N^2)$ Time}

\author[O.\,H.\,E. Philcox]{
Oliver H.\,E. Philcox$^{1,2,3}$\thanks{E-mail: \href{mailto:ohep2@cantab.ac.uk}{ohep2@cantab.ac.uk}}
\\
$^{1}$Department of Astrophysical Sciences, Princeton University, Princeton, NJ 08544, USA\\
$^{2}$Harvard-Smithsonian Center for Astrophysics, 60 Garden St., MA 02138, USA\\
$^{3}$School of Natural Sciences, Institute for Advanced Study, 1 Einstein Drive, Princeton, NJ 08540, USA\\
}

\pubyear{2020}

\begin{document}
\label{firstpage}
\pagerange{\pageref{firstpage}--\pageref{lastpage}}
\maketitle

\begin{abstract}
We present $\mathcal{O}(N^2)$ estimators for the small-scale power spectrum and bispectrum in cosmological simulations. In combination with traditional methods, these allow spectra to be efficiently computed across a vast range of scales, requiring orders of magnitude less computation time than Fast Fourier Transform based approaches alone. These methods are applicable to any tracer; simulation particles, halos or galaxies, and take advantage of the simple geometry of the box and periodicity to remove almost all dependence on large random particle catalogs. By working in configuration-space, both power spectra and bispectra can be computed via a weighted sum of particle pairs up to some radius, which can be reduced at larger $k$, leading to algorithms with decreasing complexity on small scales. These do not suffer from aliasing or shot-noise, allowing spectra to be computed to arbitrarily large wavenumbers. The estimators are rigorously derived and tested against simulations, and their covariances discussed. The accompanying code, \texttt{HIPSTER}, has been publicly released, incorporating these algorithms. Such estimators will be of great use in the analysis of large sets of high-resolution simulations.
\end{abstract}

\begin{keywords}
methods: statistical, numerical -- Cosmology: large-scale structure of Universe, theory -- galaxies: statistics
\end{keywords}



\section{Introduction}\label{sec: intro}
In the modern epoch, we have access to a large variety of cosmological datasets, both real and simulated, and through their analysis one can probe fundamental particle physics, investigate the nature of gravitation, and expose the composition of the Universe. To robustly extract such information, we require summary statistics that can be easily computed, modeled and interpreted. Considering the late Universe, the most useful tool for the analysis of surveys has historically been the isotropic two-point correlation function (2PCF) and its Fourier counterpart, the power spectrum $P(k)$. Both of these describe clustering as a function of scale, and, if the underlying density field is Gaussian, encapsulate all available cosmological information. Whilst recent galaxy survey analyses have focused on consideration of the prominent Baryon Acoustic Oscillation peak \citep[e.g.,\,][]{2005ApJ...633..560E,2014MNRAS.441...24A,2016MNRAS.460.4210G,2017MNRAS.464.3409B}, additional information is encoded in its full shape (recently demonstrated for the galaxy power spectra in \citealt{2019arXiv190905271D}, \citealt{2019arXiv190905277I,2019arXiv191208208I} and \citealt{2020arXiv200204035P}). Power spectra are not simply limited to galaxy surveys, however; additional usages can be found for example in weak lensing analyses \citep[e.g.,\,][]{2015ApJ...806....1M,2015ApJ...806....2M,2019PASJ...71...43H,2020PASJ...72...16H}.

At low redshifts, the assumption of Gaussianity fails in the real Universe, and we require statistics beyond the power spectrum, most notably the isotropic bispectrum (or the equivalent three-point correlation function; 3PCF), encoding non-linear clustering effects, and anisotropic power spectra, describing the deviation from isotropy due to redshift-space distortions \citep[RSD;\,][]{1987MNRAS.227....1K} and the Alcock-Paczyniski effect \citep{1979Natur.281..358A}. A wealth of literature exists on each quantity, describing algorithms for their measurement, theoretical descriptions and observational measurements (e.g., \citealt{1994ApJ...426...23F,2006PhRvD..74l3507T,2006PASJ...58...93Y,2008PThPh.120..609Y,2011MNRAS.415.2876B,2014MNRAS.444.1400N,2015MNRAS.453L..11B,2015PhRvD..92h3532S,2016MNRAS.455L..31S,2017JCAP...07..002H} for the power spectrum, and \citealt{1998ApJ...494L..41S,2004ApJ...605L..89S,2006PhRvD..74b3522S,1999ApJ...517..531S,2001ApJ...546..652S, 2002MNRAS.335..432V,2012PhRvD..86f3511F,2013PhRvD..88f3512S,2015MNRAS.451..539G,2015MNRAS.452.1914G,2017MNRAS.469.2059S,2018MNRAS.478.1468S,2018ApJ...862..119P,2018MNRAS.478.4500P,2019arXiv190201830H} for the bispectrum). 

Whilst powerful cosmological estimators and robust theoretical descriptions are of great use in our quest to extract information from cosmological surveys, we are missing a crucial ingredient; simulations. These have a multitude of uses; examples include testing theoretical predictions, modeling regimes where perturbation theory fails, generating accurate covariance matrices and training machine learning algorithms. Today, large simulation suites are available including \texttt{EAGLE} \citep{2016A&C....15...72M}, \texttt{BAHAMAS} \citep{2017MNRAS.465.2936M}, \texttt{AbacusCosmos} \citep{2018ApJS..236...43G}, \texttt{Aemulus} \citep{2019ApJ...875...69D} and \texttt{Quijote} \citep{2019arXiv190905273V}. These contain up to tens of thousands of simulated universes, and their number and size will only grow with time. To fully make use of these, it is necessary to compute the above summary statistics for each individual simulation, and, given the enormous volumes of data available, this clearly indicates the need for efficient estimators.

In order to measure the power spectra of discrete objects such as galaxies and simulation particles, cosmologists have long relied on methods based on Fast Fourier Transforms (FFTs), first assigning the particles to a grid of side-length $N_\mathrm{grid}$, then transforming to Fourier space in an operation whose complexity scales as $N_\mathrm{grid}\log N_\mathrm{grid}$ \citep[e.g.,\,][]{1994ApJ...426...23F,2006PASJ...58...93Y,2012PhRvD..86f3511F,2013PhRvD..88f3512S,2015PhRvD..92h3532S,2015MNRAS.453L..11B,2019MNRAS.484..364S}. This scaling is particularly favorable on large scales, and naturally translates into other observables such as correlation functions \citep[e.g.,\,][]{2016MNRAS.455L..31S}. On small scales, these computations become expensive since a large grid must be used to avoid the effects of aliasing, with a runtime scaling as $k_\mathrm{max}\log k_\mathrm{max}$.\footnote{\citet{2016MNRAS.460.3624S} provides an interesting method to ameliorate this, making use of interlaced FFT grids.} In many cases, in particular machine learning applications, we are interested in computing statistics across a broad range of scales, encompassing both the largest structures in the Universe as well as distributions of matter inside clusters. To this end, it is important to find a more efficient manner of computing small-scale spectra.

Here, we address this problem by presenting a different set of spectral estimators, first suggested in \citet{2001MNRAS.325.1389J} and \citet{2016ApJ...833..287L} and rigorously developed in \citet[hereafter \citetalias{2020MNRAS.492.1214P}]{2020MNRAS.492.1214P}. These stem from the idea that, since the power spectrum and bispectrum are fundamentally just the Fourier transforms of correlation functions, and correlation functions can be computed by simply counting groups of particles in space, we can compute spectra directly by counting groups of particles with Fourier weights $e^{i\vec k\cdot\vec r}$. Although this strictly requires counting all possible groups of particles in the survey or simulation, counts can be truncated at a finite radius with negligible impact on the small-scale spectra. Power spectrum (bispectrum) computation is thus reduced to counting pairs (triplets) of galaxies and random particles in space. For a fixed truncation radius, this does not scale with the $k$-scale considered and naturally avoids affects such as aliasing and shot-noise.

Whilst \citetalias{2020MNRAS.492.1214P} focused on the application of this method to the power spectrum of non-uniform survey data, in this paper we specialize to cosmological simulations, since these contain readily available information across a range of scales, without the added complexities afforded by observational data (e.g., fiber collisions and blending). Given that simulations (both N-body and hydrodynamical) are usually made with periodic boundary conditions, a number of simplifications are possible; in particular this obviates the need for a random particle catalog, since the random particle integrals can be performed analytically (see \citealt{2019MNRAS.486L.105P} for an analogous calculation for the 3PCF). This dramatically increases the efficiency. A large swathe of this paper is also dedicated to computation of the bispectrum; using spherical harmonic decompositions, this can also be formulated as a pair count and evaluated in comparable time to the power spectrum. In this instance, our work is similar to the work of \citet{2015MNRAS.454.4142S} and \citet{2017MNRAS.469.1738S} which considered the corresponding configuration-space 3PCF.

This paper is structured as follows. We begin in Sec.\,\ref{sec: overview} with a high-level overview of configuration-space algorithms and the main ideas of this work, before deriving the power spectrum and bispectrum estimators in full in Secs.\,\ref{sec: pk-algo}\,\&\,\ref{sec: bk-algo}. Sec.\,\ref{sec: windowed-to-true} contains a brief comment on window-function convolutions inherent to our method, before we derive the auto- and cross-covariances of our estimators in detail in Secs.\,\ref{sec: cov},\,\ref{sec: cov-pk-bk}\,\&\,\ref{sec: cov-bk}. In Sec.\,\ref{sec: application}, we discuss our implementation of the algorithms into the public \texttt{HIPSTER} code,\footnote{\href{https://HIPSTER.readthedocs.io}{HIPSTER.readthedocs.io}} and give a number of examples of its usage, including measuring the power spectrum and bispectrum on a broad range of scales. We conclude with a summary in Sec.\,\ref{sec: conclusion}, with supplementary mathematical material presented in appendices \ref{appen: E-II} to \ref{appen: Dellfn}. For the reader whose prime interest lies in understanding the algorithm rather than diving into detailed mathematics, we recommend skipping all but Secs.\,\ref{sec: overview}\,\&\,\ref{sec: application}.

\section{Overview of Configuration-Space Spectral Estimators}\label{sec: overview}
We begin with a discussion of configuration-space estimators in cosmological simulations, acting both as a summary of \citetalias{2020MNRAS.492.1214P} and a broad overview of this work. Initially, we will consider the anisotropic power spectrum $P(\vec k)$, defined as the Fourier transform of the two-point correlation function (2PCF) $\xi(\vec r)$. Using the \citet{1993ApJ...412...64L} estimator, this can be written in terms of data-data (DD), data-random (DR) and random-random (RR) counts;\footnote{In this paper, we define the forward and inverse Fourier transforms as
\beq
    \widetilde{X}(\vec k) &\equiv& \int d\vec x\,e^{-i\vec k\cdot\vec x}X(\vec x),  \qquad
    X(\vec x) \equiv \int \frac{d\vec k}{(2\pi)^3}e^{i\vec k\cdot\vec x}\widetilde{X}(\vec k)\nonumber.
\eeq
and the Dirac function $\delta_D$ via 
\beq
    \int d\vec x\,e^{i(\vec k_1-\vec k_2)\cdot\vec x} \equiv (2\pi)^3\delta_D(\vec k_1-\vec k_2).\nonumber
\eeq
The correlation function and power spectrum of the density field are defined as 
\beq
    \xi(\vec r) = \av{\delta(\vec x)\delta(\vec x+\vec r)},\qquad (2\pi)^3\delta_D(\vec k+\vec k')P(\vec k) = 
    \av{\widetilde{\delta}(\vec k)\widetilde{\delta}(\vec k')}\nonumber,
\eeq
with the power spectrum as the Fourier transform of the correlation function and higher order correlators being defined similarly.
}
\beq    
    P(\vec k) &\equiv& \int d\vec r\,e^{-i\vec k\cdot\vec r}\xi(\vec r) =  \int d\vec r\,e^{-i\vec k\cdot\vec r}\left[\frac{DD(\vec r)-2DR(\vec r)+RR(\vec r)}{RR(\vec r)}\right],
\eeq
where `data' and `randoms' refer to the cosmological particles (which can be simulation particles, halos or galaxies) and a group of randomly placed points. For a uniform simulation of volume $V$ with periodic boundary conditions and particle density $n$, we can replace the denominator by the $RR$ counts for an ideal survey, $n^2V$;
\beq
    P(\vec k) &=& \frac{1}{n^2V}\int d\vec r\,e^{-i\vec k\cdot\vec r}\left[DD(\vec r)-2DR(\vec r)+RR(\vec r)\right] \equiv \frac{1}{n^2V}\left[\widetilde{DD}(\vec k)-2\widetilde{DR}(\vec k)+\widetilde{RR}(\vec k)\right],
\eeq
where the above relation defines the functions $\widetilde{DD}$, $\widetilde{DR}$, and $\widetilde{RR}$.\footnote{Unlike \citetalias{2020MNRAS.492.1214P}, we make the simplification that the unclustered number densities are uniform and the particles are unweighted; this is a valid assumption for a cosmological simulation. This also leads to a `survey correction function' $\Phi(\vec r)$ of unity everywhere.}
Inserting the definition of the pair-counts as an integral of the number density fields over the full simulation volume, these are defined as
\beq\label{eq: general-XY-pre-Legendre}
    \widetilde{XY}(\vec k) &=& \int d\vec r\,e^{-i\vec k\cdot\vec r}XY(\vec r) = \int d\vec r\,e^{-i\vec k\cdot\vec r}\int d\vec x_1d\vec x_2\,n_X(\vec x_1)n_Y(\vec x_2)\delta_D(\vec r-(\vec x_1-\vec x_2))\\\nonumber
    &=& \int d\vec x_1d\vec x_2\,n_X(\vec x_1)n_Y(\vec x_2)e^{-i\vec k\cdot(\vec x_1-\vec x_2)}
\eeq
(where each of $X$ and $Y$ are either $D$ or $R$), where the second line follows by integrating over $\vec k$ and applying the Dirac delta $\delta_D$. For a discrete set of tracer particles (e.g., galaxies or simulation particles) the number densities can be written as a sum over Dirac deltas, giving
\beq\label{eq: discrete-xy}
    \widetilde{XY}(\vec k) &=& \sum_{i\in X}\sum_{j\in Y,\,i\neq j \text{ if } X=Y}e^{-i\vec k\cdot(\vec x_i-\vec x_j)},
\eeq
where the indices run over all particles $i,j$ in fields $X,Y$, at $\vec x_i$, $\vec x_j$.\footnote{Note that particle weights could be easily added to this formalism by introducing an additional $w_iw_j$ factor in Eq.\,\ref{eq: discrete-xy} This may be useful when considering the power spectrum of multiple species of particle for example.} For $X = Y$ we exclude self-counts (defined as pairs of particles for which $\vec x_i = \vec x_j$), since these contribute only to shot-noise and are not expected to have cosmological relevance. The computation of the power spectrum thus reduces to a summation over pairs of points with a specific weighting function. Note however that this requires a count over \textit{all} particle pairs in the data-set for any given $k$-bin, unlike in 2PCF analyses where we count only up to the maximum radial bin. To ameliorate this, we truncate the pair counts at some radius $R_0$, with a smooth window function as in \citetalias{2020MNRAS.492.1214P};
\beq\label{eq: window_defn}
    W(\vec r; R_0) &\equiv& \begin{cases} 1 & \text{if } 0\leq x < 1/2 \\ 1-8\left(2x - 1 \right)^3 + 8 \left(2x - 1\right)^4& \text{if }1/2\leq x<3/4\\ -64\left(x-1\right)^3 - 128\left(x-1\right)^4 & \text{if } 3/4\leq x< 1\\ 0 & \text {else}\end{cases}
\eeq
with $x = |\vec r|/R_0$. This is introduced for computational tractability and has negligible effect on the measured spectra for $kR_0\gg 1$. This explains why our estimator is optimized for small scales. In addition, we decompose the function into Legendre multipoles about the (local) line-of-sight which, following a somewhat involved computation, yields
\beq\label{eq: general-Pk-ell}
    \widehat{P}_\ell(k;R_0) \equiv \frac{1}{n^2V}\left[\widetilde{DD}_\ell(k,R_0)-2\widetilde{DR}_\ell(k;R_0)+\widetilde{RR}_\ell(k;R_0)\right],
\eeq
where the modified pair counts are given in discrete and continuous form by
\beq\label{eq: general-XY-counts}
    \widetilde{XY}_\ell(k;R_0) &=& (-i)^\ell(2\ell+1)\int d\vec x_1d\vec x_2\,n_X(\vec x_1)n_Y(\vec x_2)j_\ell(k|\vec x_1-\vec x_2|)L_\ell(\hat{\vec x}_{12}\cdot\hat{\vec n}_{12})W(\vec x_1-\vec x_2;R_0)\\\nonumber
    &=& (-i)^\ell(2\ell+1)\sum_{i\in X}\sum_{j\in Y,\,i\neq j \text{ if } X=Y}j_\ell(k|\vec x_i-\vec x_j|)L_\ell(\hat{\vec x}_{ij}\cdot\hat{\vec n}_{ij})W(\vec x_i-\vec x_j;R_0)
\eeq
\citepalias{2020MNRAS.492.1214P}, where $\vec x_{ab}\equiv \vec x_a-\vec x_b$ and $\vec n_{ij}$ points along the line-of-sight direction (fixed for a simulation, or equal to $(\vec x_i+\vec x_j)/2$ for survey data). This uses spherical Bessel functions $j_\ell$ and Legendre multipoles $L_\ell$. We may thus compute the power spectrum from a set of pair counts with the above weighting functions. In actual analyses it is beneficial to bin in $|\vec k|$-space; we defer this complexity to later sections. It is a key point of this paper that the counts involving randoms, $\widetilde{DR}$ and $\widetilde{RR}$, can be performed analytically. This reduces the power spectrum estimator to a simple pair-count over all tracer particles up to a maximum radius $R_0$.

A similar line of reasoning applies for the bispectrum, $B$. Here we consider only the \textit{isotropic} bispectrum, which is integrated over all orientations of the $k$-space triangle with respect to the line-of-sight, and hence does not carry RSD information. Similar conclusions apply however for the anisotropic function. Writing this in terms of its Fourier transform, the three point correlation function (3PCF) $\zeta$, and inserting the \citet{1998ApJ...494L..41S} estimator gives
\beq
    B(\vec k_1, \vec k_2) &=& \int d\vec r_1d\vec r_2\,e^{-i\vec k_1\cdot\vec r_1}e^{-i\vec k_2\cdot\vec r_2}\zeta(\vec r_1,\vec r_2)\\\nonumber
    &=& \int d\vec r_1d\vec r_2\,e^{-i\vec k_1\cdot\vec r_1}e^{-i\vec k_2\cdot\vec r_2}\left[\frac{DDD(\vec r_1,\vec r_2)-3DDR(\vec r_1,\vec r_2)+3DRR(\vec r_1,\vec r_2)-RRR(\vec r_1,\vec r_2)}{RRR(\vec r_1,\vec r_2)}\right].
\eeq
As before, we rewrite the denominator in terms of the idealized random triple counts, $6n^3V$, to give
\beq
    B(\vec k_1, \vec k_2) &=& \int d\vec r_1d\vec r_2\,e^{-i\vec k_1\cdot\vec r_1}e^{-i\vec k_2\cdot\vec r_2}\zeta(\vec r_1,\vec r_2)\\\nonumber
    &=& \frac{1}{6n^3V}\int d\vec r_1d\vec r_2\,e^{-i\vec k_1\cdot\vec r_1}e^{-i\vec k_2\cdot\vec r_2}\left[DDD(\vec r_1,\vec r_2)-3DDR(\vec r_1,\vec r_2)+3DRR(\vec r_1,\vec r_2)-RRR(\vec r_1,\vec r_2)\right]\\\nonumber
    &\equiv& \frac{1}{6n^3V}\left[\widetilde{DDD}(\vec k_1,\vec k_2)-3\widetilde{DDR}(\vec k_1,\vec k_2)+3\widetilde{DRR}(\vec k_1,\vec k_2)-\widetilde{RRR}(\vec k_1,\vec k_2)\right],
\eeq
where the modified triple counts are defined in continuous and discrete form as
\beq\label{eq: general-XYZ-counts}
    \widetilde{XYZ}(\vec k_1,\vec k_2) &=& \int d\vec x_1d\vec x_2d\vec x_3\left[n_X(\vec x_1)n_Y(\vec x_2)n_Z(\vec x_3)e^{-i\vec k_1\cdot(\vec x_1-\vec x_2)}e^{-i\vec k_2\cdot(\vec x_1-\vec x_3)} + \text{ 5 perms.}\right]\\\nonumber
    &=& \sum_{i\in X*}\sum_{j\in Y*}\sum_{k\in Z*}\left[e^{-i\vec k_1\cdot(\vec x_i-\vec x_j)}e^{-i\vec k_2\cdot\vec(\vec x_i-\vec x_k)} + \text{ 5 perms.}\right],
\eeq
where `perms.' refers to permutations of the set $\{\vec x_1,\vec x_2,\vec x_3\}$. The asterisks indicate that we exclude self-counts for identical fields \resub{hence} avoiding shot-noise contributions. Analogously to the power spectrum, we introduce pair-separation window functions, $W(\vec r;R_0)$ that will allow us to count particles only up to radius $R_0$.\footnote{Note that constraining two triangle sides to be shorter than $R_0$ constrains the third to be shorter than $2R_0$ via the triangle inequality. This does not break symmetry however, since we sum over permutations of the three sides.} Using this (and denoting the angular part of $d\vec k$ by $d\Omega_k$), the Legendre multipoles of $B$ are defined by \footnote{Note that this differs from \citetalias{2020MNRAS.492.1214P} by a factor of $(2\ell+1)$, correcting an earlier oversight.}
\beq\label{eq: general-Bkl-exp}
    B_\ell(k_1,k_2) &\equiv& (2\ell+1)\int \frac{d\Omega_{k_1}}{4\pi}\frac{d\Omega_{k_2}}{4\pi}B(\vec k_1,\vec k_2)L_\ell(\hat{\vec k}_1\cdot\hat{\vec k}_2)\\\nonumber
    \Rightarrow B_\ell(k_1,k_2;R_0) &=& \frac{1}{6n^3V}\left[\widetilde{DDD}_\ell(k_1,k_2;R_0)-3\widetilde{DDR}_\ell(k_1,k_2;R_0)+3\widetilde{DRR}_\ell(k_1,k_2;R_0)-\widetilde{RRR}_\ell(k_1,k_2;R_0)\right].
\eeq
Following some algebra, the kernels may be written as
\beq
    \widetilde{XYZ}_\ell(k_1,k_2;R_0) &=& (-1)^\ell(2\ell+1)\int d\vec x_1d\vec x_2d\vec x_3\left[j_\ell(k_1|\vec x_{12}|)j_\ell(k_2|\vec x_{13}|)L_\ell(\hat{\vec x}_{12}\cdot\hat{\vec x}_{13})W(\vec x_{12};R_0)W(\vec x_{13};R_0) + \text{ 5 perms. }\right]\\\nonumber
    &=&(-1)^\ell(2\ell+1)\sum_{i\in X*}\sum_{j\in Y*}\sum_{k\in Z*}\left[j_\ell(k_1|\vec x_{ij}|)j_\ell(k_2|\vec x_{ik}|)L_\ell(\hat{\vec x}_{ij}\cdot\hat{\vec x}_{ik})W(\vec x_{ij};R_0)W(\vec x_{ik};R_0) + \text{ 5 perms. }\right],
\eeq
where $\vec x_{ab}\equiv \vec x_a-\vec x_b$ as before \citepalias{2020MNRAS.492.1214P}. Whilst a na\"ive implementation of this estimator (and the one suggested in \citetalias{2020MNRAS.492.1214P}) is a count over triples of particles, scaling as $\mathcal{O}(N^3)$ for $N$ particles, it is in fact possible to reduce this to a pair count, making use of spherical harmonic theorems and carefully considering self-count terms. Schematically, we obtain
\beq
    \widetilde{XYZ}_\ell(k_1,k_2;R_0) = (-1)^\ell (2\ell+1)\sum_{i\in X\ast} \left [ \frac{4\pi}{2\ell+1}\sum_{m=-\ell}^{\ell} A_{\ell m}^{}(\vec x_i;k_1,R_0)A^{*}_{\ell m}(\vec x_i;k_2,R_0) - C_\ell(\vec x_i;k_1,k_2,R_0)\right],
\eeq
where the functions $A$ and $C$ can be written as a sum over all points separated by less than $R_0$ from particle $i$, and $C$ is included to capture the $j = k$ term if $Y = Z$. We further note that any term involving a random field can be computed analytically, without use of a large random catalog, though in practice, it is faster to compute one of the terms using such a catalog. We refer the reader to Sec.\,\ref{sec: bk-algo} for a detailed discussion of such effects. For the reader who is less interested in detailed mathematics, the above should be sufficient proof that such estimators exist, and we encourage them to skip directly to the applications to data in Sec.\,\ref{sec: application}.

\section{The Power Spectrum in the Periodic Limit}\label{sec: pk-algo}
For a cosmological simulation, we may substantially simplify the expressions for $P_\ell(k)$ (Eq.\,\ref{eq: general-Pk-ell}) by removing all dependence on the random particle catalogs. Firstly, since the number density of particles in the simulation is known precisely, we may use the alternative 2PCF estimator $\xi(\vec r) = DD(\vec r)/RR(\vec r) - 1$ rather than the \citet{1993ApJ...412...64L} form. (It can be shown that $DR(\vec r)\equiv RR(\vec r)$ by translational invariance).
We thus obtain the simpler power spectrum form
\beq\label{eq: simple-Pk-form}
    P_\ell(k; R_0) = \frac{1}{n^2V}\left[\widetilde{DD}_\ell(k;R_0)-\widetilde{RR}_\ell(k;R_0)\right].
\eeq
Next, we note that the modified $RR$ count (from Eq.\,\ref{eq: general-XY-counts}) becomes analytic in the limit of infinite randoms. To see this, we first write the $\widetilde{RR}_\ell$ function in continuous form using line-of-sight (LoS) vector $\hat{\vec n}$;
\beq
    \widetilde{RR}_\ell(k;R_0) &\equiv & (-1)^\ell (2\ell+1)\int d\vec x_1d\vec x_2\,n_r(\vec x_1)n_r(\vec x_2)L_\ell(\hat{\vec{x}}_{12}\cdot\hat{\vec{n}})j_\ell(k|\vec x_{12}|)W(\vec x_{12};R_0).
\eeq
Transforming variables to $\vec r = \vec x_1-\vec x_2$ and integrating over $\vec x_2$ gives
\beq
    \widetilde{RR}_\ell(k;R_0) &=& n^2V (-i)^\ell(2\ell+1)\int d\vec r\,L_\ell(\hat{\vec r}\cdot\hat{\vec n})j_\ell(kr)W(\vec r;R_0)\\\nonumber
    &=& 4\pi n^2V \delta^K_{\ell0}\int r^2dr\,j_\ell(kr)W(r;R_0) \equiv  n^2V\delta_{\ell0}^K \widetilde{W}(kR_0),
\eeq
where we have used that the angular integral of $L_\ell$ is simply $4\pi\delta_{\ell0}^K$ for Kronecker delta $\delta^K$ and that $W(\vec r;R_0)$ is isotropic with spherical Fourier transform $\widetilde{W}(kR_0)$. (Note that, for the polynomial window function used in \citetalias{2020MNRAS.492.1214P}, $W(kR_0)$ can be expressed analytically in terms of incomplete gamma functions). The full power spectrum estimator thus becomes
\beq
    \hat{P}_\ell(k;R_0) = \frac{1}{n^2V}(-i)^\ell(2\ell+1)\sum_{i\neq j}W(\vec x_i-\vec x_j;R_0)j_\ell(k|\vec x_i-\vec x_j|)L_\ell(\hat{\vec x}_{ij}\cdot\hat{\vec n}) - \delta_{\ell0}^K\widetilde{W}(kR_0),
\eeq
where $(i,j)$ run over all galaxies and we exclude self-counts to avoid shot-noise. Note that this does \textit{not} require a random catalog to compute. 

In practice, it is usually preferable to bin the power in $k$-space. For a $k$-bin $a$ with volume $v_a$, we obtain
\beq\label{eq: pk-binned}
    \hat{P}^a_\ell(R_0) = \frac{4\pi}{v_a}\int k^2dk\,\Theta^a(k)\hat{P}_\ell(k;R_0),
\eeq
where $\Theta^a(k)$ is a binning function that is unity if $k$ is in $a$ and zero else. Practically, $k$-space binning simply modifies the pair-count kernels and the analytic $RR$ term; computation proceeds identically using the bin-integrated forms. Including this, our estimator becomes
\beq
    \hat{P}^a_\ell(R_0) = \frac{1}{n^2V}(-i)^\ell(2\ell+1)\sum_{i\neq j}W(\vec x_i-\vec x_j;R_0)j^a_\ell(|\vec x_i-\vec x_j|)L_\ell(\hat{\vec x_{ij}}\cdot\hat{\vec n}) - \delta_{\ell0}^K\widetilde{W}^a(R_0),
\eeq
where the superscript $a$ indicates an average over the $k$-space bin $a$ (which has volume $v_a$). The $j_\ell^a$ functions are in fact analytic, with 
\beq\label{eq: jla-def}
    j_\ell^a(x) &\equiv& \frac{4\pi}{v_a}\int k^2dk\,\Theta^a(k)j_\ell(kx)\\\nonumber &=&\frac{3\left[D_\ell(xk_{a,\mathrm{max}})-D_\ell(xk_{a,\mathrm{min}})\right]}{x^3\left[k^3_{a,\mathrm{max}}-k^3_{a,\mathrm{min}}\right]},
\eeq
where $D_\ell(u) \equiv \int u^2du\,j_\ell(u)$ can be evaluated via a recursion relation, as shown in Appendix \ref{appen: Dellfn}.
Using this notation, the binned pair-separation windows are thus
\beq\label{eq: wka-def}
    \widetilde{W}^a(R_0) &\equiv& \frac{4\pi}{v_a}\int k^2dk\,\Theta^a(k)\widetilde{W}(kR_0) =4\pi\int_0^{R_0} r^2dr\,j_0^a(r)W(r;R_0).
\eeq

\section{The Bispectrum in the Periodic Limit}\label{sec: bk-algo}
A less trivial extension of the above is to the bispectrum. We begin by rewriting the windowed bispectrum of Eq.\,\ref{eq: general-Bkl-exp};
\beq
    B_\ell(k_1,k_2;R_0) = \frac{1}{6Vn^3}\left[\widetilde{DDD}_\ell(k_1,k_2;R_0)-3\widetilde{DDR}_\ell(k_1,k_2;R_0)+3\widetilde{DRR}_\ell(k_1,k_2;R_0)-\widetilde{RRR}_\ell(k_1,k_2;R_0)\right].
\eeq
As for the power spectrum, this expression can be simplified by noting that, due to translation invariance, the triple counts $DRR$ are equivalent to $RRR$. To see this rigorously, consider the triple counts in spatial bins $a$, $b$ with associated volumes $\delta\vec x_a, \delta\vec x_b$
\beq
    RRR_{ab} &\equiv& \int d\vec x_1d\vec x_2d\vec x_3\,n_r(\vec x_1)n_r(\vec x_2)n_r(\vec x_3)\left[\Theta^a(\vec x_1-\vec x_2)\Theta^b(\vec x_2-\vec x_3) + \text{ 5 perms.}\right]\\\nonumber
    &=&6n^3\int d\vec x_1d\vec x_2d\vec x_3\Theta^a(\vec x_1-\vec x_2)\Theta^b(\vec x_2-\vec x_3) = 6n^3V\delta\vec x_a\delta\vec x_b\\\nonumber
    DRR_{ab} &\equiv& \int d\vec x_1d\vec x_2d\vec x_3\,n_r(\vec x_1)n_r(\vec x_2)n_g(\vec x_3)\left[\Theta^a(\vec x_1-\vec x_2)\Theta^b(\vec x_2-\vec x_3) + \text{ 5 perms.}\right]\\\nonumber
    &=& 2n^2\int d\vec x_1d\vec x_2d\vec x_3\,n_g(\vec x_3)\left[2\Theta^a(\vec x_1-\vec x_2)\Theta^b(\vec x_2-\vec x_3) +\Theta^a(\vec x_1-\vec x_3)\Theta^b(\vec x_1-\vec x_3)\right]\\\nonumber
    &=& 2n^2 \int d\vec x_3n_g(\vec x_3)\left[\delta\vec x_a\delta\vec x_b+2\delta\vec x_a\delta\vec x_b\right] = 6n^3V\delta\vec x_a\delta\vec x_b \equiv RRR_{ab},
\eeq
using the symmetries of the expressions under permutations of $\{a,b\}$ and $\{\vec x_1,\vec x_2,\vec x_3\}$ and noting that $\int d\vec r\,\Theta^a(\vec r-\vec s)= \delta\vec r_a$, and $\int d\vec r\, n_g(\vec r) = nV$. This naturally applies also to the modified pair counts with the $k$-space kernels, allowing us to combine the $\widetilde{DRR}$ and $\widetilde{RRR}$ terms in the bispectrum estimator. Note that it is not correct to write the 3PCF estimator as $\zeta(\vec r_1,\vec r_2) = DDD(\vec r_1,\vec r_2)/RRR(\vec r_1,\vec r_2) - 1$ (as one might expect from the simplified 2PCF estimator); this will contain additional contributions from the 2PCF.

\subsection{$RRR$ Term}
We proceed by evaluating the modified $RRR$ count analytically, in the limit of infinite randoms;
\beq
    \widetilde{RRR}_\ell(k_1,k_2;R_0) &\equiv& 6(-1)^\ell(2\ell+1) \int d\vec x_1d\vec x_2d\vec x_3\,n_r(\vec x_1)n_r(\vec x_2)n_r(\vec x_3)j_\ell(k_1|\vec x_{12}|)j_\ell(k_2|\vec x_{13}|)W(|\vec x_{12}|;R_0)W(|\vec x_{13}|;R_0)L_\ell(\hat{\vec x}_{13}\cdot\hat{\vec x}_{23})\\\nonumber
    &=& 6Vn^3(-1)^\ell(2\ell+1) \int d\vec r_1d\vec r_2\,j_\ell(k_1r_1)j_\ell(k_2r_2)W(r_1;R_0)W(r_2;R_0)L_\ell(\hat{\vec r}_1\cdot\hat{\vec r}_2)\\\nonumber
    &=&6Vn^3(-1)^\ell (2\ell+1)\left[4\pi\int r_1^2dr_1\,j_\ell(k_1r_1)W(r_1;R_0)\right]\left[4\pi\int r_2^2dr_2\,j_\ell(k_2r_2)W(r_2;R_0)\right]\delta^K_{\ell0}\\\nonumber
    &=& 6Vn^3\widetilde{W}(k_1R_0)\widetilde{W}(k_2R_0)\delta_{\ell0}^K,
\eeq
where we have inserted the $k$-space kernel with $\vec x_{pq}\equiv \vec x_p-\vec x_q$ and noted that all six permutations of $\{\vec x_1, \vec x_2, \vec x_3\}$ give the same result. In the second line we have transformed variables and integrated over $\vec x_1$ (which is trivial due to the periodicity), and in the third, we note that $\int d\Omega_{r}L_\ell(\hat{\vec r}\cdot\hat{\vec s}) = 4\pi \delta_{\ell0}^K$. Given this simplification, the bispectrum may be written in somewhat friendlier form;
\beq
    B_\ell(k_1,k_2;R_0) = \frac{1}{6Vn^3}\left[\widetilde{DDD}_\ell(k_1,k_2;R_0)-3\widetilde{DDR}_\ell(k_1,k_2;R_0)\right] + 2\delta_{\ell0}^K\widetilde{W}(k_1R_0)\widetilde{W}(k_2R_0).
\eeq

\subsection{$DDD$ Term}
The $DDD$ triple counts are formally given as the weighted triple count over the survey;
\beq
    \widetilde{DDD}_\ell(k_1,k_2;R_0) &=& (-1)^\ell(2\ell+1)\sum_{i\neq j\neq k}\left[j_\ell(k_1|\vec x_{ij}|)j_\ell(k_2|\vec x_{ik}|)W(|\vec x_{ij}|;R_0)W(|\vec x_{ik}|;R_0)L_\ell(\hat{\vec x}_{ij}\cdot\hat{\vec x}_{ik}) + \text{ 5 perms.}\right]\\\nonumber
    &=& 6 \sum_{i\neq j\neq k}A_\ell(\vec x_{ij},\vec x_{ik};k_1,k_2;R_0),
\eeq
(as in Eq.\,\ref{eq: general-XYZ-counts}), defining the kernel $A_\ell$, where the sum runs over all distinct triples of galaxies at $(\vec x_i,\vec x_j,\vec x_k)$ and the six permutations are identical by symmetry. A na\"{i}ve estimate of this term would involve counting all weighted triplets of particles with $i-j$ and $i-k$ separations up to $R_0$. However, in this case, we may do considerably better and reduce this to a simple pair count, following an analogous computation to that performed for the 3PCF in \citet{2015MNRAS.454.4142S}. We first isolate the $j=k$ term of the summation;
\beq
    \widetilde{DDD}_\ell(k_1,k_2;R_0) &=& 6(-1)^\ell(2\ell+1)\sum_i\sum_{j\neq i}\sum_{k\neq i}j_\ell(k_1|\vec x_{ij}|)j_\ell(k_2|\vec x_{ik}|)W(|\vec x_{ij}|;R_0)W(|\vec x_{ik}|;R_0)L_\ell(\hat{\vec x}_{ij}\cdot\hat{\vec x}_{ik})\\\nonumber
    &&\,-\, 6(-1)^\ell(2\ell+1)\sum_i\sum_{j\neq i}j_\ell(k_1|\vec x_{ij}|)j_\ell(k_2|\vec x_{ij}|)W^2(|\vec x_{ij}|;R_0),
\eeq
where the second line follows as $L_\ell\left(\hat{\vec x}_{ij}\cdot\hat{\vec x}_{ij}\right)\equiv 1$. The first term may be rewritten by expanding $L_\ell$ in terms of spherical harmonics as
\beq
    L_\ell(\hat{\vec a}\cdot\hat{\vec b})=\frac{4\pi}{2\ell+1}\sum_{m=-\ell}^\ell Y_{\ell m}^{}(\hat{\vec a})Y^{*}_{\ell m}(\hat{\vec b})
\eeq
\citep[Eq.\,14.30.9]{nist_dlmf} and separating out the $j$ and $k$ summations;
\beq
    \widetilde{DDD}_\ell(k_1,k_2;R_0) &=& 6(-1)^\ell(2\ell+1)\frac{4\pi}{2\ell+1}\sum_{m=-\ell}^{\ell}\sum_i\sum_{j\neq i}\left[j_\ell(k_1|\vec x_{ij}|)W(|\vec x_{ij}|;R_0)Y^{}_{\ell m}(\hat{\vec x}_{ij})\right]\sum_{k\neq i}\left[j_\ell(k_2|\vec x_{ik}|)W(|\vec x_{ik}|;R_0)Y^{*}_{\ell m}(\hat{\vec x}_{ik})\right]\\\nonumber
    &&\,-\, 6(-1)^\ell(2\ell+1)\sum_i\sum_{j\neq i}j_\ell(k_1|\vec x_{ij}|)j_\ell(k_2|\vec x_{ij}|)W^2(|\vec x_{ij}|;R_0)\\\nonumber
    &=& 6(-1)^\ell(2\ell+1)\sum_{i}\left[\frac{4\pi}{2\ell+1}\sum_{m=-\ell}^\ell A^{}_{\ell m}(\vec x_i;k_1,R_0)A^{*}_{\ell m}(\vec x_i;k_2,R_0) - C_\ell(\vec x_i;k_1,k_2;R_0)\right],
\eeq
defining
\beq\label{eq: AC-def}
    A_{\ell m}(\vec x_i;k,R_0) &=& \sum_{j\neq i}j_{\ell}(k|\vec x_{ij}|)W(|\vec x_{ij}|;R_0)Y_{\ell m}(\hat{\vec x}_{ij})\\\nonumber
    C_{\ell}(\vec x_i;k_1,k_2,R_0) &=& \sum_{j\neq i}j_{\ell}(k_1|\vec x_{ij}|)j_{\ell}(k_2|\vec x_{ij}|)W^2(|\vec x_{ij}|;R_0).
\eeq
To estimate the $\widetilde{DDD}$ term, we thus need to compute (and sum) the $A_{\ell m}$ and $C_\ell$ functions at the location of each particle in the simulation. Since both are simply sums over one set of points, we are essentially only counting \textit{pairs} of particles rather than triples; this section of the algorithm thus has complexity $\mathcal{O}(NnR_0^3)$ rather than $\mathcal{O}(Nn^2R_0^6)$ (as would be expected for a triple count). This affords a significant speed boost.

\subsection{$DDR$ Term}
For the $DDR$ counts, simplification is possible by analytically performing the integral over the random field, though this is non-trivial. Firstly, we rewrite the term as
\beq
    \widetilde{DDR}_\ell(k_1,k_2;R_0) &=& \int d\vec x_1d\vec x_2d\vec x_3\,n_g(\vec x_1)n_g(\vec x_2)n_r(\vec x_3)\left[A_\ell(\vec x_{12},\vec x_{13};k_1,k_2;R_0) + \text{ 5 perms.}\right]\\\nonumber
    &=& 2\int d\vec x_1d\vec x_2d\vec x_3\,n_g(\vec x_1)n_g(\vec x_2)n_r(\vec x_3)\left[2A_\ell(\vec x_{12},\vec x_{13};k_1,k_2;R_0) + A_\ell(\vec x_{13},\vec x_{23};k_1,k_2;R_0)\right]\\\nonumber
    &\equiv& 4\widetilde{DDR}^{I}_\ell(k_1,k_2;R_0)+2\widetilde{DDR}^{II}_\ell(k_1,k_2;R_0),
\eeq
where we have separated out terms with different structures under permutation. Physically the two terms arise from whether the two $k$ vectors correspond to (a) one data-data side of the triangle and one data-random side or (b) two data-random sides. Both terms may be written in terms of a sum over galaxy pairs (due to the two $n_g$ fields) with an integral over the field of randoms; the first gives
\beq
    \widetilde{DDR}^{I}_\ell(k_1,k_2;R_0) &=& \sum_{i\neq j}\int d\vec x_3\,n_r(
    \vec x_3)A_\ell(\vec x_{ij},\vec x_{i3};k_1,k_2;R_0)\\\nonumber
    &=& \frac{1}{2}n(-1)^\ell(2\ell+1)\sum_{i\neq j}\int d\vec x_3\,j_\ell(k_1|\vec x_{ij}|)j_\ell(k_2|\vec x_{i3}|)W(|\vec x_{ij}|;R_0)W(|\vec x_{i3}|;R_0)L_\ell(\hat{\vec x}_{ij}\cdot\hat{\vec x}_{i3}) + (k_1\leftrightarrow k_2)\\\nonumber
    &=& \frac{1}{2}n(-1)^\ell(2\ell+1) \sum_{i\neq j}j_\ell(k_1|\vec x_{ij}|)W(|\vec x_{ij}|;R_0)\mathcal{E}^{I}_\ell(\vec x_i,\vec x_j;k_2;R_0) + (k_1\leftrightarrow k_2).
\eeq
This is simply a count of all pairs in the survey separated by distances up to $R_0$, weighted by a kernel $\mathcal{E}^I$. This function may be rewritten as
\beq
    \mathcal{E}^{I}_\ell(\vec x_i,\vec x_j;k_2;R_0) &=& \int d\vec x_3\,j_\ell(k_2|\vec x_{i3}|)W(|\vec x_{i3}|;R_0)L_\ell(\hat{\vec x}_{ij}\cdot\hat{\vec{x}}_{i3})\\\nonumber
    &=& \int d\vec r\,j_\ell(k_2|\vec r|)W(|\vec r|;R_0)L_\ell(\hat{\vec x}_{ij}\cdot\hat{\vec r})\\\nonumber
    &=& 4\pi\delta_{\ell0}^K\int r^2dr\,j_\ell(k_2r)W(r;R_0) = \delta^K_{\ell0}\widetilde{W}(k_2R_0) 
\eeq
where we have substituted $\vec r = \vec x_{i3}\equiv \vec x_i-\vec x_3$ in the second line (possible by periodicity) and integrated over angle in the third, noting that this is simply the spherical Fourier transform of $W(r;R_0)$. Thus 
\beq
    \widetilde{DDR}_\ell^I(k_1,k_2;R_0) = \frac{1}{2}n\delta^K_{\ell0}\widetilde{W}(k_2R_0)\sum_{i\neq j}j_0(k_1|\vec x_{ij}|)W(|\vec x_{ij}|;R_0)  + (k_1\leftrightarrow k_2).
\eeq

Similar treatment is possible for $\widetilde{DDR}^{II}$, giving
\beq\label{eq: EII-intro}
    \widetilde{DDR}^{II}_\ell(k_1,k_2;R_0) &=& \sum_{i\neq j}\int d\vec x_3\,n_r(
    \vec x_3)A_\ell(\vec x_{i3},\vec x_{j3};k_1,k_2;R_0)\\\nonumber
    &=& n(-1)^\ell(2\ell+1)\sum_{i\neq j}\int d\vec x_3\,j_\ell(k_1|\vec x_{i3}|)j_\ell(k_2|\vec x_{j3}|)W(|\vec x_{i3}|;R_0)W(|\vec x_{j3}|;R_0)L_\ell(\hat{\vec x}_{i3}\cdot\hat{\vec x}_{j3})\\\nonumber
    &=& n(-1)^\ell(2\ell+1)\sum_{i\neq j}\mathcal{E}^{II}_\ell(\vec x_i,\vec x_j;k_1,k_2;R_0).
\eeq
The form of this is somewhat different to $\widetilde{DDR}^{I}$. In particular, we note that the sum over tracer particles $i$ and $j$ is \textit{not} restricted to $|\vec x_i-\vec x_j|$ by a pair-separation window $W$, though, due to the triangle inequality and the constraints that $|\vec x_i-\vec x_3|\leq R_0$ and $|\vec x_j-\vec x_3|\leq R_0$, $\mathcal{E}_\ell^{II}=0$ for $|\vec x_i-\vec x_j|\geq 2R_0$. Computation thus reduces to a weighted sum over all pairs up to a radius $2R_0$. It is possible to substantially simplify the kernel $\mathcal{E}^{II}$ into a function depending only on the separation between $i$ and $j$, the derivation of which is discussed in Appendices \ref{appen: E-II}\,\&\,\ref{appen: alternative-cal-E-II}. Since the pair-separation function used in this paper (Eq.\,\ref{eq: window_defn}) is simply a piecewise polynomial, we expect that $\mathcal{E}^{II}$ is analytic (since it involves only products of Bessel functions and polynomials); however, the full solution is exceedingly complex, especially when $k$-space binning is included. For this reason, it is simpler to use numerical integration in a pre-processing step, then interpolate at run-time.

In practice, the $\widetilde{DDR}^{II}$ term is extremely laborious to compute. Since the pair-count computation scales as $R_0^3$, we must sample eight times as many pairs for this term compared to the other bispectrum contributions. Furthermore, the $\mathcal{E}^{II}_\ell$ function needs to be pre-computed for each of the $n_\mathrm{bins}(n_\mathrm{bins}+1)n_\ell/2$ combination of $n_\mathrm{bins}$ radial and $n_\ell$ Legendre bins and interpolated, which significantly degrades the computational efficiency, especially when the number of radial bins is large. For this reason, we will adopt a different method to compute $\widetilde{DDR}^{II}_\ell$, making use of a random catalog.

First, we rewrite the term in integral form, separating out the $i=j$ term and using the expansion formula for Legendre polynomials, as for the $DDD$ counts;
\beq
    \widetilde{DDR}^{II}_\ell(k_1,k_2;R_0) &=& (-1)^\ell(2\ell+1)\sum_{i\neq j}\int d\vec x_3\,n_r(
    \vec x_3)j_\ell(k_1|\vec x_{i3}|)j_\ell(|k_2|\vec x_{j3}|)W(|\vec x_{i3}|;R_0)W(|\vec x_{j3}|;R_0)L_\ell(\hat{\vec x}_{i3}\cdot\hat{\vec x}_{j3})\\\nonumber
    &=& 4\pi(-1)^\ell\int d\vec x_3\,n_r(\vec x_3)\left[\sum_ij_\ell(k_1|\vec x_{i3})W(|\vec x_{i3}|;R_0)Y_{\ell m}^{}(\hat{\vec x}_{i3})\right]\left[\sum_jj\ell(k_2|\vec x_{j3})W(|\vec x_{j3}|;R_0)Y_{\ell m}^{*}(\hat{\vec x}_{j3})\right]\\\nonumber
    &&\,-\,(-1)^\ell(2\ell+1)\int d\vec x_3\,n_r(\vec x_3)\left[\sum_i j_\ell(k_1|\vec x_{i3}|)j_\ell(k_2|\vec x_{i3}|)W^2(|\vec x_{i3}|;R_0)\right].
\eeq
We now assume that we have a random catalog with $N_\mathrm{rand}$ particles positioned at $\{\vec r_r\}$ (which are drawn from a uniform distribution encompassing the simulation volume). It is generally advisable to use more randoms than tracer particles, thus we set $N_\mathrm{rand} = f_\mathrm{rand}N_\mathrm{data}$ for $N_\mathrm{data}$ tracer particles and $f_\mathrm{rand} \gtrsim 1$. With this simplification, the integral may be written as a summation;
\beq
    \widetilde{DDR}^{II}_\ell(k_1,k_2;R_0) &\approx& (-1)^\ell(2\ell+1)\frac{1}{f_\mathrm{rand}}\sum_{r}\left[\frac{4\pi}{2\ell+1}\sum_{m=-\ell}^\ell A^{}_{\ell m}(\vec x_r;k_1,R_0)A^{*}_{\ell m}(\vec x_r;k_2,R_0) - C_\ell(\vec x_r;k_1,k_2;R_0)\right],
\eeq
where $A_{\ell m}$ and $C_\ell$ are the functions defined in Eq.\,\ref{eq: AC-def} and we divide by the ratio of randoms to data-points; $f_\mathrm{rand}$. In this form the $\widetilde{DDR}^{II}$ term can be computed analogously to the $\widetilde{DDD}$ term, except that we compute the $A_{\ell m}$ and $C_\ell$ coefficients from the position of each random-particle, rather than each tracer-particle position. Whilst we note that this is strictly only an approximation, it becomes exact in the limit of infinite randoms, and we find the term to be well approximated with $f_\mathrm{rand}$ of order a few. Other choices of window function $W(\vec x;R_0)$ may lead to more tractable forms for the kernel $\mathcal{E}^{II}_\ell$ and hence $\widetilde{DDR}^{II}$, especially if one deals only with unbinned estimators.

\subsection{$k$-space Binning}
One consideration remains; the effects of finite $k$ bins. As in \citetalias{2020MNRAS.492.1214P} and analogous to the power spectrum (Eq.\,\ref{eq: pk-binned}), we consider a bispectrum in $k$-space bins $a,b$ as 
\beq
    B_\ell^{ab}(R_0) \equiv \frac{4\pi}{v_a}\frac{4\pi}{v_b}\int k_1^2dk_1\,k_2^2dk_2\,\Theta^a(k_1)\Theta^b(k_2)B_\ell(k_1,k_2;R_0),
\eeq
where $v_x$ is the volume of $k$-bin $x$. Practically, the binning simply modifies the integration kernels used in the pair-count summations (since we add $k$-space binning directly). In particular $\widetilde{RRR}_\ell$ and $\widetilde{DDD}_\ell$ become
\beq
    \widetilde{RRR}_\ell^{ab}(R_0) &=& 6Vn^3\widetilde{W}^a(R_0)\widetilde{W}^b(R_0)\delta_{\ell0}^K\\\nonumber
    \widetilde{DDD}_\ell^{ab}(R_0) &=& 6(-1)^\ell(2\ell+1)\sum_i\left[\frac{4\pi}{2\ell+1}\sum_{m=-\ell}^\ell A^a_{\ell m}(\vec x_i;R_0)A_{\ell m}^{b*}(\vec x_i;R_0) - C_\ell^{ab}(\vec x_i;R_0)\right],
\eeq
where the superscripts $(a,b)$ indicate the relevant $k$-space bin $a,b$, following the definitions of Eqs.\,\ref{eq: jla-def}\,\&\,\ref{eq: wka-def} and
\beq
    A_{\ell m}^a(\vec x_i;R_0) &=& \sum_{j\neq i} j^a_\ell(|\vec x_{ij}|)W(|\vec x_{ij}|;R_0)Y_{\ell m}(\hat{\vec x}_{ij})\\\nonumber
    C_\ell^{ab}(\vec x_i;R_0) &=& \sum_{j\neq i} j^a_\ell(|\vec x_{ij}|)j^b_\ell(|\vec x_{ik}|)W^2(|\vec x_{ij}|;R_0).
\eeq

For $\widetilde{DDR}$ we obtain
\beq
    \widetilde{DDR}_\ell^{I,ab}(R_0) &=& \frac{1}{2}n\delta_{\ell0}^K\left[\widetilde{W}^a(R_0)\sum_{i\neq j}j_0^b(k_1|\vec x_i-\vec x_j|)W(|\vec x_i-\vec x_j|;R_0)+(a\leftrightarrow b)\right]\\\nonumber
    \widetilde{DDR}^{II,ab}_\ell(R_0) &\approx& (-1)^\ell(2\ell+1)\frac{1}{f_\mathrm{rand}}\sum_{r}\left[\frac{4\pi}{2\ell+1}\sum_{m=-\ell}^\ell A^{a}_{\ell m}(\vec x_r,R_0)A^{b*}_{\ell m}(\vec x_r,R_0) - C^{ab}_\ell(\vec x_r;R_0)\right].
\eeq

\subsection{Summary}
We conclude with a summary. The small-scale windowed bispectrum estimator may be written as a combination of pair counts over particles (or galaxies) in the simulation box as
\beq
    \widehat{B}_\ell^{ab}(R_0) &=& \frac{1}{6Vn^3}\left[\widetilde{DDD}_\ell^{ab}(R_0)-3\widetilde{DDR}^{ab}_\ell(R_0)\right] + 2\delta_{\ell0}^K\widetilde{W}^a(R_0)\widetilde{W}^b(R_0)\\\nonumber
    &=& \frac{(-1)^\ell(2\ell+1)}{Vn^3}\left\{\sum_i\left[ \frac{4\pi}{2\ell+1}\sum_{m=-\ell}^\ell A^a_{\ell m}(\vec x_i;R_0)A^{b*}_{\ell m}(\vec x_i;R_0) - C_\ell^{ab}(\vec x_i;R_0)\right]\right.\\\nonumber
    &&\qquad\qquad\qquad\left.-\frac{1}{f_\mathrm{rand}}\sum_r\left[ \frac{4\pi}{2\ell+1}\sum_{m=-\ell}^\ell A^a_{\ell m}(\vec x_r;R_0)A^{b*}_{\ell m}(\vec x_r;R_0) - C_\ell^{ab}(\vec x_r;R_0)\right]\right\}\\\nonumber
    &&- \frac{(-1)^\ell(2\ell+1)}{Vn^2}\sum_{i\neq j}\left[\delta_{\ell0}^K\widetilde{W}^a(R_0)j_0^b(|\vec x_i-\vec x_j|)W(|\vec x_i-\vec x_j|;R_0)+(a\leftrightarrow b)\right]+2\delta_{\ell0}^K\widetilde{W}^a(R_0)\widetilde{W}^b(R_0),
\eeq
where we sum over all pairs of particles $i,j$ in the box with separations up to radius $R_0$ and denote the random particles by the index $r$. Note that, due to the analytic random integrals and the $DDD$ simplifications, our full estimator has complexity $\mathcal{O}(NnR_0^3)$, far superior to the $\mathcal{O}(Nn^2R_0^6)$ scaling claimed in \citetalias{2020MNRAS.492.1214P} (for fixed volume and $R_0$, this is simply $\mathcal{O}(N^2)$). Our power spectrum and bispectrum estimators thus require similar computation time, and we note that this is analogous to the $\mathcal{O}(N^2)$ 3PCF estimators of \citet{2015MNRAS.454.4142S}.

\section{Relating Windowed and True Spectra}\label{sec: window-vs-true-spectra}\label{sec: windowed-to-true}
To provide physical interpretation of our spectra, we must understand the impact of the pair-separation window function $W(\vec r;R_0)$. One application of this is the comparison of data to models; although the windowed spectrum matches its unwindowed form on very small scales, we may wish to compute a window-convolved theoretical model that works on all scales. For simplicity we will work with unbinned spectra, though similar conclusions hold in the binned case.

For the power spectrum, recall the initial definition as the Fourier transform of the windowed 2PCF;
\beq\label{eq: Pk-theory-conv}
    P(\vec k;R_0) &=& \int d\vec r\,\xi(\vec r)W(\vec r;R_0)e^{-i\vec k\cdot\vec r}.
\eeq
Since this is simply the Fourier transform of a product of functions, it can be rewritten as a \textit{convolution} of the true power $P(\vec k)$ with the Fourier transform of the window function $\widetilde{W}(\vec k;R_0)$, i.e.
\beq\label{eq: Pk-convolution}
    P(\vec k;R_0) = \left[P\ast \widetilde{W}\right](\vec k;R_0) \equiv \int\frac{d\vec p}{(2\pi)^3}P(\vec p)\widetilde{W}(\vec k-\vec p;R_0)
\eeq
\citepalias[Eq.\,3.4]{2020MNRAS.492.1214P}. To form the multipoles of the windowed power, it is more convenient to start with Eq.\,\ref{eq: Pk-theory-conv}, and note that, since $W$ is an isotropic function, the multipoles can be related to the multipoles of $\xi$ as
\beq
    P_\ell(k;R_0) = 4\pi(-i)^\ell\int r^2dr\,\xi_\ell(r)W(r;R_0)j_\ell(kr)
\eeq
\citepalias[Eq.\,3.9]{2020MNRAS.492.1214P}, which is simply the power spectrum multipoles of a configuration-space function whose multipoles are $\xi_\ell(r)W(r;R_0)$. To obtain a windowed theory prediction, we simply insert a theoretical 2PCF function into the above equation, which can be computed efficiently via Hankel transforms.

In a similar vein, the windowed bispectrum can be written as a convolution of the true bispectrum $B$;
\beq\label{eq: Bkk-convolution}
    B(\vec k_1,\vec k_2;R_0) = \int \frac{d\vec p_1d\vec p_2}{(2\pi)^6}B(\vec p_1,\vec p_2)\widetilde{W}(\vec k_1-\vec p_1;R_0)\widetilde{W}(\vec k_2-\vec p_2;R_0)
\eeq
\citepalias[Eq.\,7.14]{2020MNRAS.492.1214P}, or in terms of the 3PCF multipoles, $\zeta_\ell$;
\beq
    B(\vec k_1,\vec k_2;R_0) &=& \int d\vec x_1d\vec x_2\,\zeta(\vec x_1,\vec x_2)W(\vec x_1;R_0)W(\vec x_2;R_0)e^{-i\vec k_1\cdot\vec x_1}e^{-i\vec k_2\cdot\vec x_2}\\\nonumber
    &=& \sum_{\ell L_1 L_2}(2\ell+1)(2L_1+1)(2L_2+1)i^{L_1+L_2}\int d\vec x_1d\vec x_2\,\zeta_\ell(x_1,x_2)W(x_1;R_0)W(x_2;R_0)\\\nonumber
    &&\times\,j_{L_1}(k_1x_1)j_{L_2}(k_2x_2)L_\ell(\hat{\vec x}_1\cdot\hat{\vec x}_2)L_{L_1}(\hat{\vec k}_1\cdot\hat{\vec x}_1)L_{L_2}(\hat{\vec k}_2\cdot\hat{\vec x}_2)\\\nonumber
    &=& (4\pi)^2\sum_\ell(-1)^\ell (2\ell+1)\int x_1^2dx_1\,x_2^2dx_2\,\zeta_\ell(x_1,x_2)j_\ell(k_1x_1)j_\ell(k_2x_2)W(x_1;R_0)W(x_2;R_0)L_\ell(\hat{\vec k}_1\cdot\hat{\vec k}_2),
\eeq
using the plane wave expansion \citep[Eq.\,16.63]{arfken2013mathematical}, spherical harmonic addition \citep[Eq.\,14.30.9]{nist_dlmf} and noting that $\int d\hat{\vec x}\,L_\ell(\hat{\vec x}\cdot\hat{\vec y})L_L(\hat{\vec x}\cdot\hat{\vec z}) = 4\pi\delta^K_{\ell L}L_\ell(\hat{\vec y}\cdot\hat{\vec z})/(2\ell+1)$ via orthogonality  \citep[Eq.\,14.17.6]{nist_dlmf}. The bispectrum multipoles are thus given by
\beq\label{eq: bk-from-3pcf}
    B_\ell(k_1,k_2;R_0) &=& (2\ell+1)\int \frac{d\Omega_{k_1}}{4\pi}\frac{d\Omega_{k_2}}{4\pi}B(\vec k_1,\vec k_2;R_0)L_\ell(\hat{\vec k}_1\cdot\hat{\vec k}_2)\\\nonumber
    &=& (4\pi)^2(-1)^\ell\int x_1^2dx_1\,x_2^2dx_2\,\zeta_\ell(x_1,x_2)j_\ell(k_1x_1)j_\ell(k_2x_2)W(x_1;R_0)W(x_2;R_0),
\eeq
which are simply the Fourier space multipoles of a function with 3PCF $\zeta_\ell(x_1,x_2)W(x_1;R_0)W(x_2;R_0)$. Following this technique we may thus compare our windowed configuration-space power spectrum and bispectrum estimates to arbitrary theoretical models.

\section{Power Spectrum Covariances}\label{sec: cov}
Given the above estimators for small-scale spectra, it is instructive to consider their theoretical covariance. This is similar in form to the treatment of \citet{2019arXiv190806234S}, though we additionally include a treatment of Legendre multipoles and the window function. 
In general, covariance is produced by three effects;
\begin{enumerate}
    \item \textbf{Intrinsic covariance}: the covariance of the underlying density field, which contains both Gaussian and non-Gaussian terms, modified by the window function $W(\vec r;R_0)$. This is the covariance one would obtain for an infinite sample of tracer particles (i.e. the covariance of Eq.\,\ref{eq: Pk-theory-conv}). It is also the covariance of the matter power spectrum, subject to a double convolution with $W$.
    \item \textbf{Poisson covariance}: the covariance arising from the finite number of particles sampled. We expect this to dominate for small samples of particles (low $n$) and at large $k$, where the intrinsic covariance is small.
    \item \textbf{Super-sample covariance}: the covariance induced by modes larger than the simulation volume, which modulate the background density of the box \citep{2013PhRvD..87l3504T}. This can be modeled by considering the power spectrum `response' to long wavelength modes \citep[e.g.,][]{2018JCAP...02..022L}. Since this is not usually included in cosmological simulations (though see \citealt{2014PhRvD..89h3519L} for an example) and its modeling is specific to the field in question (e.g., it differs for biased and unbiased tracers), it will be ignored in this work. It is however an important source of covariance for cosmological surveys. 
\end{enumerate}
For simulations, each of the above terms can be derived in terms of the power spectrum in a tractable format. In real surveys however, this is often far more difficult, since the non-trivial survey geometries can have a non-negligible impact, and require careful consideration. In practice this can be done by assuming a separable covariance \citep{2019arXiv191002914W} or in terms of stochastic integrals over the survey selection function \citep{2016MNRAS.462.2681O,2019MNRAS.487.2701O,2019MNRAS.490.5931P,2020MNRAS.491.3290P}. We do not consider these complexities here.

We begin by discussing the covariance of the power spectrum in full, considering both intrinsic and Poissonian contributions. To see how the latter terms appear, it is instructive to begin by rewriting the power spectrum estimator (Eq.\,\ref{eq: simple-Pk-form}, before $\vec k$-space binning) as an integral over the discrete tracer field and the continuous random field (as in Eq.\,\ref{eq: general-XY-pre-Legendre});
\beq\label{eq: Pk-integral-form}
    \hat{P}(\vec k;R_0) &=& \frac{1}{n^2V}\left[\widetilde{DD}(\vec k;R_0) - \widetilde{RR}(\vec k;R_0)\right]\\\nonumber
    &=& \frac{1}{n^2V}\int_{\vec x_1\neq \vec x_2} d\vec x_1d\vec x_2\,\hat{n}_D(\vec x_1)\hat{n}_D(\vec x_2)e^{-i\vec k\cdot(\vec x_1-\vec x_2)}W(\vec x_1-\vec x_2;R_0)-\frac{1}{V}\int_{\vec x_1\neq \vec x_2} d\vec x_1d\vec x_2\,e^{-i\vec k\cdot(\vec x_1-\vec x_2)}W(\vec x_1-\vec x_2;R_0)\\\nonumber
    &\equiv& \frac{1}{V}\int_{\vec x_1\neq \vec x_2} d\vec x_1d\vec x_2\,\left[\frac{\hat{n}_D(\vec x_1)\hat{n}_D(\vec x_2)}{n^2} - 1\right]K(\vec x_1-\vec x_2;\vec k,R_0),
\eeq
defining the kernel $K(\vec x_1-\vec x_2)$ for simplicity, where we have written the discrete number density of tracer particles as $\hat{n}_D$ (which is just a sum over Dirac deltas) and used that the background number density $n$ is uniform. Note that we have imposed that $\vec x_1\neq \vec x_2$; this results from avoiding self-counts in our discrete sums over tracer particles. Before continuing, let us consider the statistics of the tracer field $\hat{n}_D$;
\beq
    \av{\hat{n}_D(\vec x)} &=& \av{n_D(\vec x)} = n\\\nonumber
    \av{\hat{n}_D(\vec x)\hat{n}_D(\vec y)} &=& \av{n_D(\vec x)n_D(\vec y)} + \av{n_D(\vec x)}\delta_D(\vec x-\vec y)\\\nonumber
    &=& n^2\left[1+\xi(\vec x-\vec y)\right]+n\delta_D(\vec x-\vec y),
\eeq
where we denote the underlying density field as $n_D$ (without a hat), and note that the Dirac delta $\delta_D$ arises from the discrete nature of the field. Inserting the above into Eq.\,\ref{eq: Pk-integral-form} confirms that $\av{\hat{P}(\vec k;R_0)} = P(\vec k;R_0)$ (as in Sec.\,\ref{sec: window-vs-true-spectra}), with the $\vec x_1\neq \vec x_2$ restriction removing the Poissonian shot noise term.

Given the above form, the covariance may be written as\footnote{Note that this is simply a continuous form of the covariance presented in \citetalias{2020MNRAS.492.1214P}.}
\beq\label{eq: cov-initial}
    \operatorname{cov}\left(\hat{P}(\vec k;R_0),\hat{P}(\vec k';R_0)\right) &\equiv& \av{\hat{P}(\vec k;R_0)\hat{P}(\vec k';R_0)}-\av{\hat{P}(\vec k;R_0)}\av{\hat{P}(\vec k';R_0)}\\\nonumber
    &=& \frac{1}{V^2}\int_{\vec x_1\neq\vec x_2}\int_{\vec x_3\neq \vec x_4}d\vec x_1d\vec x_2d\vec x_3d\vec x_4\,K(\vec x_1-\vec x_2;\vec k,R_0)K(\vec x_3-\vec x_4;\vec k',R_0)\\\nonumber
    &&\times\frac{1}{n^4}\left\{\left\langle\left[\hat{n}_D(\vec x_1)\hat{n}_D(\vec x_2)-n^2\right]\left[\hat{n}_D(\vec x_3)\hat{n}_D(\vec x_4)-n^2\right]\right\rangle-\left\langle\left[\hat{n}_D(\vec x_1)\hat{n}_D(\vec x_2)-n^2\right]\right\rangle\left\langle\left[\hat{n}_D(\vec x_3)\hat{n}_D(\vec x_4)-n^2\right]\right\rangle\right\}.
\eeq
The term in curly brackets may be written
\beq
    \{...\} &=& \av{\nd{1}\nd{2}\nd{3}\nd{4}}-\av{\nd{1}\nd{2}}\av{\nd{3}\nd{4}}\\\nonumber
    &=& \av{\np{1}\np{2}\np{3}\np{4}}-\av{\np{1}\np{2}}\av{\np{3}\np{4}}\\\nonumber
    && + 4\delta_D(\vec x_1-\vec x_4)\av{\np{1}\np{2}\np{3}}\\\nonumber
    && + 2\delta_D(\vec x_1-\vec x_3)\delta_D(\vec x_2-\vec x_4)\av{\np{1}\np{2}},
\eeq
where we have performed Poisson averaging, subject to the $\vec x_1\neq\vec x_2$, $\vec x_3\neq\vec x_4$ conditions. We have additionally grouped terms symmetric under $\vec x_1\leftrightarrow\vec x_2$ and $\vec x_3\leftrightarrow\vec x_4$ since the $K$ kernels are symmetric under these permutations (except for a phase inversion, which will vanish upon Legendre multipole binning). Note that we have decomposed the covariance into terms depending on four, three and two copies of the density field due to the Poissonian expansion. To proceed, we note that $n_D(\vec x) = n(1+\delta(\vec x))$ and take statistical averages, introducing the $n$-point correlation functions (again imposing symmetry constraints);
\beq
    \{...\} &=& n^4\left[4\xi(\vec x_1-\vec x_3) + 4\zeta(\vec x_1,\vec x_2,\vec x_3) + 2\xi(\vec x_1-\vec x_3)\xi(\vec x_2-\vec x_4) + \xi^{(4)}(\vec x_1,\vec x_2,\vec x_3,\vec x_4)\right]\\\nonumber
    &&+ 4n^3\delta_D(\vec x_1-\vec x_4)\left[1+\xi(\vec x_1-\vec x_2)+\xi(\vec x_1-\vec x_3)+\xi(\vec x_2-\vec x_3)+\zeta(\vec x_1,\vec x_2,\vec x_3)\right]\\\nonumber
    &&+ 2n^2\delta_D(\vec x_1-\vec x_3)\delta_D(\vec x_2-\vec x_4)\left[1+\xi(\vec x_1-\vec x_2)\right],
\eeq
where $\zeta$ and $\xi^{(4)}$ are the 3PCF and 4PCF respectively, whose arguments must sum to zero. We have additionally used Wick's theorem to write the connected four-point correlator as $\av{\delta(\vec x_1)\delta(\vec x_2)\delta(\vec x_3)\delta(\vec x_4)}_c = \xi(\vec x_1-\vec x_2)\xi(\vec x_3-\vec x_4) + \text{2 perms.}$ 

We can now insert the above into the covariance (Eq.\,\ref{eq: cov-initial}), giving a sum of four-, three- and two-point terms;
\beq\label{eq: cov-234-expansion}
    \operatorname{cov}\left(\hat{P}(\vec k;R_0),\hat{P}(\vec k';R_0)\right) &\equiv& {}^4\mathcal{C}(\vec k,\vec k';R_0)+{}^3\mathcal{C}(\vec k,\vec k';R_0)+{}^2\mathcal{C}(\vec k,\vec k';R_0)\\\nonumber
    {}^4\mathcal{C}(\vec k,\vec k';R_0) &=& \frac{1}{V^2}\int d\vec x_1d\vec x_2d\vec x_3d\vec x_4\,K(\vec x_1-\vec x_2;\vec k,R_0)K(\vec x_3-\vec x_4;\vec k',R_0)\\\nonumber
    &&\times \left[4\xi(\vec x_1-\vec x_3) + 4\zeta(\vec x_1,\vec x_2,\vec x_3) + 2\xi(\vec x_1-\vec x_3)\xi(\vec x_2-\vec x_4) + \xi^{(4)}(\vec x_1,\vec x_2,\vec x_3,\vec x_4)\right]\\\nonumber
    {}^3\mathcal{C}(\vec k,\vec k';R_0) &=& \frac{4}{nV^2}\int d\vec x_1d\vec x_2d\vec x_3\,K(\vec x_1-\vec x_2;\vec k,R_0)K(\vec x_3-\vec x_1;\vec k',R_0)\\\nonumber
    &&\times \left[1+\xi(\vec x_1-\vec x_2)+\xi(\vec x_1-\vec x_3)+\xi(\vec x_2-\vec x_3)+\zeta(\vec x_1,\vec x_2,\vec x_3)\right]\\\nonumber
    {}^2\mathcal{C}(\vec k,\vec k';R_0) &=& \frac{2}{n^2V^2}\int d\vec x_1d\vec x_2\,K(\vec x_1-\vec x_2;\vec k,R_0)K(\vec x_1-\vec x_2;\vec k',R_0)\left[1+\xi(\vec x_1-\vec x_2)\right],
\eeq
where we have dropped the $\vec x_1\neq\vec x_2$ and $\vec x_3\neq \vec x_4$ which are no longer necessary since the fields are now continuous thanks to the Poisson averaging. Here, the separation of the two types of covariance is clear; the four-point term is independent of the number density of tracers and is thus an intrinsic covariance, whilst the three- and two-point terms scale as $n^{-1}$ and $n^{-2}$ respectively and are a Poisson covariance. In the limit of an infinitely sampled density field, these terms go to zero. We now proceed to evaluate the individual terms.

\subsection{Intrinsic Covariance}
For the four-point intrinsic covariance, a number of simplifications are possible. First, we note that the terms involving a single 2PCF and 3PCF are identically zero for a periodic survey. To see this for the 2PCF term, we rewrite the covariance using variables $\vec x = \vec x_1-\vec x_2$, $\vec y = \vec x_3-\vec x_4$ and $\vec z = \vec x_1-\vec x_3$ (invoking translation invariance);
\beq
    {}^4\mathcal{C}(\vec k,\vec k';R_0) &\supset& \frac{4}{V^2}\int d\vec x_2d\vec xd\vec yd\vec z\,K(\vec x;\vec k,R_0)K(\vec y;\vec k',R_0)\xi(\vec z)\\\nonumber
    &=& \frac{4}{V}\left[\int d\vec x\,K(\vec x;\vec k,R_0)\right]\left[\int d\vec y\,K(\vec y;\vec k',R_0)\right]\left[\int d\vec z\,\xi(\vec z)\right].
\eeq
Since the 2PCF is defined as the over-random probability for two galaxies to be separated by $\vec r$, $\int d\vec z\,\xi(\vec z) = 0$, so this term does not contribute to the covariance. A similar line of reasoning applies to the 3PCF term (setting $\vec x = \vec x_1-\vec x_3$, $\vec y = \vec x_1-\vec x_3$, $\vec z = \vec x_3-\vec x_4$);
\beq
    {}^4\mathcal{C}(\vec k,\vec k';R_0) &\supset& \frac{4}{V^2}\int d\vec x_3d\vec xd\vec yd\vec z\,K(\vec x;\vec k,R_0)K(\vec z;\vec k',R_0)\zeta(\vec x,\vec y)\\\nonumber
    &=& \frac{4}{V}\left[\int d\vec z\,K(\vec z;\vec k,R_0)\right]\left[\int d\vec x\,K(\vec x;\vec k',R_0)\int d\vec y\,\zeta(\vec x,\vec y)\right] = 0,
\eeq
since the integral of the 3PCF over one of its arguments is zero.\footnote{This follows by noting that the over-random probability of finding three particles in a triangle described by the vectors $\vec x$ and $\vec y$ is defined in terms of the 2PCF and 3PCF as $\xi(\vec x)+\xi(\vec y)+\xi(\vec x-\vec y)+\zeta(\vec x, \vec y)$. Averaging over $\vec y$, recalling that $\int d\vec y\,\xi(\vec y) = \int d\vec y\,\xi(\vec x-\vec y) = 0$, gives $\xi(\vec x)+\int d\vec y\,\zeta(\vec x,\vec y)/V$. By definition, this should be equal to the \textit{two-point} over-random probability $\xi(\vec x)$, giving $\int d\vec y\,\zeta(\vec x,\vec y) = 0$.} We thus obtain
\beq\label{eq: intrinsic-cov-initial}
    {}^4\mathcal{C}(\vec k,\vec k';R_0) &=& {}^4\mathcal{C}_\mathrm{G}(\vec k,\vec k';R_0) + {}^4\mathcal{C}_\mathrm{NG}(\vec k,\vec k';R_0)\\\nonumber
    {}^4\mathcal{C}_\mathrm{G}(\vec k,\vec k';R_0) &=& \frac{2}{V^2}\int d\vec x_1d\vec x_2d\vec x_3d\vec x_4 W(\vec x_1-\vec x_2;R_0)W(\vec x_3-\vec x_4;R_0)e^{-i\vec k\cdot(\vec x_1-\vec x_2)}e^{-i\vec k'\cdot(\vec x_3-\vec x_4)}\xi(\vec x_1-\vec x_3)\xi(\vec x_2-\vec x_4)\\\nonumber
    {}^4\mathcal{C}_\mathrm{NG}(\vec k,\vec k';R_0) &=& \frac{1}{V^2}\int d\vec x_1d\vec x_2d\vec x_3d\vec x_4 W(\vec x_1-\vec x_2;R_0)W(\vec x_3-\vec x_4;R_0)e^{-i\vec k\cdot(\vec x_1-\vec x_2)}e^{-i\vec k'\cdot(\vec x_3-\vec x_4)}\xi^{(4)}(\vec x_1,\vec x_2,\vec x_3,\vec x_4),
\eeq
separating out the Gaussian and non-Gaussian components and inserting the definition of the kernel $K$.

We proceed by changing variables and integrating over one volume once (for the Gaussian part);
\beq
    {}^4\mathcal{C}_\mathrm{G}(\vec k,\vec k';R_0) &=& \frac{1}{V^2}\int d\vec rd\vec r'd\vec xd\vec x'\left[\xi(\vec x-\vec x')\xi(\vec x-\vec x'+\vec r-\vec r')+\xi(\vec x-\vec x'-\vec r')\xi(\vec x+\vec r-\vec x')\right]W(\vec r;R_0)W(\vec r';R_0)e^{-i\vec k\cdot\vec r}e^{-i\vec k'\cdot\vec r'}\\\nonumber
    &=& \frac{1}{V}\int d\vec rd\vec r'\left[\left[\xi\ast\xi\right](\vec r-\vec r')+\left[\xi\ast\xi\right](\vec r+\vec r')\right]W(\vec r;R_0)W(\vec r';R_0)e^{-i\vec k\cdot\vec r}e^{-i\vec k'\cdot\vec r'}\\\nonumber
    {}^4\mathcal{C}_\mathrm{NG}(\vec k,\vec k';R_0) &=& \frac{1}{V^2}\int d\vec rd\vec r'd\vec xd\vec x'\,\xi^{(4)}(\vec x, \vec x',\vec x+\vec r,\vec x'+\vec r')W(\vec r;R_0)W(\vec r';R_0)e^{-i\vec k\cdot\vec r}e^{-i\vec k'\cdot\vec r'},
\eeq
where we have used the convolution operator $\ast$ on the product of two 2PCFs. Note that we have separated the two 2PCF terms previously assumed to be symmetric; this will allow us to robustly show the (previously assumed) symmetry properties for even multipoles. We now write all quantities in terms of their Fourier-space counterparts and apply the convolution theorem;
\beq
    {}^4\mathcal{C}_\mathrm{G}(\vec k,\vec k';R_0)&=& \frac{1}{V}\int d\vec rd\vec r'\frac{d\vec p_1d\vec p_2d\vec p_3}{(2\pi)^9}P^2(\vec p_1)\widetilde{W}(\vec p_2;R_0)\widetilde{W}(\vec p_3;R_0)\left[e^{i\vec p_1\cdot(\vec r-\vec r')}+e^{i\vec p_1\cdot(\vec r+\vec r')}\right]e^{i(\vec p_2-\vec k)\cdot\vec r}e^{i(\vec p_3-\vec k')\cdot\vec r'}\\\nonumber
    &=& \frac{1}{V}\int\frac{d\vec p}{(2\pi)^3}P^2(\vec p)\widetilde{W}(\vec k-\vec p;R_0)\left[\widetilde{W}(\vec k'+\vec p;R_0)+\widetilde{W}(\vec k'-\vec p;R_0)\right]\\\nonumber
    {}^4\mathcal{C}_\mathrm{NG}(\vec k,\vec k';R_0) &=& \frac{1}{V^2}\int d\vec rd\vec r'd\vec xd\vec x'\left[\prod_{i=1}^6\frac{d\vec p_i}{(2\pi)^3}\right]T(\vec p_1,\vec p_2,\vec p_3,\vec p_4)\delta_D(\vec p_1+\vec p_2+\vec p_3+\vec p_4)\widetilde{W}(\vec p_5;R_0)\widetilde{W}(\vec p_6;R_0)\\\nonumber
    &&\times\, e^{i(\vec p_1+\vec p_3)\cdot\vec x}e^{i(\vec p_2+\vec p_4)\cdot\vec x'}e^{i(\vec p_3+\vec p_5-\vec k)\cdot\vec r}e^{i(\vec p_4+\vec p_6-\vec k')\cdot\vec r'}\\\nonumber
    &=& \frac{1}{V}\int\frac{d\vec p_1d\vec p_2}{(2\pi)^6}T(\vec p_1,-\vec p_1,\vec p_2,-\vec p_2)\widetilde{W}(\vec k+\vec p_1;R_0)W(\vec k'+\vec p_2;R_0),
\eeq
where $\widetilde{W}$ is the Fourier transform of $W$ and we introduce the trispectrum $T$, integrating over the exponentials and resulting Dirac deltas.

In the limit $R_0\rightarrow\infty$ (corresponding to no windowing), $\widetilde{W}(\vec p)\rightarrow (2\pi)^3\delta_D(\vec p)$, thus the covariances tend to the familiar form \citep[e.g.,\,][]{1999ApJ...527....1S};
\beq
    \lim_{R_0\rightarrow\infty}\mathcal{C}_\mathrm{G}(\vec k,\vec k';R_0) &=& \frac{1}{V}P^2(\vec k)(2\pi)^3\left[\delta_D(\vec k+\vec k')+\delta_D(\vec k-\vec k')\right]\\\nonumber
    \lim_{R_0\rightarrow\infty}\mathcal{C}_\mathrm{NG}(\vec k,\vec k';R_0) &=& \frac{1}{V}T(\vec k,-\vec k,\vec k',-\vec k').
\eeq

From this, we may additionally consider the four-point covariance of the Legendre multipoles of $P(\vec k;R_0)$, via
\beq
     \operatorname{cov}\left(P_\ell(k;R_0),P_{\ell'}(k';R_0)\right) &\supset& {}^4\mathcal{C}^\mathrm{G}_{\ell\ell'}(k,k';R_0)+{}^4\mathcal{C}^\mathrm{NG}_{\ell\ell'}(k,k';R_0)\\\nonumber
     &=& (2\ell+1)(2\ell'+1)\int \frac{d\Omega_k}{4\pi}\frac{d\Omega_{k'}}{4\pi}L_\ell(\hat{\vec k}\cdot\hat{\vec n})L_{\ell'}(\hat{\vec k}'\cdot\hat{\vec n})\times {}^4\mathcal{C}(\vec k,\vec k';R_0),
\eeq
where $\Omega_k$ is the angular part of $\vec k$ and $\hat{\vec n}$ is the (fixed) line-of-sight vector. Considering the first-part of the Gaussian covariance, and writing $\widetilde{W}$ in terms of its inverse Fourier transform;
\beq
    {}^4\mathcal{C}^\mathrm{G-I}_{\ell\ell'}(k,k';R_0) = \frac{(2\ell+1)(2\ell'+1)}{V}\int\frac{d\vec p}{(2\pi)^3}P^2(\vec p)\int d\vec rd\vec r'W(\vec r;R_0)W(\vec r';R_0)e^{i\vec p\cdot(\vec r-\vec r')}
    \int \frac{d\Omega_kd\Omega_{k'}}{(4\pi)^2}e^{-i\vec k\cdot\vec r}e^{-i\vec k'\cdot\vec r'}L_\ell(\hat{\vec k}\cdot\hat{\vec n})L_{\ell'}(\hat{\vec k}'\cdot\hat{\vec n}).
\eeq
This is simplified with the relation
\beq
    \int\frac{d\Omega_k}{4\pi}e^{-i\vec k\cdot\vec r}L_\ell(\hat{\vec k}\cdot\hat{\vec n}) &=& \sum_L(2L+1)(-i)^Lj_L(kr)\int\frac{d\Omega_k}{4\pi}L_L(\hat{\vec k}\cdot\hat{\vec r})L_\ell(\hat{\vec k}\cdot\hat{\vec n})\\\nonumber
    &=& \sum_L(2L+1)(-i)^Lj_L(kr)\left[\delta^K_{\ell L}\frac{4\pi}{2\ell+1}L_\ell(\hat{\vec r}\cdot\hat{\vec n})\right] = (-i)^\ell j_\ell(kr)L_\ell(\hat{\vec r}\cdot\hat{\vec n}),
\eeq
via the plane wave expansion and Legendre polynomial orthogonality. This implies
\beq
    {}^4\mathcal{C}^\mathrm{G-I}_{\ell\ell'}(k,k';R_0) = \frac{(2\ell+1)(2\ell'+1)}{V}(-i)^{\ell+\ell'}\int\frac{d\vec p}{(2\pi)^3}P^2(\vec p)\int d\vec rd\vec r'\,W(\vec r;R_0)W(\vec r';R_0)e^{i\vec p\cdot(\vec r-\vec r')}j_\ell(kr)j_{\ell'}(k'r')L_{\ell}(\hat{\vec r}\cdot\hat{\vec n})L_{\ell'}(\hat{\vec r}'\cdot\hat{\vec n}).
\eeq
setting $k=k'$. Following similar logic, we can compute the second Gaussian term (with $\vec k'\rightarrow-\vec k'$) which is identical except for $i^{\ell'}\rightarrow (-i)^{\ell'}$.

We may proceed by performing the angular integrals over $\vec r$ and $\vec r'$ (noting that $W$ is isotropic), which have a similar form to the above;
\beq\label{eq: integ-L-exp}
    \int d\Omega_r e^{i\vec p\cdot\vec r}L_\ell(\hat{\vec r}\cdot\hat{\vec n}) = 4\pi i^\ell j_\ell(pr)L_\ell(\hat{\vec p}\cdot\hat{\vec n}),
\eeq
leaving us with
\beq
    {}^4\mathcal{C}^\mathrm{G}_{\ell\ell'}(k,k';R_0) = \frac{(2\ell+1)(2\ell'+1)}{V}\left[(-1)^{\ell'}+1\right]\int\frac{d\vec p}{(2\pi)^3}P^2(\vec p)\omega_\ell(p;k,R_0)\omega_{\ell'}(p;k',R_0)L_{\ell}(\hat{\vec p}\cdot\hat{\vec n})L_{\ell'}(\hat{\vec p}\cdot\hat{\vec n}),
\eeq
defining the $\omega_\ell$ function;
\beq\label{eq: omega-ell-def}
    \omega_\ell(p;k,R_0) = 4\pi\int_0^{R_0} r^2dr\,j_\ell(kr)j_\ell(pr)W(r;R_0),
\eeq
noting that this is sharply peaked at $p\sim k$ for large $R_0$.\footnote{Due to the form of the window function, $\omega_\ell$ is analytic, and may be computed by inserting the definition of $W(r;R_0)$ then integrating over the resulting polynomial-weighted pair of spherical Bessel functions using the indefinite integral results given in \citet{2017arXiv170306428B}.} Note that this recovers the aforementioned symmetry for even $\ell'$ (or by symmetry $\ell$).

The next step is to integrate over the angular part of $\vec p$, though this requires knowledge of the angular dependence of $P^2(\vec p)$. To do this, we use the fact that the multipoles of $P^2$ are related to those of $P$ via
\beq
    \left[P^2\right]_L(p) &=& \int \frac{d\Omega_p}{4\pi}L_L(\hat{\vec p}\cdot\hat{\vec n})P^2(\vec p)= (2L+1)\sum_{\ell_1\ell_2}\tj{\ell_1}{\ell_2}{L}{0}{0}{0}^2P_{\ell_1}(p)P_{\ell_2}(p),
\eeq
where the term in parentheses is a Wigner 3j symbol \citep[Eq.\,34.2.4]{nist_dlmf}, assuming $\{\ell_1,\ell_2,L\}$ to obey triangle conditions, using the result
\beq\label{eq: triple-legendre-integral}
    \int d\Omega_p L_{\ell_1}(\hat{\vec p}\cdot\hat{\vec n})L_{\ell_2}(\hat{\vec p}\cdot\hat{\vec n})L_{\ell_3}(\hat{\vec p}\cdot\hat{\vec n}) = 4\pi \tj{\ell_1}{\ell_2}{\ell_3}{0}{0}{0}^2,
\eeq
via the Gaunt integral \citep[Eq.\,34.3.21]{nist_dlmf}. Inserting this relation and integrating over $\vec p$ (again via Eq.\,\ref{eq: triple-legendre-integral}) gives the final result
\beq
    {}^4\mathcal{C}^\mathrm{G}_{\ell\ell'}(k,k';R_0) &=& 2\frac{(2\ell+1)(2\ell'+1)}{V}\int \frac{p^2dp}{2\pi^2}\,\omega_\ell(p;k,R_0)\omega_{\ell'}(p;k',R_0)Q_{\ell\ell'}(p)
\eeq
(assuming $\ell,\ell'$ to be even), where the symmetric function $Q_{\ell\ell'}$ is defined in terms of the multipoles of $P(\vec k)$ by
\beq
    Q_{\ell\ell'}(p) = \sum_{L\ell_1\ell_2}(2L+1)\tj{\ell}{\ell'}{L}{0}{0}{0}^2\tj{\ell_1}{\ell_2}{L}{0}{0}{0}^2P_{\ell_1}(p)P_{\ell_2}(p).
\eeq
Whilst this may seem complex, the various symmetries required for a non-zero 3j symbol mean that the expression is relatively manageable at small $\ell,\ell'$, with important values including
\beq
    Q_{00}(p) &=& P_0^2(p)+\frac{1}{5}P_2^2(p)+...\\\nonumber
    Q_{02}(p) &=& \frac{2}{5}P_0(p)P_2(p)+\frac{2}{35}P_2^2(p)+...\\\nonumber
    Q_{22}(p) &=& \frac{1}{5}P_0^2(p)+\frac{4}{35}P_0(p)P_2(p)+\frac{3}{35}P_2^2(p)+...
\eeq
ignoring terms above the quadrupole for brevity. If we assume an isotropic field, such that $P_\ell=0$ for $\ell>0$, we obtain $Q_{\ell\ell'}(k) = (2\ell+1)^{-1}\delta^K_{\ell\ell'}P_0(k)$. Taking the $R_0\rightarrow\infty$ limit (i.e. replacing $\omega_\ell$ with Eq.\,\ref{eq: omega-ell-lim}) yields
\beq
    \lim_{R_0\rightarrow\infty}{}^4\mathcal{C}^\mathrm{G}_{\ell\ell'}(k,k';R_0) &=& 2\frac{(2\ell+1)(2\ell'+1)}{V}Q_{\ell\ell'}(k)\times \frac{2\pi^2}{k^2}\delta_D(k-k')\\\nonumber
    &=& \left[2\delta^K_{\ell\ell'}\frac{2\ell+1}{V}P^2_0(k)+...\right]\times \frac{2\pi^2}{k^2}\delta_D(k-k'),
\eeq
reproducing familiar results \citep[e.g.,\,][]{2019arXiv190806234S,2019JCAP...01..016L,2019arXiv191002914W}, with the final line obtained by assuming the power to be dominated by the monopole $P_0$.

For the non-Gaussian part, a similar derivation is possible, again expressing the $\widetilde{W}$ functions in terms of their inverse Fourier transforms and integrating over the Legendre polynomials. Following some algebra, we obtain the result
\beq
    {}^4\mathcal{C}^\mathrm{NG}_{\ell\ell'}(k,k';R_0) &=& \frac{(2\ell+1)(2\ell'+1)}{V}(-1)^{\ell+\ell'}\int \frac{p_1^2dp_1}{2\pi^2}\frac{p_2^2dp_2}{2\pi^2}\overline{T}_{\ell\ell'}(p_1,p_2)\omega_\ell(p_1;k,R_0)\omega_{\ell'}(p_2;k',R_0),
\eeq
where $\overline{T}_{\ell\ell'}$ are the multipole moments of a collapsed trispectrum, defined as
\beq
    \overline{T}_{\ell\ell'}(p_1,p_2) = \int \frac{d\Omega_{p_1}}{4\pi}\frac{d\Omega_{p_2}}{4\pi}T(\vec p_1,-\vec p_1,\vec p_2,-\vec p_2)L_\ell(\hat{\vec p}_1\cdot\hat{\vec n})L_{\ell'}(\hat{\vec p}_2\cdot\hat{\vec n}).
\eeq
As $R_0\rightarrow\infty$, this has the limit
\beq
    \lim_{R_0\rightarrow\infty}{}^4\mathcal{C}^\mathrm{NG}_{\ell\ell'}(k,k';R_0) = \frac{(2\ell+1)(2\ell'+1)}{V}(-1)^{\ell+\ell'}\overline{T}_{\ell\ell'}(k,k'),
\eeq
again agreeing with the standard form.

\subsection{Poisson Covariance}
As previously mentioned, the four-point term is not the only contributor to the covariance; the limited number of tracer particles give two- and three-point terms, which vanish in the limit of infinite $n$. We start with the three-point term of Eq.\,\ref{eq: cov-234-expansion}, relabeling variables by translation invariance;
\beq
    {}^3\mathcal{C}(\vec k,\vec k';R_0) &=& \frac{4}{nV^2}\int d\vec x_1d\vec xd\vec y\,K(\vec x;\vec k,R_0)K(\vec y;\vec k',R_0)\left[1+\xi(\vec x)+\xi(\vec y)+\xi(\vec x-\vec y)+\zeta(\vec x,\vec y)\right].
\eeq
Several terms are simplified by noting
\beq
    \int d\vec x\,K(\vec x;\vec k,R_0) &=& \int d\vec x\,e^{-i\vec k\cdot\vec x}W(\vec x;R_0) \equiv \widetilde{W}(\vec k;R_0)\\\nonumber
    \int d\vec x\,K(\vec x;\vec k,R_0)\xi(\vec x) &=& \int d\vec x\,e^{-i\vec k\cdot\vec x}W(\vec x;R_0)\xi(\vec x) \equiv P(\vec k;R_0)\\\nonumber
    \int d\vec x\,d\vec y\,K(\vec x;\vec k,R_0)K(\vec y;\vec k',R_0)\zeta(\vec x,\vec y) &=& \int d\vec xd\vec ye^{-i\vec k\cdot\vec x}e^{-i\vec k'\cdot\vec y}W(\vec x;R_0)W(\vec y;R_0)\zeta(\vec x,\vec y) \equiv B(\vec k,\vec k';R_0),
\eeq
thus
\beq
    {}^3\mathcal{C}(\vec k,\vec k';R_0) &=& \frac{4}{nV}\left[\widetilde{W}(\vec k;R_0)\widetilde{W}(\vec k';R_0)+\widetilde{W}(\vec k;R_0)P(\vec k';R_0)+\widetilde{W}(\vec k';R_0)P(\vec k;R_0)+ B(\vec k,\vec k';R_0)\right.\\\nonumber
    && + \quad\left.\int d\vec xd\vec y\,K(\vec x;\vec k,R_0)K(\vec y;\vec k',R_0)\xi(\vec x-\vec y)\right].
\eeq
We expect the bispectrum term to be subdominant to the power spectrum terms, thus this can usually be ignored, though we keep it in the below for completeness. For the first four terms, it is straightforward to extract the Legendre multipoles;\footnote{Only the bispectrum term is non-trivial; this may be computed by expanding $B(\vec k,\vec k';R_0) = \sum_L B_L(k,k';R_0)L_L(\hat{\vec k}\cdot\hat{\vec k}')$ and using spherical harmonic theorems.}
\beq
    {}^3\mathcal{C}_{\ell\ell'}(k,k';R_0) &\supset& \frac{4}{nV}\left[\widetilde{W}(k;R_0)\widetilde{W}(k';R_0)\delta^K_{\ell0}\delta^K_{\ell'0}+\widetilde{W}(k;R_0)P_{\ell'}(k';R_0)\delta^K_{\ell0}+\widetilde{W}(k';R_0)P_\ell(k;R_0)\delta^K_{\ell'0}+B_\ell(k,k';R_0)\delta^K_{\ell\ell'}\right].
\eeq
The remaining three-point term may be rewritten by expressing the 2PCF in Fourier space;
\beq
    {}^3\mathcal{C}_{\ell\ell'}(k,k';R_0) &\supset& \frac{4}{nV}(2\ell+1)(2\ell'+1)\int \frac{d\vec p}{(2\pi)^3}P(\vec p)\int d\vec xd\vec y\,W(\vec x;R_0)W(\vec y;R_0)\int \frac{d\Omega_kd\Omega_{k'}}{(4\pi)^2}L_\ell(\hat{\vec k}\cdot\hat{\vec n})L_\ell(\hat{\vec k}'\cdot\hat{\vec n})\,e^{-i(\vec k-\vec p)\cdot\vec x}e^{-i(\vec k'+\vec p)\cdot\vec y}.
\eeq
Next we use the result that
\beq
    \int d\vec x\,W(\vec x;R_0)\int \frac{d\Omega_k}{4\pi}L_\ell(\hat{\vec k}\cdot\hat{\vec n})e^{-i(\vec k-\vec p)\cdot\vec x} &=& \int d\vec x\,W(\vec x;R_0)e^{i\vec p\cdot\vec x}\sum_L(-i)^Lj_L(kx)\left[\int \frac{d\Omega_k}{4\pi}(2L+1)L_\ell(\hat{\vec k}\cdot\hat{\vec n})L_L(\hat{\vec k}\cdot\hat{\vec x})\right]\\\nonumber
    &=& \int d\vec x\,W(\vec x;R_0)e^{i\vec p\cdot\vec x}(-i)^\ell j_\ell(kx)L_\ell(\hat{\vec x}\cdot\hat{\vec n})\\\nonumber
    &=& \int x^2dx\,W(x;R_0)j_\ell(kx)\sum_L j_L(px)i^{L-\ell}\left[\int d\Omega_x(2L+1)L_\ell(\hat{\vec x}\cdot\hat{\vec n})L_L(\hat{\vec p}\cdot\hat{\vec x})\right]\\\nonumber
    &=& 4\pi\int x^2dx\,W(x;R_0)j_\ell(kx)j_\ell(px)L_\ell(\hat{\vec p}\cdot\hat{\vec n}) = \omega_\ell(p;k,R_0)L_\ell(\hat{\vec p}\cdot\hat{\vec n})
\eeq
utilizing Legendre polynomial completeness and the plane wave expansion, giving
\beq
    {}^3\mathcal{C}_{\ell\ell'}(k,k';R_0) &\supset& \frac{4}{nV}(2\ell+1)(2\ell'+1)(-1)^{\ell'}\int \frac{d\vec p}{(2\pi)^3}P(\vec p)\omega_\ell(p;k,R_0)\omega_{\ell'}(p;k',R_0)L_\ell(\hat{\vec p}\cdot\hat{\vec n})L_{\ell'}(\hat{\vec p}\cdot\hat{\vec n})\\\nonumber
    &=& \frac{4}{nV}(2\ell+1)(2\ell'+1)(-1)^\ell \int \frac{p^2dp}{2\pi^2}\sum_LP_L(p)\omega_\ell(p;k,R_0)\omega_{\ell'}(p;k',R_0)\left[\int \frac{d\Omega_p}{4\pi}L_L(\hat{\vec p}\cdot\hat{\vec n})L_\ell(\hat{\vec p}\cdot\hat{\vec n})L_{\ell'}(\hat{\vec p}\cdot\hat{\vec n})\right]\\\nonumber
    &=& \frac{4}{nV}(2\ell+1)(2\ell'+1)(-1)^\ell \sum_L\tj{L}{\ell}{\ell'}{0}{0}{0}^2\int \frac{p^2dp}{2\pi^2}P_L(p)\omega_\ell(p;k,R_0)\omega_\ell(p;k',R_0),
\eeq
where we have expressed $P(\vec p)$ in terms of its multipoles and used the result of Eq.\,\ref{eq: triple-legendre-integral} to evaluate the integral over three Legendre polynomials. Note that, for large $R_0$, the $\omega$ functions will be sharply peaked, enforcing $p\approx k\approx k'$. In the limit of $R_0\rightarrow\infty$, we obtain
\beq
    \lim_{R_0\rightarrow\infty}{}^3\mathcal{C}_{\ell\ell'}(k,k';R_0) = \frac{4}{nV}(2\ell+1)(2\ell'+1)(-1)^\ell \frac{2\pi^2}{k^2}\delta_D(k-k')\sum_L\tj{L}{\ell}{\ell'}{0}{0}{0}^2 P_L(k)+\frac{4}{nV}\delta_{\ell\ell'}^KB_\ell(k,k';R_0),
\eeq
noting that the other three-point terms only contribute to the zero-lag covariance ($k=0$ and/or $k'=0$) which has been ignored.

Computing the two-point term proceeds similarly, starting from
\beq
    {}^2\mathcal{C}(\vec k,\vec k';R_0) = \frac{4}{n^2V^2}\int d\vec x_1d\vec x\,K(\vec x;\vec k,R_0)K(\vec x;\vec k',R_0)\left[1+\xi(\vec x)\right],
\eeq
thus
\beq
    {}^2\mathcal{C}_{\ell\ell'}(k,k';R_0) &=& \frac{4}{n^2V}(2\ell+1)(2\ell'+1)\int d\vec x\int \frac{d\Omega_kd\Omega_{k'}}{4\pi^2}e^{-i(\vec k+\vec k')\cdot\vec x}W^2(\vec x;R_0)\left[1+\xi(\vec x)\right]L_\ell(\hat{\vec k}\cdot\hat{\vec n})L_{\ell'}(\hat{\vec k'}\cdot\hat{\vec n})\\\nonumber
    &=& \frac{4}{n^2V}(2\ell+1)(2\ell'+1)(-i)^{\ell+\ell'}\int x^2dx\,W^2(x;R_0)j_\ell(kx)j_{\ell'}(k'x)\int d\Omega_x\,L_\ell(\hat{\vec x}\cdot\hat{\vec n})L_{\ell'}(\hat{\vec x}\cdot\hat{\vec n})\left[1+\xi(\vec x)\right],
\eeq
using Eq.\,\ref{eq: integ-L-exp}. The integral over $\Omega_x$ can be performed via Legendre polynomial completeness (for the first term) and using Eq.\,\ref{eq: triple-legendre-integral} for the second, expressing $\xi(\vec x)$ in terms of its multipoles $\xi_\ell(x)$. This yields
\beq
    {}^2\mathcal{C}_{\ell\ell'}(k,k';R_0) &=& \frac{4}{n^2V}(2\ell+1)(2\ell'+1)(-i)^{\ell+\ell'}4\pi\int x^2dx\,W^2(x;R_0)j_\ell(kx)j_{\ell'}(k'x)\left[\frac{\delta^K_{\ell\ell'}}{2\ell+1}+\sum_L\tj{L}{\ell}{\ell'}{0}{0}{0}^2\xi_L(x)\right].
\eeq
Whilst the $R_0\rightarrow\infty$ limit of this expression is not informative, the unbinned version is more so;
\beq
    \lim_{R_0\rightarrow\infty}{}^2\mathcal{C}(\vec k,\vec k';R_0) &=& \frac{4}{n^2V}\left[(2\pi)^3\delta_D(\vec k+\vec k')+P(\vec k+\vec k')\right],
\eeq
consisting of a constant term on the $\vec k +\vec k'=\vec 0$ diagonal and a (subdominant) off-diagonal term.

\subsection{$k$-space Binning and Summary}
For proper comparison with data, we ought to consider the covariance of the binned power spectra, $P^a_\ell(R_0)$, which is related to the standard covariance via
\beq
    \operatorname{cov}\left(P^a_\ell(R_0),P^b_{\ell'}(R_0)\right) &=& \frac{4\pi}{v_a}\int k^2dk\,\Theta^a(k)\frac{4\pi}{v_b}\int k'^2dk'\,\Theta^b(k')\times \operatorname{cov}\left(P_\ell(k;R_0),P_{\ell'}(k',R_0)\right).
\eeq
Since $k$ and $k'$ only enter the covariance through the $\omega_\ell$ functions, introducing binning is equivalent to replacing $\omega_\ell(p;k,R_0)$ with $\omega_\ell^a(p;R_0)$ (and similarly for $k'$), where $\omega_\ell^a$ is defined as
\beq\label{eq: omega-ell-a-def}
    \omega_\ell^a(p;R_0) &\equiv& 4\pi\int_0^{R_0}r^2dr\,j_\ell(pr)j_\ell^a(r)W(r;R_0)
\eeq
For the window function of Eq.\,\ref{eq: window_defn}, this is analytic (and computable in terms of Bessel function recursion relations, as in \citealt{2017arXiv170306428B}), yet complex. The final intrinsic covariance is hence
\beq\label{eq: cov-int}
    \left.\operatorname{cov}\left(P^a_\ell(R_0),P^b_{\ell'}(R_0)\right)\right|_\mathrm{intrinsic}&\equiv&{}^4\mathcal{C}_{\ell\ell'}^{\mathrm G,ab}(R_0)+{}^4\mathcal{C}_{\ell\ell'}^{\mathrm{NG},ab}(R_0)\\\nonumber
    {}^4\mathcal{C}_{\ell\ell'}^{\mathrm G,ab}(R_0) &=& 2\frac{(2\ell+1)(2\ell'+1)}{V}\int \frac{p^2dp}{2\pi^2}\omega_l^a(p;R_0)\omega_{\ell'}^b(p;R_0)Q_{\ell\ell'}(p)\\\nonumber
    {}^4\mathcal{C}_{\ell\ell'}^{\mathrm{NG},ab}(R_0) &=& \frac{(2\ell+1)(2\ell'+1)}{V}(-1)^{\ell+\ell'}\int \frac{p_1^2dp_1}{2\pi^2}\int\frac{p_2^2dp_2}{2\pi^2}\overline{T}_{\ell\ell'}(p_1,p_2)\omega^a_\ell(p_1;R_0)\omega^b_{\ell'}(p_2;R_0).
\eeq
In the limit of $R_0\rightarrow\infty$, we obtain
\beq\label{eq: cov-int-ideal}
    \lim_{R_0\rightarrow\infty}{}^4\mathcal{C}_{\ell\ell'}^{\mathrm G,ab}(R_0) &=& 2(2\pi)^3\delta^K_{ab}\frac{(2\ell+1)(2\ell'+1)}{Vv_a}Q^a_{\ell\ell'}\\\nonumber
    &\approx& 2(2\pi)^3\delta^K_{ab}\delta^K_{\ell\ell'}\frac{2\ell+1}{Vv_a}\left[P_0^2\right]^a\\\nonumber
    \lim_{R_0\rightarrow\infty}{}^4\mathcal{C}_{\ell\ell'}^{\mathrm{NG},ab}(R_0) &=& \frac{(2\ell+1)(2\ell'+1)}{V}(-1)^{\ell+\ell'}\overline{T}^{ab}_{\ell\ell'},
\eeq
where $Q^a_{\ell\ell'}$, $\left[P^2_0\right]^a$ and $\overline{T}^{ab}_{\ell\ell'}$ are the binned forms of $Q_{\ell\ell'}(k)$, $P^2_0(k)$ and $\overline{T}_{\ell\ell'}(k,k')$ respectively. The approximate form of the Gaussian covariance is derived assuming the power to be dominated by the monopole. This matches standard results \citep[e.g.,\,][]{1999ApJ...527....1S,2019arXiv190806234S,2019arXiv191002914W,2019JCAP...01..016L}, noting that $Vv_a/(2\pi)^3$ is equal to the number of $k$-space modes in the bin.

For the Poisson covariance, computation is similar, yielding
\beq\label{eq: cov-poiss}
    \left.\operatorname{cov}\left(P^a_\ell(R_0),P^b_{\ell'}(R_0)\right)\right|_\mathrm{Poisson} &\equiv& {}^3\mathcal{C}_{\ell\ell'}^{ab}(R_0)+{}^2\mathcal{C}_{\ell\ell'}^{ab}(R_0)\\\nonumber
    {}^3\mathcal{C}_{\ell\ell'}^{ab}(R_0) &=& \frac{4}{nV}\left[\widetilde{W}^a(R_0)\widetilde{W}^b(R_0)\delta^K_{\ell0}\delta^K_{\ell'0}+\widetilde{W}^a(R_0)P^b_{\ell'}(R_0)\delta^K_{\ell0}+\widetilde{W}^b(R_0)P^a_{\ell}(R_0)\delta^K_{\ell'0}+B^{ab}_\ell(R_0)\delta_{\ell\ell'}^{K}\right]\\\nonumber
    &&+ \frac{4}{nV}(2\ell+1)(2\ell'+1)(-1)^\ell\sum_L\tj{L}{\ell}{\ell'}{0}{0}{0}^2\int \frac{p^2dp}{2\pi^2}P_L(p)\omega^a_\ell(p;R_0)\omega^b_{\ell'}(p;R_0)\\\nonumber
    {}^2\mathcal{C}_{\ell\ell'}^{ab}(R_0) &=& \frac{4}{n^2V}(2\ell+1)(2\ell'+1)(-i)^{\ell+\ell'}4\pi\int x^2dx\,W^2(x;R_0)j_\ell^a(x)j_{\ell'}^b(x)\left[\frac{\delta^K_{\ell\ell'}}{2\ell+1}+\sum_L\tj{L}{\ell}{\ell'}{0}{0}{0}^2\xi_L(x)\right].
\eeq
In the limit of $R_0\rightarrow\infty$, and ignoring zero-lag terms, we obtain
\beq\label{eq: cov-poiss-ideal}
    \lim_{R_0\rightarrow\infty}{}^3\mathcal{C}_{\ell\ell'}^{ab}(R_0) &=& \frac{4}{nV}\frac{(2\pi)^3}{v_a}\delta^K_{ab}(2\ell+1)(2\ell'+1)(-1)^\ell\sum_L\tj{L}{\ell}{\ell'}{0}{0}{0}^2P_L^a+\frac{4}{nV}B^{ab}_\ell\delta^K_{\ell\ell'}\\\nonumber
    \lim_{R_0\rightarrow\infty}{}^2\mathcal{C}_{\ell\ell'}^{ab}(R_0) &=& \frac{4}{n^2V}\frac{(2\pi)^3}{v_a}\delta^K_{ab}\delta^K_{\ell\ell'}(2\ell+1)(-1)^\ell\\\nonumber
    &&+\frac{4}{n^2V}(2\ell+1)(2\ell'+1)(-i)^{\ell+\ell'}4\pi\int x^2dx\,j_\ell^a(x)j_\ell^b(x)\sum_L\tj{L}{\ell}{\ell'}{0}{0}{0}^2\xi_L(x).
\eeq
This completes the covariance computation.

\section{Cross-Covariance of the Power Spectrum and Bispectrum}\label{sec: cov-pk-bk}
Before considering the covariance of the bispectrum itself, we give brief results pertaining to the cross-covariance of the power spectrum multipoles $P_\ell(k)$ and the isotropic bispectrum multipoles $B_\ell(k_1,k_2)$. Due to the greater mathematical complexities of this compared to the power spectrum covariance, we will consider only Gaussian terms (both intrinsic and Poissonian) and work in the $R_0\rightarrow\infty$ limit (additionally assuming infinite randoms). The principal effect of finite $R_0$ is to add off-diagonal contributions (as well as to modify the low-$k$ covariances), but we expect it to be small for wide $k$-bins and moderate $R_0$. We further avoid complexity by ignoring $k$-space binning in this section.

In the $R_0\rightarrow\infty$ limit, our covariances will be identical to those obtained from conventional analyses. For this reason, we take our starting point to be the  cross-covariance presented in \citet[Sec.\,2.4]{2019arXiv190806234S}, which include contributions from shot-noise but not the projection onto the Legendre multipole basis. In our notation, and dropping all non-Gaussian terms, we can write
\newcommand{\PN}{P^{(\mathrm{N})}}
\newcommand{\BN}{B^{(\mathrm{N})}}
\newcommand{\TN}{T^{(\mathrm{N})}}
\beq\label{eq: pb-cov}
    \mathrm{cov}\left(\hat{P}(\vec k),\hat{B}(\vec k_1,\vec k_2)\right) &=& \frac{2}{V}(2\pi)^3\delta_D(\vec k+\vec k_1)\PN(\vec k_1)\BN(\vec k_1,\vec k_2,\vec k_3) + \text{2 cyc.}
\eeq
defining $\vec k_3=-\vec k_1-\vec k_2$, with cyclic interchanges performed over $\{\vec k_1',\vec k_2',\vec k_3'\}$. This uses the asymmetric definitions
\beq
    \PN(\vec k) &=& P(\vec k)+\frac{1}{n}\\\nonumber
    \BN(\vec k_1,\vec k_2,\vec k_3) &=& \frac{1}{n}\left(P(\vec k_2)+P(\vec k_3)\right)
\eeq
\citep{2019arXiv190806234S}, where we have again dropped all non-Gaussian terms. In this case, we note that there is no intrinsic Gaussian covariance, and further, that Eq.\,\ref{eq: pb-cov} mixes together terms of different orders in $n^{-1}$ for brevity. 

Projecting onto multipoles, we obtain
\beq
    \mathrm{cov}\left(\hat{P}_\ell(k),\hat{B}_{\ell'}(k_1,k_2)\right) &\equiv& (2\ell+1)(2\ell'+1)\int_{\Omega_{k_1}\Omega_{k_2}\Omega_{k}} L_\ell(\hat{\vec k}\cdot\hat{\vec n})L_{\ell'}(\hat{\vec k}_1\cdot\hat{\vec k}_2)\,\mathrm{cov}\left(\hat{P}(\vec k),\hat{B}(\vec k_1,\vec k_2)\right),
\eeq
for (fixed) line-of-sight vector $\hat{\vec n}$, writing $\int_{\Omega_k}\equiv \int \frac{d\Omega_k}{4\pi}$ for brevity. Inserting Eq.\,\ref{eq: pb-cov}, we find two terms with different structures under cyclic permutation;
\beq
    \mathrm{cov}\left(\hat{P}_\ell(k),\hat{B}_{\ell'}(k_1,k_2)\right) &\equiv& \mathcal{C}^A_{\ell\ell'}(k,k_1,k_2)+\mathcal{C}^B_{\ell\ell'}(k,k_1,k_2)\\\nonumber
      \frac{\mathcal{C}^A_{\ell\ell'}(k,k_1,k_2)}{(2\ell+1)(2\ell'+1)} &=& 
    \frac{2}{nV}\int_{\Omega_{k_1}\Omega_{k_2}\Omega_k}(2\pi)^3\delta_D(\vec k_1+\vec k_2-\vec k)L_\ell(\hat{\vec k}\cdot\hat{\vec n})L_{\ell'}(\hat{\vec k}_1\cdot\hat{\vec k}_2)\PN(\vec k)\left[P(\vec k_1)+P(\vec k_2)\right]\\\nonumber
     \frac{\mathcal{C}^B_{\ell\ell'}(k,k_1,k_2)}{(2\ell+1)(2\ell'+1)} &=& \frac{2}{nV}\delta_D^{k_1k}\int_{\vec k_3}\int_{\Omega_{k_1}\Omega_{k_2}}(2\pi)^3\delta_D(\vec k_1+\vec k_2+\vec k_3)(-1)^\ell L_\ell(\hat{\vec k}_1\cdot\hat{\vec n})L_{\ell'}(\hat{\vec k}_1\cdot\hat{\vec k}_2)\PN(-\vec k_1)\left[P(\vec k_2)+P(\vec k_3)\right]+(k_1\leftrightarrow k_2),
\eeq
where we have introduced $\vec k_3$ via a Dirac function, and denoted $\delta_D^{k_1k} \equiv 2\pi^2/(k_1k)\times \delta_D(k_1-k)$, i.e. the radial part of the Dirac function. To simplify this we can expand the spectra in terms of their multipoles, i.e. $P(\vec k)\equiv \sum_LP_L(k)L_L(\hat{\vec k}\cdot\hat{\vec n})$ and compute the integrals over Legendre polynomials analytically. Whilst the full calculation is somewhat lengthy, it yields the following results
\beq
    \frac{\mathcal{C}^A_{\ell\ell'}(k,k_1,k_2)}{(2\ell+1)(2\ell'+1)} &=& \frac{2}{nV}\sum_{LL'P}(-1)^{\ell'+P-L'}\mathcal{R}_{L'P\ell'}(k,k_1,k_2)\tj{\ell'}{P}{L'}{0}{0}{0}^2\tj{L'}{\ell}{L}{0}{0}{0}^2\PN_L(k)P_{L'}(k_1)+(k_1\leftrightarrow k_2)\\\nonumber
    \frac{\mathcal{C}^B_{\ell\ell'}(k,k_1,k_2)}{(2\ell+1)(2\ell'+1)} &=& \frac{2}{nV}\delta_D^{k_1k}\left[\sum_{LL'P}(2P+1)(-1)^{\ell+(\ell'+P)/2}\tj{L}{\ell}{L'}{0}{0}{0}^2\tj{\ell'}{P}{L'}{0}{0}{0}^2\mathcal{T}_{L'\ell';L'}(k_1,k_2)\PN_L(k_1)\right.\\\nonumber
    &&\,\quad\qquad +\left.\frac{1}{2\ell'+1}\sum_L(-1)^{\ell+L}\tj{\ell}{\ell'}{L}{0}{0}{0}^2\PN_L(k_1)P_{L'}(k_2)\right]+(k_1\leftrightarrow k_2),
\eeq
using $\int d\vec k\,P(\vec k)e^{i\vec k\cdot\vec x}\equiv \xi(\vec x)$ and defining
\beq\label{eq: T-def}
    \mathcal{T}_{\ell_1\ell_2;\ell_3}(k_1,k_2) &=& 4\pi\int x^2dx\,j_{\ell_1}(k_1x)j_{\ell_2}(k_2x)\xi_{\ell_3}(x).
\eeq
In the simple case of a $1/r^2$ 2PCF, this may be evaluated using \citet[Eqs.\,6.574.1-3]{2007tisp.book.....G}. In the general case, Eq.\,\ref{eq: T-def} can be evaluated via the prescriptions of \citet{2017JCAP...11..054A} or \citet{2019arXiv191200065S}.

\section{Idealized Bispectrum Covariance}\label{sec: cov-bk}
We now present a brief discussion of the bispectrum auto-covariance of the bispectrum, leading on from the preceding sections. Since the mathematics of this section is significantly more involved than the above, we will assume Gaussianity and the $R_0\rightarrow\infty$ limit (as for the cross-spectrum) and additionally work in real-space (which is equivalent to assuming that the power spectra are dominated by their monopole contributions). We defer consideration of the full covariance including all the above effects to future work. Starting from the results of \citet[Sec.\,2.5]{2019arXiv190806234S}, we obtain
\beq\label{eq: sugiyama-bk}
    \mathrm{cov}\left(\hat{B}(\vec k_1,\vec k_2),\hat{B}(\vec k_1',\vec k_2')\right) &\equiv& 
    \mathcal{C}^{PPP}(\vec k_1,\vec k_2,\vec k_1',\vec k_2')+\mathcal{C}^{BB}(\vec k_1,\vec k_2,\vec k_1',\vec k_2')+\mathcal{C}^{PT}(\vec k_1,\vec k_2,\vec k_1',\vec k_2')\\\nonumber
    \mathcal{C}^{PPP}(\vec k_1,\vec k_2,\vec k_1',\vec k_2') &=& \frac{1}{V}\PN(k_1)\PN(k_2)\PN(k_3)\left[(2\pi)^3\delta_D(\vec k_1+\vec k_1')(2\pi)^3\delta_D(\vec k_2+\vec k_2') + \text{5 perms.}\right]\\\nonumber
    \mathcal{C}^{BB}(\vec k_1,\vec k_2,\vec k_1',\vec k_2') &=& \frac{1}{V}\left[\BN(\vec k_1,\vec k_2,\vec k_3)\BN(\vec k_1',\vec k_2',\vec k_3')(2\pi)^3\delta_D(\vec k_1-\vec k_1') + \text{8 perms.}\right]\\\nonumber
    \mathcal{C}^{PT}(\vec k_1,\vec k_2,\vec k_1',\vec k_2') &=& \frac{1}{V}\left[\PN(k_1)\TN(\vec k_2,\vec k_3,\vec k_2',\vec k_3')(2\pi)^3\delta_D(\vec k_1+\vec k_1') + \text{8 perms.}\right],
\eeq
with permutations performed over $\{\vec k_1',\vec k_2',\vec k_3'\}$ as before. This uses the additional (symmetrized) definition
\beq
    \TN(\vec k_1,\vec k_2,\vec k_1',\vec k_2') &=& \frac{1}{2n^2}\left[P(|\vec k_1+\vec k_1'|)+P(|\vec k_1+\vec k_2'|+P(|\vec k_2+\vec k_1'|)+P(|\vec k_2+\vec k_2'|)\right].
\eeq
The multipole covariance follows from
\beq
    \mathrm{cov}\left(\hat{B}_\ell(k_1,k_2),\hat{B}_{\ell'}(k_1',k_2')\right) &\equiv&  (2\ell+1)(2\ell'+1)\int_{\Omega_{k_1}\Omega_{k_2}\Omega_{k_1'}\Omega_{k_2'}}L_\ell(\hat{\vec k}_1\cdot\hat{\vec k}_2)L_{\ell'}(\hat{\vec k}'_1\cdot\hat{\vec k}'_2)\,\mathrm{cov}\left(\hat{B}(\vec k_1,\vec k_2),\hat{B}(\vec k_1',\vec k_2')\right).
\eeq
We now consider the individual terms in this expansion.

\subsection{$PPP$ Term}
The computation of the first term of the multipole covariance proceeds similarly to that of the 3PCF discussed in \citet[Sec.\,6]{2015MNRAS.454.4142S}. We begin by explicitly including the momentum conserving Dirac deltas $\delta_D(\vec k_1+\vec k_2+\vec k_3)+\delta_D(\vec k_1'+\vec k_2'+\vec k_3')$ in the six-point covariance
\beq
    \frac{\mathcal{C}_{\ell\ell'}^{PPP}(k_1,k_2,k_1',k_2')}{(2\ell+1)(2\ell'+1)} &=& \frac{1}{V}\int_{\vec k_3\vec k_3'}\int_{\Omega_{k_1}\Omega_{k_2}\Omega_{k_1'}\Omega_{k_2'}}\PN(k_1)\PN(k_2)\PN(k_3)L_\ell(\hat{\vec k}_1\cdot\hat{\vec k}_2)L_{\ell'}(\hat{\vec k}'_1\cdot\hat{\vec k}'_2)(2\pi)^3\delta_D(\vec k_1+\vec k_2+\vec k_3)\\\nonumber
    &&\,\times \left[(2\pi)^3\delta_D(\vec k_1'+\vec k_1)(2\pi)^3\delta_D(\vec k_2+\vec k_2')(2\pi)^3\delta_D(\vec k_3+\vec k_3') + \text{5 perms.}\right].
\eeq
We now separate this into the sum of two components with different structures under permutation and apply the Dirac deltas;
\beq
    \mathcal{C}^{PPP}_{\ell\ell'}(k_1,k_2,k_1',k_2') &\equiv& \mathcal{C}^{PPP,A}_{\ell\ell'}(k_1,k_2,k_1',k_2')+\mathcal{C}^{PPP,B}_{\ell\ell'}(k_1,k_2,k_1',k_2')\\\nonumber
    \frac{\mathcal{C}^{PPP,A}_{\ell\ell'}(k_1,k_2,k_1',k_2')}{(2\ell+1)(2\ell'+1)} &=& \frac{1}{2V}\int\frac{k_3^2dk_3}{2\pi^2}\int_{\Omega_{k_1}\Omega_{k_2}\Omega_{k_3}}\PN(k_1)\PN(k_2)\PN(k_3)L_\ell(\hat{\vec k}_1\cdot\hat{\vec k}_2)(2\pi)^3\delta_D(\vec k_1+\vec k_2+\vec k_3)\\\nonumber
    &&\,\times\,\left[L_{\ell'}(\hat{\vec k}_1\cdot\hat{\vec k}_2)\delta_D^{k_1k_1'}\delta_D^{k_2k_2'}+\text{3 sym.}\right]\\\nonumber
    \frac{\mathcal{C}^{PPP,B}_{\ell\ell'}(k_1,k_2,k_1',k_2')}{(2\ell+1)(2\ell'+1)} &=& \frac{1}{V}\int_{\Omega_{k_1}\Omega_{k_2}\Omega_{k_2'}}\PN(k_1)\PN(k_2)L_\ell(\hat{\vec k}_1\cdot\hat{\vec k}_2)\\\nonumber
    &&\,\times\,\left[(2\pi)^3\delta_D(\vec k_1+\vec k_2+\vec k_2')\PN(k_2')L_{\ell'}(\hat{\vec k}_1\cdot\hat{\vec k}_2')\delta_D^{k_1k_1'}+\text{3 sym.}\right]
\eeq
summing over $k_1\leftrightarrow k_1'$, $k_2\leftrightarrow k_2'$. To evaluate these, we consider the angular parts;
\beq
    \int_{\Omega_{k_1}\Omega_{k_2}} L_\ell(\hat{\vec k}_1\cdot\hat{\vec k}_2)L_{\ell'}(\hat{\vec k}_1\cdot\hat{\vec k}_2) &=& \frac{\delta_{\ell\ell'}^K}{2\ell+1}\\\nonumber
    \int_{\Omega_{k_1}\Omega_{k_2}\Omega_{k_3}} L_\ell(\hat{\vec k}_1\cdot\hat{\vec k}_2)L_{\ell'}(\hat{\vec k}_1\cdot\hat{\vec k}_2)(2\pi)^3\delta_D(\vec k_1+\vec k_2+\vec k_3) &=& \sum_L (-1)^L(2L+1)\tj{\ell}{\ell'}{L}{0}{0}{0}^2 \mathcal{R}_{LL0}(k_1,k_2,k_3)\\\nonumber
    \int_{\Omega_{k_1}\Omega_{k_2}\Omega_{k_2'}} L_\ell(\hat{\vec k}_1\cdot\hat{\vec k}_2)L_{\ell'}(\hat{\vec k}_1\cdot\hat{\vec k}'_2)(2\pi)^3\delta_D(\vec k_1+\vec k_2+\vec k_2') &=& \sum_L (-1)^{(\ell+\ell'+L)/2}(2L+1)\tj{\ell}{\ell'}{L}{0}{0}{0}^2 \mathcal{R}_{L\ell\ell'}(k_1,k_2,k_2')
\eeq
using \citet[Eq.\,59]{2015MNRAS.454.4142S} where the parentheses indicate a 3j symbol 
and
\beq
    \mathcal{R}_{\ell_1\ell_2\ell_3}(k_1,k_2,k_3) \equiv 4\pi \int x^2dx\,j_{\ell_1}(k_1x)j_{\ell_2}(k_2x)j_{\ell_3}(k_3x).
\eeq
Using the method of \citet{fabrikant} (summarized in \citealt[][Appendix B]{2017JCAP...11..039F}), this can be written in terms of the derivatives of the Gamma function $\Gamma$ as
\beq\label{eq: triple-sbf-expansion}
    \mathcal{R}_{\ell_1\ell_2\ell_3}(k_1,k_2,k_3) &=& \pi(-1)\cos\frac{\pi}{2}\left(\ell_{123}+1\right)\prod_{i=1}^3\left[(-1)^{\ell_i}k_i^{\ell_i}\frac{\partial^{\ell_i}}{\left(k_i\partial k_i\right)^{\ell_i}}\right]\frac{\Gamma(-\ell_{123})}{k_1k_2k_3}\\\nonumber
    &&\times\left[|k_3+k_1-k_2|^{\ell_{123}}\operatorname{sgn}(k_3+k_1-k_2) + \text{2 cyc.} - (k_1+k_2+k_3)^{\ell_{123}}\right]
\eeq
where `$\operatorname{sgn}$' is the sign function, and $\ell_{123}\equiv \ell_1+\ell_2+\ell_3$. Useful special cases are summarized in \citet[Appendix B.4]{2018arXiv180512394F}.

For the first intrinsic covariance component, it is useful to switch the order of integral over $k_3$ and $x$ via Fubini's theorem (separating out the $1/n$ term in $\PN(k_3)$) and note that
\beq
    \int \frac{k_3^2dk_3}{2\pi^2}j_0(k_3x)P(k_3) \equiv \xi(x)
\eeq
since this is merely a spherical Fourier transform. Simplification thus yields
\beq
    \mathcal{C}^{PPP,A}_{\ell\ell'}(k_1,k_2,k_1',k_2') &=& \frac{(2\ell+1)(2\ell'+1)}{2V}\PN(k_1)\PN(k_2)\left[\delta_D^{k_1k_1'}\delta_D^{k_2k_2'} + \text{3 sym.}\right]\\\nonumber
    &&\,\times\left\{\sum_L(-1)^L(2L+1)\tj{\ell}{\ell'}{L}{0}{0}{0}^2\,\mathcal{S}_{LL}(k_1,k_2)+\frac{1}{n}\frac{\delta_{\ell\ell'}^K}{2\ell+1}\right\}\\\nonumber
    \mathcal{C}^{PPP,B}_{\ell\ell'}(k_1,k_2,k_1',k_2') &=& \frac{(2\ell+1)(2\ell'+1)}{V}\PN(k_1)\PN(k_2)\sum_L(-1)^{(\ell+\ell'+L)/2}(2L+1)\tj{\ell}{\ell'}{L}{0}{0}{0}^2\\\nonumber
    &&\,\times\,\left[\PN(k_2')\mathcal{R}_{L\ell\ell'}(k_1,k_2,k_2')\delta_D^{k_1k_1'} + \text{3 sym.}\right],
\eeq
where we define
\beq
    \mathcal{S}_{\ell_1\ell_2}(k_1,k_2) = 4\pi\int x^2dx\,j_{\ell_1}(k_1x)j_{\ell_2}(k_2x)\xi(x).
\eeq
which is a special case of Eq.\,\ref{eq: T-def}. Note that the covariance expressions given above are relatively simple to compute since each depends only on one numerical integral. 

\subsection{$BB$ Term}
Calculation of the $BB$ term proceeds similarly, starting with the form
\beq
    \mathcal{C}^{BB}_{\ell\ell'}(k_1,k_2,k_1',k_2') &=& \frac{(2\ell+1)(2\ell'+1)}{V}\int_{\vec k_3\vec k_3'}\int_{\Omega_{k_1}\Omega_{k_2}\Omega_{k_1'}\Omega_{k_2'}}L_\ell(\hat{\vec k}_1\cdot\hat{\vec k}_2)L_{\ell'}(\hat{\vec k}'_1\cdot\hat{\vec k}'_2)(2\pi)^6\delta_D(\vec k_1+\vec k_2+\vec k_3)\delta_D(\vec k_1'+\vec k_2'+\vec k_3')\\\nonumber
    &&\,\times\,\left[\BN(\vec k_1,\vec k_2,\vec k_3)\BN(\vec k_1',\vec k_2',\vec k_3')(2\pi)^3\delta_D(\vec k_1-\vec k_1') + \text{8 perms.}\right].\\\nonumber
\eeq
Splitting into three terms with different permutative structures and inserting the definition of $\BN$ gives;
\beq
    \frac{\mathcal{C}^{BB,A}_{\ell\ell'}(k_1,k_2,k_1',k_2')}{(2\ell+1)(2\ell'+1)} &=& \frac{1}{n^2V}\int_{\Omega_{k_1}\Omega_{k_2}\Omega_{k_1'}\Omega_{k_2'}}L_\ell(\hat{\vec k}_1\cdot\hat{\vec k}_2)L_{\ell'}(\hat{\vec k}'_1\cdot\hat{\vec k}'_2)(2\pi)^3\delta_D(\vec k_1+\vec k_2-\vec k_1'-\vec k_2')\\\nonumber
    &&\,\times\,\left[(P(k_1)+P(k_2))(P(k_1')+P(k_2'))\right]\\\nonumber
    \frac{\mathcal{C}^{BB,B}_{\ell\ell'}(k_1,k_2,k_1',k_2')}{(2\ell+1)(2\ell'+1)} &=& \frac{1}{n^2V}\int_{\vec k_3\vec k_3'}\int_{\Omega_{k_1}\Omega_{k_2}\Omega_{k_2'}}L_\ell(\hat{\vec k}_1\cdot\hat{\vec k}_2)L_{\ell'}(\hat{\vec k}_1\cdot\hat{\vec k}'_2)(2\pi)^3\delta_D(\vec k_1+\vec k_2+\vec k_3')(2\pi)^3\delta_D(\vec k_1+\vec k_2'+\vec k_3')\\\nonumber
    &&\,\times\,\left[(P(k_2)+P(k_3))(P(k_2')+P(k_3'))\right]\delta_D^{k_1k_1'} + \text{3 perms.}\\\nonumber
    \frac{\mathcal{C}^{BB,C}_{\ell\ell'}(k_1,k_2,k_1',k_2')}{(2\ell+1)(2\ell'+1)} &=& \frac{1}{n^2V}\int_{\vec k_3'}\int_{\Omega_{k_1}\Omega_{k_1'}\Omega_{k_2}\Omega_{k_2'}}L_\ell(\hat{\vec k}_1\cdot\hat{\vec k}_2)L_{\ell'}(\hat{\vec k}'_1\cdot\hat{\vec k}'_2)(2\pi)^3\delta_D(\vec k_1+\vec k_2+\vec k_2')(2\pi)^3\delta_D(\vec k_1'+\vec k_2'+\vec k_3')\\\nonumber
    &&\,\times\,\left[(P(k_1)+P(k_2))(P(k_1')+P(k_3'))\right] + \text{3 perms.}\\\nonumber
\eeq
Following a lengthy calculation, similar in form to the above, we obtain
\beq
    \frac{\mathcal{C}^{BB,A}_{\ell\ell'}(k_1,k_2,k_1',k_2')}{(2\ell+1)(2\ell'+1)} &=& \frac{1}{n^2V}\left[(P(k_1)+P(k_2))(P(k_1')+P(k_2'))\right](-1)^{\ell+\ell'}\mathcal{R}_{\ell\ell\ell'\ell'}(k_1,k_2,k_1',k_2')\\\nonumber
    \frac{\mathcal{C}^{BB,B}_{\ell\ell'}(k_1,k_2,k_1',k_2')}{(2\ell+1)(2\ell'+1)} &=& \frac{1}{n^2V}\left[(-1)^\ell \mathcal{S}_{\ell\ell}(k_1,k_2)+P(k_2)\delta_{\ell 0}^K\right]\left[(-1)^{\ell'}\mathcal{S}_{\ell'\ell'}(k_1',k_2')+P(k_2')\delta_{\ell'0}^K\right]\delta_D^{k_1k_1'} + \text{3 perms.}\\\nonumber
    \frac{\mathcal{C}^{BB,C}_{\ell\ell'}(k_1,k_2,k_1',k_2')}{(2\ell+1)(2\ell'+1)} &=& \left[P(k_1)+P(k_2)\right]\left[(-1)^{\ell+\ell'}\mathcal{R}_{\ell\ell0}(k_1,k_2,k_2')\mathcal{S}_{\ell'\ell'}(k_1',k_2')+\delta_{\ell0}^K\delta_{\ell'0}^KP(k_1')\right] + \text{3 perms.}
\eeq
where we define
\beq
    \mathcal{R}_{\ell_1\ell_2\ell_3\ell_4}(k_1,k_2,k_3,k_4) &=& 4\pi\int x^2dx\,j_{\ell_1}(k_1x)j_{\ell_2}(k_2x)j_{\ell_3}(k_3x)j_{\ell_4}(k_4x).
\eeq
This can be computed analytically, following the prescription of \citet{2009arXiv0909.0494M} and \citet{fabrikant}.

\subsection{$PT$ Term}
We finally turn to the $PT$ term of the bispectrum covariance. This has the form
\beq
    \mathcal{C}^{PT}_{\ell\ell'}(k_1,k_2,k_1',k_2') &=& \frac{(2\ell+1)(2\ell'+1)}{V}\int_{\vec k_3\vec k_3'}\int_{\Omega_{k_1}\Omega_{k_2}\Omega_{k_1'}\Omega_{k_2'}}L_\ell(\hat{\vec k}_1\cdot\hat{\vec k}_2)L_{\ell'}(\hat{\vec k}'_1\cdot\hat{\vec k}'_2)(2\pi)^6\delta_D(\vec k_1+\vec k_2+\vec k_3)\delta_D(\vec k_1'+\vec k_2'+\vec k_3')\\\nonumber
    &&\,\times\,\left[\PN(k_1)\TN(\vec k_2,\vec k_3,\vec k_2',\vec k_3')(2\pi)^3\delta_D(\vec k_1+\vec k_1') + \text{8 perms.}\right].
\eeq
Here, we find three terms with different structures;
\beq
    \frac{\mathcal{C}^{PT,A}_{\ell\ell'}(k_1,k_2,k_1',k_2')}{(2\ell+1)(2\ell'+1)} &=& \frac{1}{2n^2V}\int_{\vec k_3\vec k_3'}\int_{\Omega_{k_1}\Omega_{k_2}\Omega_{k_2'}}L_\ell(\hat{\vec k}_1\cdot\hat{\vec k}_2)L_{\ell'}(\hat{\vec k}_1\cdot\hat{\vec k}'_2)(-1)^{\ell'}(2\pi)^6\delta_D(\vec k_1+\vec k_2+\vec k_3)\delta_D(\vec k_2'+\vec k_3'-\vec k_1)\\\nonumber
    &&\,\times\,\left[\delta_D^{k_1k_1'}\PN(k_1)\left(P(|\vec k_2+\vec k_2'|)+ P(|\vec k_2+\vec k_3'|) + P(|\vec k_3+\vec k_2'|) + P(|\vec k_3+\vec k_3'|)\right)+\text{ 3 perms.}\right]\\\nonumber
    \frac{\mathcal{C}^{PT,B}_{\ell\ell'}(k_1,k_2,k_1',k_2')}{(2\ell+1)(2\ell'+1)} &=& \frac{1}{2n^2V}\int_{\vec k_3}\int_{\Omega_{k_1}\Omega_{k_2}\Omega_{k_1'}\Omega_{k_2'}}L_\ell(\hat{\vec k}_1\cdot\hat{\vec k}_2)L_{\ell'}(\hat{\vec k}'_1\cdot\hat{\vec k}'_2)(2\pi)^6\delta_D(\vec k_1+\vec k_2+\vec k_3)\delta_D(\vec k_1'+\vec k_2'-\vec k_1)\\\nonumber
    &&\,\times\,\left[\PN(k_1)\left(P(|\vec k_2+\vec k_1'|)+ P(|\vec k_2+\vec k_2'|) + P(|\vec k_3+\vec k_1'|) + P(|\vec k_3+\vec k_2'|)\right)+\text{1 perm.}\right]+\text{1 sym.}\\\nonumber
    \frac{\mathcal{C}^{PT,C}_{\ell\ell'}(k_1,k_2,k_1',k_2')}{(2\ell+1)(2\ell'+1)} &=&
    \frac{1}{2n^2V}\int_{\vec k_3}\int_{\Omega_{k_1}\Omega_{k_2}\Omega_{k_1'}\Omega_{k_2'}}L_\ell(\hat{\vec k}_1\cdot\hat{\vec k}_2)L_{\ell'}(\hat{\vec k}'_1\cdot\hat{\vec k}'_2)(2\pi)^6\delta_D(\vec k_1+\vec k_2+\vec k_3)\delta_D(\vec k_1'+\vec k_2'-\vec k_3)\\\nonumber
    &&\,\times\,\left[\PN(k_3)\left(P(|\vec k_1+\vec k_1'|)+ P(|\vec k_1+\vec k_2'|) + P(|\vec k_2+\vec k_1'|) + P(|\vec k_2+\vec k_2'|)\right)\right].
\eeq
These may be simplified in a similar way to before, writing the power spectra in terms of correlation functions and computing the angular integrals. Following a substantial calculation, this yields
\beq
    \frac{\mathcal{C}^{PT,A}_{\ell\ell'}(k_1,k_2,k_1',k_2')}{(2\ell+1)(2\ell'+1)} &=& \frac{1}{n^2V}\delta_D^{k_1k_1'}\PN(k_1)\\\nonumber
    &&\,\times\,\left[\frac{\delta_{\ell\ell'}^K}{2\ell+1}\mathcal{S}_{\ell\ell}(k_2,k_2')+\sum_L(-1)^{(\ell+\ell'+L)/2}(2L+1)\tj{\ell}{\ell'}{L}{0}{0}{0}^2\mathcal{S}_{\ell\ell'L}(k_2',k_2,k_1)\right]+\text{3 perms.}\\\nonumber
    \frac{\mathcal{C}^{PT,B}_{\ell\ell'}(k_1,k_2,k_1',k_2')}{(2\ell+1)(2\ell'+1)} &=& \frac{1}{n^2V}\PN(k_1)\\\nonumber
    &&\,\times\,\left[(-1)^\ell\mathcal{S}_{\ell\ell}(k_2,k_2')\sum_L(-1)^{(\ell'-\ell+L)/2}(2L+1)\tj{\ell}{\ell'}{L}{0}{0}{0}^2\mathcal{R}_{\ell\ell'L}(k_1,k_1',k_2')+(k_1'\leftrightarrow k_2')\right]+\text{1 perm.}+\text{1 sym.}\\\nonumber
    \frac{\mathcal{C}^{PT,C}_{\ell\ell'}(k_1,k_2,k_1',k_2')}{(2\ell+1)(2\ell'+1)} &=& \frac{1}{2n^2V}\sum_{LL'PP'Q}(2L+1)(2L'+1)(2P+1)(2P'+1)(2Q+1)(-1)^{L+L'}(-1)^{(P+Q-L-\ell')/2}\\\nonumber
    &&\,\quad\times\,\tj{\ell}{L'}{P}{0}{0}{0}^2\tj{\ell'}{P}{P'}{0}{0}{0}^2\tj{L}{P'}{Q}{0}{0}{0}^2\mathcal{R}_{QPL\ell'}(k_1,k_2,k_1',k_2')\mathcal{S}_{LL}(k_1,k_1')\mathcal{S}_{L'L'}(k_1,k_2)+\text{3 sym.}
\eeq
where we additionally define
\beq
    \mathcal{S}_{\ell_1\ell_2\ell_3}(k_1,k_2,k_3) &=& 4\pi\int x^2dx\,j_{\ell_1}(k_1x)j_{\ell_2}(k_2x)j_{\ell_3}(k_3x)\xi(x).
\eeq
As for Eq.\,\ref{eq: T-def}, this can be computed via the prescription of \citet{2019arXiv191200065S} (or \citealt{fabrikant} for a $1/r^2$ 2PCF).

\section{Practical Application}\label{sec: application}
We now turn to a practical application of the power spectrum algorithm (Sec.\,\ref{sec: pk-algo}) and bispectrum algorithm (Sec.\,\ref{sec: bk-algo}) to simulated data. Both of these have been incorporated into the \texttt{HIPSTER} C++ package, which is highly optimized and allows for fast estimation of small-scale spectra, by exhaustive weighted pair counting up to some maximum radius $R_0$. Previously the package included only an algorithm for power spectra in arbitrary geometries; we augment it with the efficient periodic-box power spectrum and bispectrum algorithms developed in this work. The code is publicly available online, with extensive documentation.\footnote{\href{https://HIPSTER.readthedocs.io}{HIPSTER.readthedocs.io}.} In \texttt{HIPSTER}, the spherical harmonics needed for the bispectrum estimator are computed efficiently by working in Cartesian space, as in \citet{2015MNRAS.454.4142S}. The limiting step of our algorithm is in repeated calculation of the Bessel function weights $j_\ell^a$; we expect runtime could be substantially reduced by making effective use of GPUs. Note that, via simple modifications to \texttt{HIPSTER}, one may also compute isotropic bispectra in arbitrary geometries, as well as \textit{anisotropic} bispectra. 

\subsection{Discussion of Algorithm Scalings and Hyperparameters}
Before diving into numerical results, we present a brief discussion of the leading dependencies and scalings of our configuration-space spectral code \texttt{HIPSTER}. Denoting the computation time by $T$, we have the following;
\begin{itemize}
    \item $T\propto Nn$ for $N$ tracer particles with number density $n$, or $T\propto N^2$ at fixed volume. This occurs since both the power spectrum and bispectrum estimators can be written as a count over all pairs of particles up to some maximum radius, the number of which scales as $Nn$. In particular, we note that the computation times for the power spectrum and bispectrum are comparable, unlike conventional methods.
    \item $T\propto R_0^3$ for pair-count truncation scale $R_0$. The number of particles within $R_0$ from a given primary is proportional to $nR_0^3$, explaining this scaling.
    \item $T\sim n_kn_\ell$ (power spectrum) or $T\propto n_k^2n_\ell$ (bispectrum) for $n_k$ $k$-bins and $n_\ell$ Legendre bins. For each pair of particles we must compute their contribution to every possible $k$ and $\ell$ bin. Note that this is not an exact scaling, since many of the computational processes (e.g., choosing particles) are independent of the binning. We further stress that the speed is sensitive only the \textit{number} of $k$-bins used, and not their magnitudes.
    \item $T\propto (1+f_\mathrm{rand})$ for the bispectrum, when using a random catalog $f_\mathrm{rand}$ times larger than the data. This is because the number density $n_\mathrm{rand}$ of randoms satisfies $n_\mathrm{rand} = nf_\mathrm{rand}$ and we must count all data-data and random-data pairs (but not the random-random pairs since these are computed analytically). We find $f_\mathrm{rand}\sim 3$ to be sufficient in practice.
    \item $T\propto N_\mathrm{cpu}^{-1}$ when computed on $N_\mathrm{cpu}$ CPU cores. Since the algorithm is a pair count it can be trivially parallelized.
\end{itemize}

A key hyperparameter of the code is the truncation scale $R_0$, which strongly influences the runtime. Although a larger $R_0$ implies a count over more pairs of particles, choosing an optimal $R_0$ is not as simple as minimizing the variance of the estimators at fixed runtime (which is the way in which the number of random particles used in 2PCF and 3PCF estimators are optimized), since lowering $R_0$ reduces only the number of \textit{large separation} pairs counted, which have minimal impact on high-$k$ estimates. As shown in Sec.\,\ref{sec: window-vs-true-spectra} and \citetalias{2020MNRAS.492.1214P}, the primary effect of $W(r;R_0)$ is to convolve the true spectra with a characteristic scale $\Delta k \sim 3/R_0$, leading to a reduction in the variance of a given mode (shown in Sec.\,\ref{subsec: cov-data}). Practically, the truncation radius is set by the minimum $k$ bin used, since we observe spectral distortions above percent-level for for $k\lesssim 10\Delta k$, giving $R_0\sim k_\mathrm{min}^{-1}$. For $R_0 = 100\Mpch$ ($50\Mpch$), $k\gtrsim 0.25\hMpc$ ($0.5\hMpc$) can be measured robustly using \texttt{HIPSTER}, without consideration of the convolution window. Using $R_0\sim k_\mathrm{min}^{-1}$ gives $T\sim k_\mathrm{min}^{-3}$, making this method optimal for high-$k$ spectral measurements. This further sets the optimal $k$-binning, requiring $\Delta k \gtrsim 3/R_0$ to minimize the covariance between bins. In the below, we principally use linear bins with $\Delta k = 0.05$, satisfying this condition, reducing the correlation between nearby bins. Note that the above argument assumes that we do \textit{not} perform window-convolution of the theory model; this would partially ameliorate these distortions, though a full Fisher analysis is needed to assess the impact on cosmological observables due to the truncation.

For conventional Fourier-transform estimators, the principal scaling is with the size of the discrete particle grid, $N_\mathrm{grid}$, with the complexity scaling as $\mathcal{O}(N_\mathrm{grid}\log N_\mathrm{grid})$ thanks to the efficient FFT algorithm. In practice, $N_\mathrm{grid}$ is set by the Nyquist frequency $k_\mathrm{Nyq} = \pi N_\mathrm{grid}/L$ of the length $L$ box; we require $k\lesssim k_\mathrm{Nyq}/2$ to avoid the affects of aliasing. $N_\mathrm{grid}$ thus scales linearly with the maximum $k$-vector used, giving $T\sim k_\mathrm{max}\log k_\mathrm{max}$, opposite to \texttt{HIPSTER}. Assuming the most computationally intensive segment of a spectral algorithm to be performing the FFTs, power spectrum computation is roughly independent of the number of bins (unlike the configuration-space estimators), though the bispectrum estimators, which are proportional to $n_k^2$ as discussed below, are not superior to \texttt{HIPSTER} in this regard. Further benefits of our estimators are that they do not suffer from shot-noise \resub{(except in the covariance)}, and they require very little memory, since only the sample of particles has to be held and not large FFT grids.

Given that our estimators have a strong scaling with the number of particles, $N$, not seen in FFT-based approaches, it is often useful to \textit{subsample} the data by using only a random fraction $1/f_\mathrm{sub}$ of the full catalog of tracer particles. Whilst this is not necessary for the computation of halo spectra, it is of great importance for matter spectra, where $n$ is large. This does not affect the intrinsic covariance (which is independent of $n$), but boosts the two- and three-point Poisson terms. This is discussed further in Sec.\,\ref{subsec: cov-data}.

\subsection{Simulated Power Spectra}
\subsubsection{Comparison of Methods}\label{subsec: pk-comparison}
To determine whether our power spectrum estimator is accurate, we compute the $z = 0$ power spectrum of 100 N-body simulations, taken from the \texttt{Quijote} project \citep{2019arXiv190905273V}; a suite of over 40,000 simulations run using the \texttt{GADGET-III} TreePM + SPH code \citep{2005MNRAS.364.1105S}. Here, we use only a subset of the fiducial cosmology simulations, each of which have boxsize $L = 1h^{-1}\mathrm{Gpc}$, and contain $512^3$ cold dark matter (CDM) particles evolved starting from $z = 127$, with initial conditions generated from second-order Lagrangian perturbation theory (2LPT). For the majority of this section, we focus on halo power spectra, though we note the algorithms are equally applicable to matter power spectra. Halo catalogs are computed with the friends-of-friends algorithm \citep{1985ApJ...292..371D} with a linking length of $b = 0.2$. Furthermore, we add redshift-space-distortions using the velocity of the simulated halos in combination with the Hubble expansion parameter.

For each simulation, power spectra are estimated (a) via \texttt{HIPSTER} as detailed above, and (b) via the \texttt{nbodykit} code \citep{2018AJ....156..160H} which computes spectra in the conventional method via FFTs. For both codes, this is performed for $\ell\in\{0,2,4\}$ over $k$-bins with fixed width $\Delta k = 0.05\hMpc$ for $k\in[0,2]\hMpc$. In the FFT-based method, spectra are computed with triangle-shaped-cell interpolation using grid-sizes of $N_\mathrm{grid}\in\{512,1024\}$, giving Nyquist frequencies of $k_\mathrm{Nyq} = \pi N_\mathrm{grid}/L\in\{1.6,3.2\}\hMpc$. We subtract the shot-noise contribution $P_\mathrm{shot} = n^{-1}$ from the spectrum monopole, where $n$ is the number density of tracers $\sim 4\times 10^{-4}h^3\,\mathrm{Mpc}^{-3}$. For \texttt{HIPSTER}, we use truncation scales $R_0\in\{50,100\}\Mpch$, and do not need to remove shot-noise. Considering computation time, for a galaxy catalog containing $\sim 4\times 10^5$ particles with this choice of $k$-bins and Legendre multipoles, \texttt{HIPSTER} with $R_0 = 100\Mpch$ truncation requires $\sim 60$ core-minutes to compute each spectrum, whilst the analogous computation with \texttt{nbodykit} needs $\sim 20$ core-minutes at $N_\mathrm{grid} = 1024$. We note however that \texttt{HIPSTER} can compute spectra up to much higher $k$ with no additional cost (recalling that the computation time scales with the number of bins), and that the computation time can be reduced with smaller $R_0$ or by subsampling the data.

\begin{figure}
  \centering
  \begin{minipage}[t]{0.48\textwidth}
    \includegraphics[width=\textwidth]{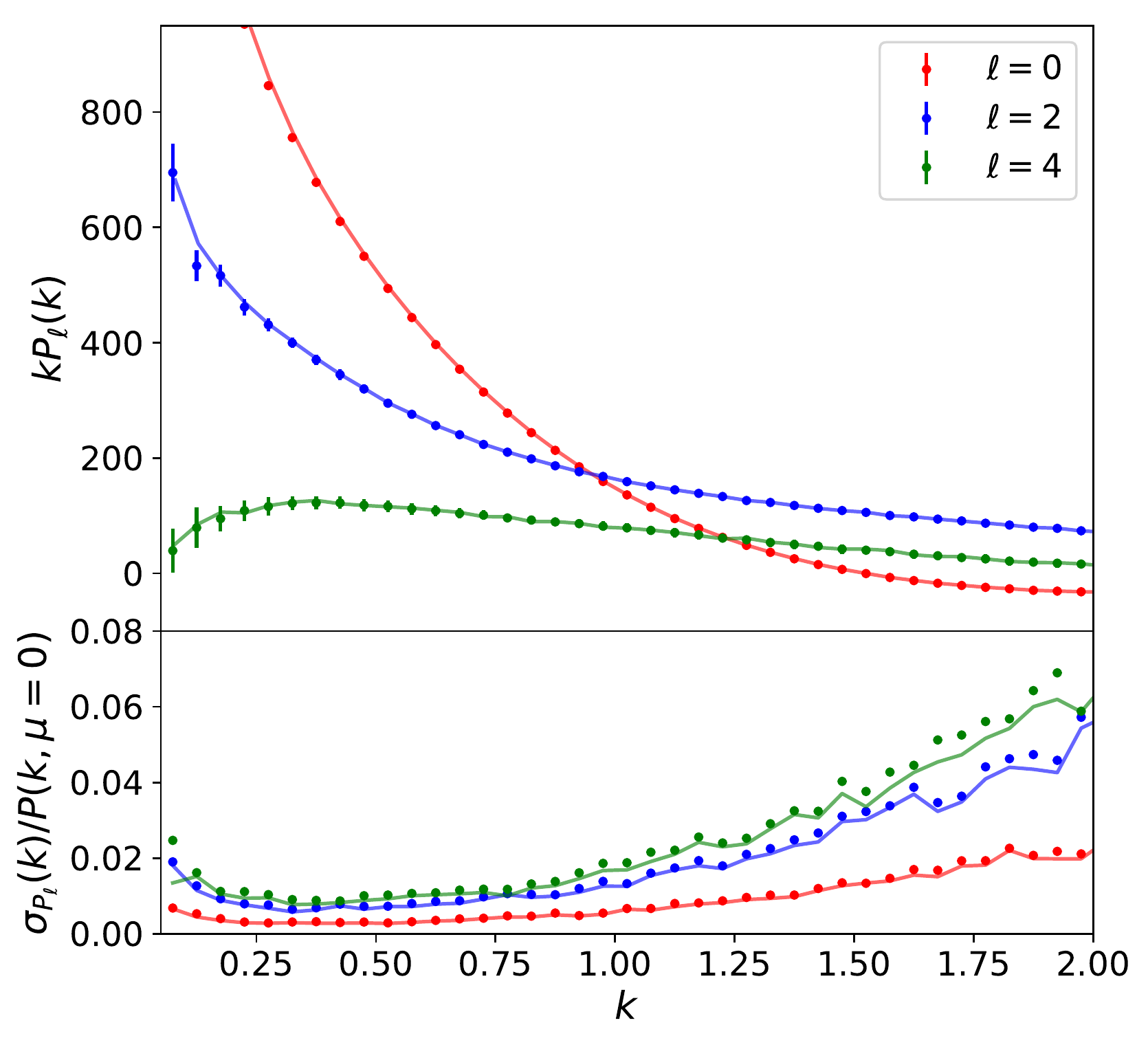}
    \caption{Comparison of the power spectrum multipoles computed from the configuration-space code \texttt{HIPSTER} (points, described in this work) and conventional FFT-based methods (lines, implemented via \texttt{nbodykit}). The top plot shows the mean and standard deviation (without normalizing by the number of mocks) of the spectra computed from a set of 100 halo catalogs drawn from N-body simulations, with the bottom showing the statistical error in each measurement relative to the sum of the multipoles at $\mu = 0$ (equal to the power in the absence of redshift-space distortions). \texttt{HIPSTER} measurements are computed using a truncation scale $R_0=100\Mpch$, and we use a $1024^3$ grid to perform the FFTs. The ratio of these measurements, and their dependence on hyperparameters, is shown in Fig.\,\ref{fig: pk_ratio}.}\label{fig: pk_multipoles}
  \end{minipage}
  \hfill
  \begin{minipage}[t]{0.48\textwidth}
    \includegraphics[width=\textwidth]{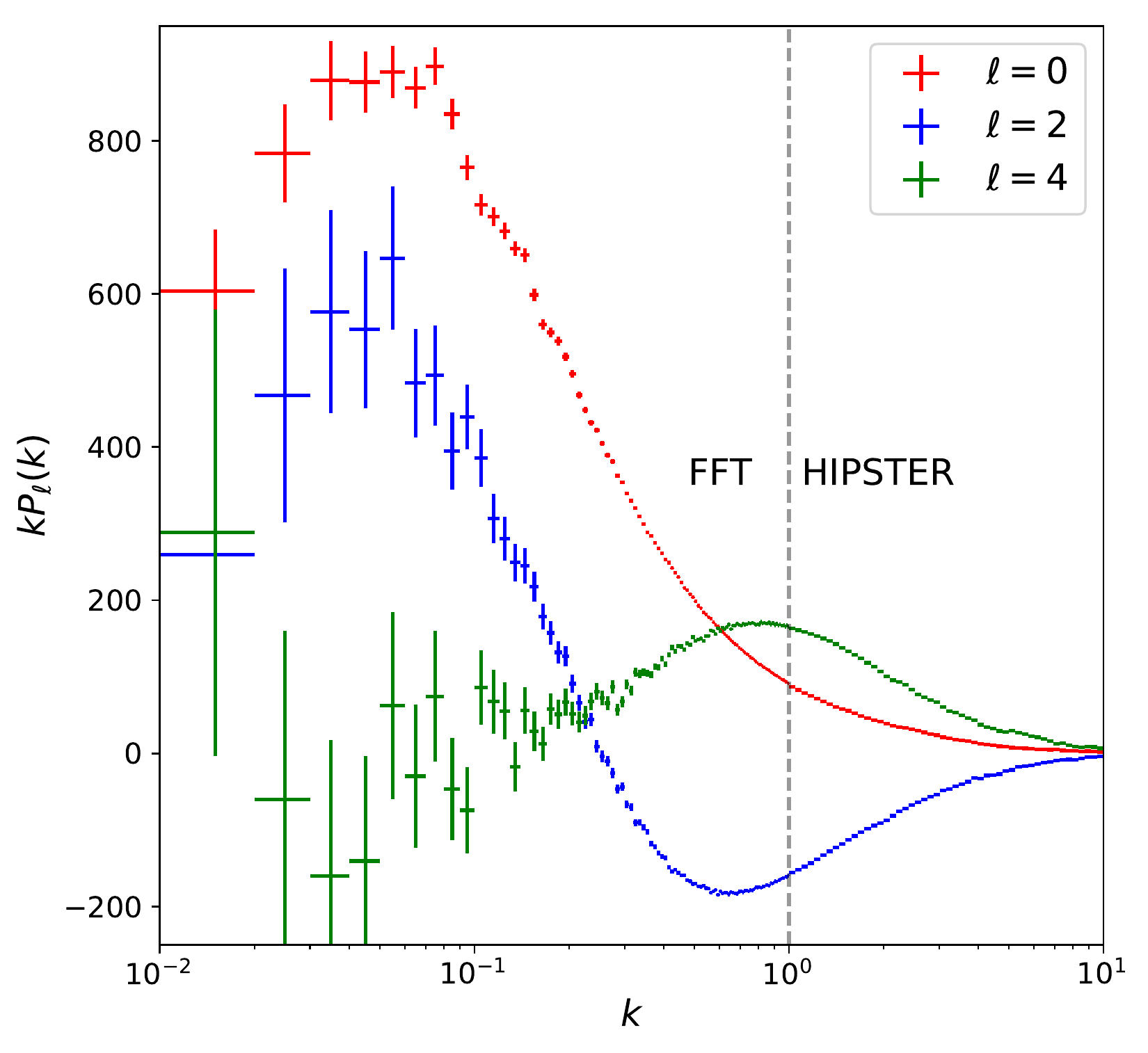}
    \caption{Power spectrum multipoles computed from the redshift-space matter distribution of a single N-body simulation across a broad range of $k$-bins, using conventional FFT-based methods for $k<1\hMpc$ and configuration-spaced estimators (described in this work) for larger $k$. Both of the spectra take a few core-hours to compute, and the joint approach allows one to use a small sampling grid ($N_\mathrm{grid}=512$) for the FFTs and a small truncation scale $R_0 = 30\Mpch$ for the configuration-space counts. The N-body simulation contains $512^3$ particles, with a 1\% subsample used to compute the \texttt{HIPSTER} spectra. Errorbars represent the expected standard deviation of the spectrum and are obtained using the formulae derived in Sec.\,\ref{sec: cov}.}
    \label{fig: huge-pk}
  \end{minipage}
\end{figure}

The corresponding spectra are shown in Fig.\,\ref{fig: pk_multipoles}, and we immediately note that the two codes appear highly consistent in both their means and errors, though one might note minor deviations in the smallest $k$-bins, and slight differences in the error bars from the largest $k$-bins. As discussed in Sec.\,\ref{sec: window-vs-true-spectra}, the measured configuration-space power spectrum is a convolution of the true spectrum with a window function; this causes the slight distortions at low-$k$ (see also \citetalias[Fig.\,2]{2020MNRAS.492.1214P}). At large scales, where $k$ approaches the Nyquist frequency of the box, we do not expect FFT-based estimates to be accurate.

\begin{figure}
    \centering
    \includegraphics[width=0.8\textwidth]{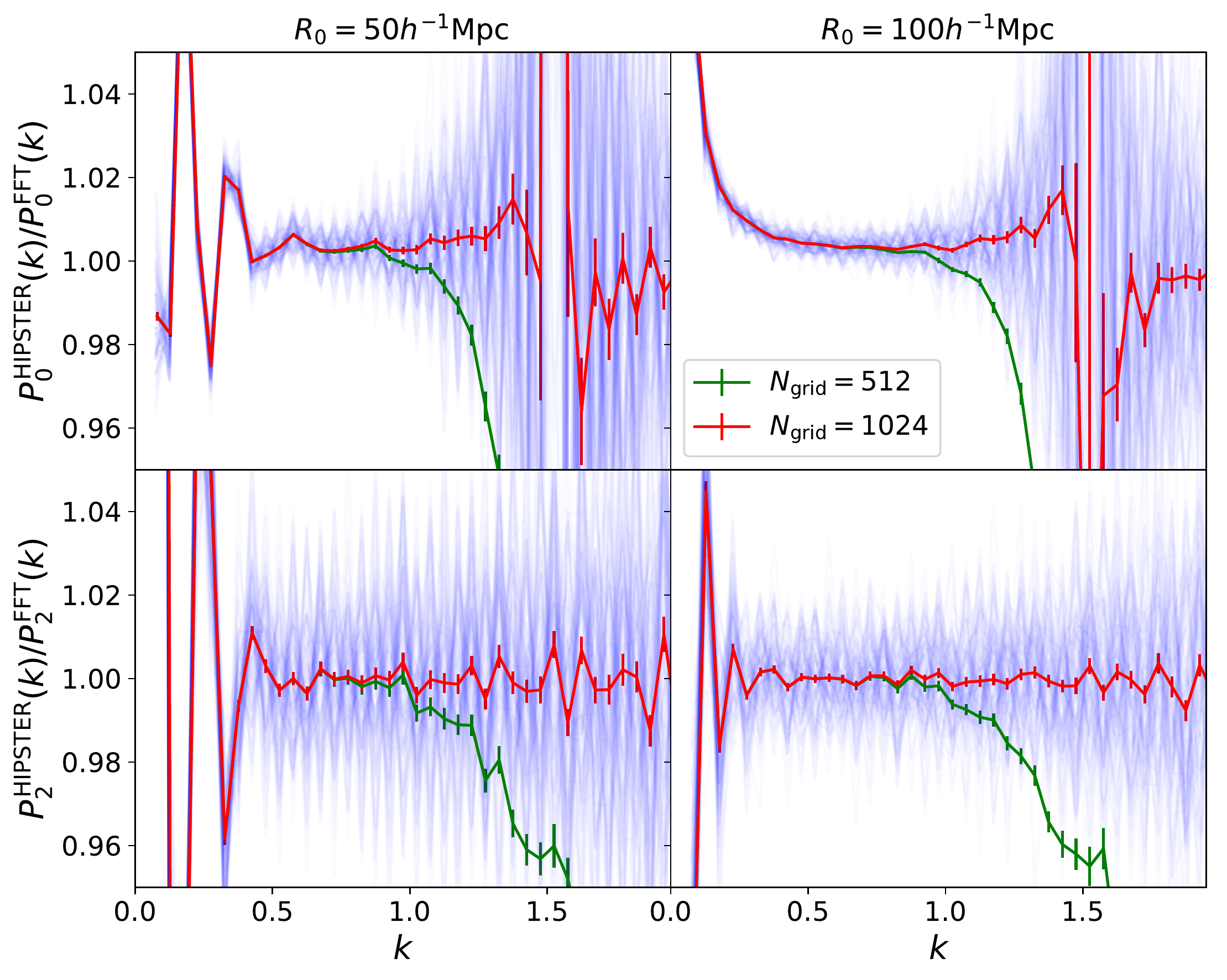}
    \caption{Ratio of power spectra computed from the configuration-space \texttt{HIPSTER} and FFT-based \texttt{nbodykit} codes, for the monopole (upper) and quadrupole (lower) power. Left and right panels show the ratio computed using \texttt{HIPSTER} pair count truncation widths of $R_0 = 50\Mpch$ and $100\Mpch$ respectively. For each panel, we plot the ratio obtained using $512^3$ (green) and $1024^3$ (red) FFT grids showing the mean and standard error in the mean across 100 mocks. The blue curves show the individual ratios from the 100 mocks. These spectra are shown in full in Fig.\,\ref{fig: pk_multipoles}, and we note large amplitude oscillations at low $k$ due to the pair-separation window, though this is negligible at large $R_0$ and $k$.} 
    \label{fig: pk_ratio}
\end{figure}

To investigate these effects in detail, in Fig.\,\ref{fig: pk_ratio} we plot the ratio of \texttt{HIPSTER} and \texttt{nbodykit} power spectra as a function of the pair-count truncation scale $R_0$ and grid-size $N_\mathrm{grid}$. Considering the average over 100 mocks, we find sub-percent agreement between the two methods for $k>0.5\hMpc$ ($k>0.25\hMpc$) for $R_0 = 50\Mpch$ ($100\Mpch$), indicating that our estimators are unbiased. At smaller $k$, we observe oscillatory behavior in the average ratio which can be attributed to the inclusion of a pair-count window function in \texttt{HIPSTER}. This effect is clearly amplified at larger $\ell$ and smaller $R_0$, as noted in \citetalias{2020MNRAS.492.1214P}, but negligible at large $k$. For the smaller value of $N_\mathrm{grid}$, there is a clear divergence near the Nyquist frequency ($\sim 1.6\hMpc$), representing the fundamental limit of FFT-based approaches. Whilst this is not observed for the $N_\mathrm{grid} = 1024$ sample, we expect deviations to start appearing near $k\approx 2\hMpc$ based on the $N_\mathrm{grid} = 512$ data. An initial glance at the monopole ratio near $k\approx 1.5\hMpc$ would suggest that \texttt{HIPSTER} is inaccurate there; in fact this is caused by the monopole power crossing zero around this wavenumber (and hence the ratio being between two small stochastic quantities), and is not found in the quadrupole.

Notably, we observe significant stochastic variations in the \texttt{HIPSTER}-to-FFT ratio across the datasets. If the two estimators were measuring the same quantity from identical data we would not anticipate such variation; however, this is not expected in our scenario since, due to the finite truncation radius \texttt{HIPSTER} only `sees' a subset of all particle pairs, and thus will exhibit different noise properties. The fluctuations in this ratio are amplified for larger $\ell$ and smaller $R_0$, whixh follows by noting that (a) the higher multipole measurements carry an intrinsically larger variance (which scales as $2\ell+1$) and (b) modified pair-counts with larger $R_0$ `see' more of the survey, and thus are expected to be less variable. The full covariance of the measurements is discussed below.

\subsubsection{Jointly Estimating $P_\ell(k)$ on All Scales}
Whilst we have shown configuration-space estimators to be accurate and unbiased estimators of the galaxy power spectra on small-scales, we do not anticipate that they will replace Fourier-transform methods, primarily due to the large $R_0$ that would be needed for unbiased measurements of the low-$k$ power. However, given that FFT-based methods are optimal at small $k$, and configuration-space estimators perform best at large $k$, a natural approach would be combine the two techniques, allowing accurate spectra to computed across a wide range of wavenumbers, without the problems of aliasing or shot-noise (from the FFTs) and convolution-distortions (from \texttt{HIPSTER}). Further, this allows us to use small $N_\mathrm{grid}$ and small $R_0$, giving an efficient computation both in terms of time and memory allocation.

To demonstrate this, we measure the power spectrum multipoles, $P_\ell(k)$, from a single \texttt{Quijote} simulation, this time using the full distribution of $512^3$ matter particles in redshift-space at $z = 0$. This is done for a set of 150 $k$-bins in the range $[0,10]\hMpc$, using 100 linearly spaced bins for $k<1\hMpc$ and 50 logarithmically spaced bins for larger $k$. For $k<1\hMpc$, we compute the spectra via FFTs using \texttt{nbodykit} with $N_\mathrm{grid} = 512$ (which Fig.\,\ref{fig: pk_ratio} has shown to be percent-level accurate in this regime) and for $k>1\hMpc$, \texttt{HIPSTER} is used, truncating at $R_0 = 30\Mpch$ for speed, which will have negligible impact on the high-$k$ power. We further subsample the \texttt{HIPSTER} particle catalog by $100$ times to keep computation times manageable.\footnote{From Sec.\,\ref{sec: cov}, we note that this will affect only the covariance of the shot-noise-induced 2- and 3-point terms. Since the number of particles is large, these terms are inherently small, thus this subsampling does not have a huge impact on the measured covariances.}

Fig.\,\ref{fig: huge-pk} shows the resulting spectra; there is no discernable difference between the FFT spectra at $k<1\hMpc$ and the \texttt{HIPSTER} spectra at $k>1\hMpc$, giving an efficiently computed spectrum across a broad region in $k$-space. However, whilst the measurements themselves are consistent, there is a slight change in the variances (which are computed using the formulae of Sec.\,\ref{sec: cov} for the FFT and \texttt{HIPSTER} sections separately) across $k = 1\hMpc$. This is primarily attributed to the switch from linear to logarithmic bins across $k = 1\hMpc$, as well as the addition of subsampling in \texttt{HIPSTER}, which boosts the Poissonian noise. If one had used a binning scheme that varied smoothly across the boundary and a smaller subsampling ratio, we would not expect to observe this effect.

With the above choice of binning and hyperparameters, both codes take a few core-hours to compute their regions of the power spectrum. If one wished to use only FFTs to compute the same measurement, $N_\mathrm{grid} \gtrsim 6400$ would be needed (setting $k_\mathrm{Nyq}>2k_\mathrm{max}$), requiring a huge amount more computation time and memory than the combination of an $N_\mathrm{grid} = 512$ FFT and an $R_0 = 30\hMpc$ \texttt{HIPSTER} analysis. This motivates the conclusion that a combination of FFT-based and configuration-space algorithms is optimal for computing power spectra from cosmological simulations.

\subsection{Sample Covariance Matrices}\label{subsec: cov-data}
In Sec.\,\ref{sec: cov}, the theoretical covariance of the power spectrum estimates $P_\ell(k;R_0)$ was discussed; here we discuss their accuracy by comparing model predictions to the sample covariances obtained from the 100 redshift-space power spectra used in Sec.\,\ref{subsec: pk-comparison}. For simplicity we will consider only the diagonal terms here (i.e. the variances), though we note that the pair-count window $W$ has a characteristic $k$-space width of $\Delta k \sim 3/R_0$ implying that we expect a significant non-zero contribution from the covariance matrix elements adjacent to the diagonal.

\begin{figure}
    \centering
    \includegraphics[width=\textwidth]{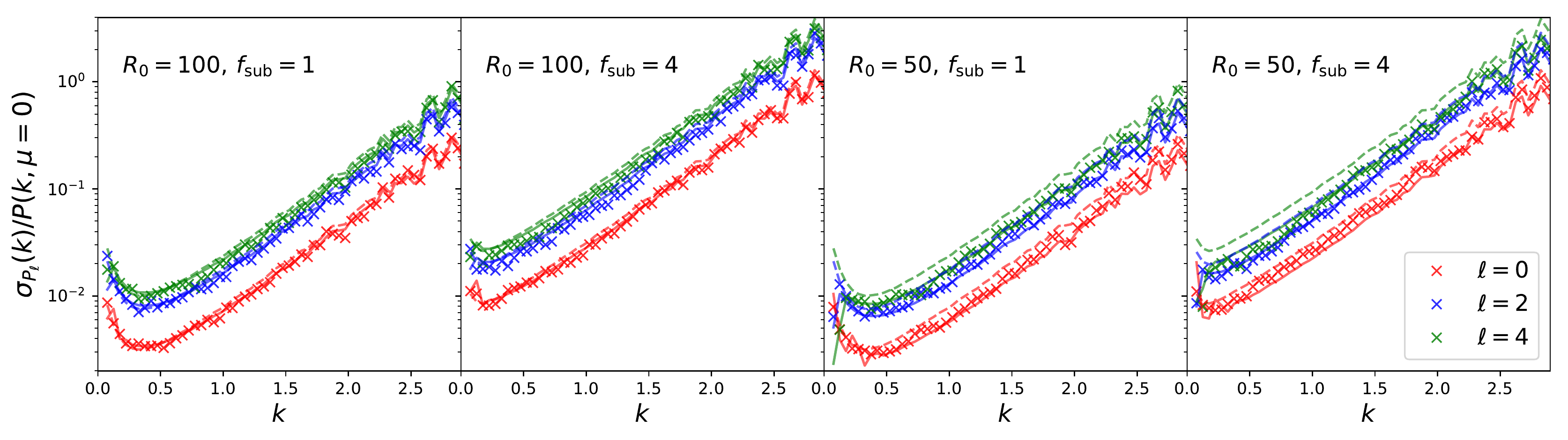}
    \caption{Standard deviation of power spectrum multipoles $P_\ell(k)$ from 100 N-body simulations (points), the theory model developed in Sec.\,\ref{sec: cov} of this paper (solid lines) and theory in the limit of $R_0\rightarrow\infty$ (dashed lines). The latter curves correspond to the expected errors if all pairs of points in the survey were counted, not just those within the truncation scale $R_0$. This uses the power spectrum multipoles from Fig.\,\ref{fig: pk_multipoles}, using truncation scales of $R_0 = 50\Mpch$ and $100\Mpch$. $f_\mathrm{sub}$ indicates the fraction by which the data is subsampled before the spectra are estimated. This does not affect the intrinsic covariance, but increases the Poissonian parts.}
    \label{fig: pk-cov}
\end{figure}

In Fig.\,\ref{fig: pk-cov} we plot the ratio of the standard deviation in $P_\ell(k)$ (across all spectra) to the real-space power $P(k,\mu=0) = \sum_{\ell=0} P_\ell(k)$, for various choices of the pair-count truncation scale $R_0$ and the subsampling parameter $f_\mathrm{sub}$.\footnote{Recall that subsampling by $f_\mathrm{sub}$ corresponds to using a random subset of $1/f_\mathrm{sub}$ of the dataset, and thus decreasing the number density by a factor $f_\mathrm{sub}$.} Alongside, we plot the estimates obtained from Eq.\,\ref{eq: cov-int}\,\&\,\ref{eq: cov-poiss} and their corresponding $R_0\rightarrow\infty$ limits, which are simply computed using numerical integration methods. For this purpose, we ignore the bispectrum and trispectrum terms, as well as the second two-point covariance term, since the former are difficult to model up to large $k$ and the latter is expected to be small. This is found to be a good approximation in practice.

From the figure, we firstly note that the model and true covariances are in very good agreement across the range of wavenumbers tested, even though we have ignored the non-Gaussian terms. In this instance, we are strongly dominated by the Poissonian two- and three-point covariances, since the halo number density ($n\sim 4\times 10^{-5}h^3\,\mathrm{Mpc}^{-3}$) is small. This is clearly seen by comparing the $f_\mathrm{sub} = 1$ and $f_\mathrm{sub} = 4$ plots, with the latter having a much large covariance due to the lower density ($n\rightarrow n/f_\mathrm{sub}$). For denser samples (e.g., for matter power spectra), the relative importance of the Poissonian terms is weaker, thus we are able to subsample the data without significantly altering the covariance at moderate $k$, although we always expect to be Poisson- or non-Gaussianity-dominated at large $k$. 

An additional interesting feature is that the measured standard deviations for finite $R_0$ are \textit{less} than those in the $R_0\rightarrow\infty$ limit and this discrepancy increases as $R_0$ falls. Whilst this may appear counter-intuitive as spectra computed with smaller $R_0$ naturally use less pairs of particles, it is fully explained by the additional off-diagonal covariance at smaller $R_0$ due to the window-function convolution. In practice, this convolution limits the useful bin-width to $\Delta k \sim 3/R_0$. We conclude that the standard deviations are well described by the covariance matrix model given in Sec.\,\ref{sec: cov} for a range of truncation scales and $k$-bins.

\subsection{Simulated Bispectra}

\subsubsection{Comparison of Methods}
To test the validity of our new bispectrum algorithm, we follow a similar procedure to the above, computing the isotropic bispectra from both conventional FFT-based approaches and \texttt{HIPSTER}. Since the computational effort involved in computing bispectra usually significantly exceeds that of power spectra, we consider only a single redshift-space halo catalog here, containing $\sim 4\times 10^5$ particles. The \texttt{HIPSTER} bispectra are computed using the algorithm discussed in Sec.\,\ref{sec: bk-algo}, with a total of 40 linearly spaced $k$-bins up to $k_\mathrm{max} = 0.8\hMpc$. We use the first seven Legendre multipoles $\ell = 0$ to $\ell = 6$, giving a total of $5740$ non-trivial bispectrum bins in $(k_1,k_2,\ell)$-space. In terms of hyperparameters, we adopt a ratio of randoms-to-galaxies of $f_\mathrm{rand} = 3$ (found to be sufficient in initial testing) and truncate particle pair counts at $R_0 = 100\Mpch$ which requires $\sim 20$ core-hours to compute with no data subsampling. (An analogous computation with $R_0 = 50\Mpch$ takes only $\sim 3$ core-hours on the same machine.)

For the conventional FFT-based calculation, we use the routines provided in the \texttt{Pylians} code \citep{2018ascl.soft11008V}, which is based on \citet{2017MNRAS.472.2436W}, involving loops of forward and reverse FFTs. We refer the reader to \citet{2005PhDT........23S}, \citet{2010PhDT.........4J} and \citet{2017MNRAS.472.2436W} for a full discussion of this procedure. FFT-based algorithms conventionally compute $B(k_1,k_2,k_3)$ (where $k_1,k_2,k_3$ satisfy triangle inequalities) or $B(k_1,k_2,\mu_{12})$, where $\mu_{12} = \hat{\vec k}_1\cdot\hat{\vec k}_2$, whilst our estimators require $B_\ell(k_1,k_2)$ (to make use of spherical harmonic decompositions and to aid visualization); conversion between the two is non-trivial, but possible via the integral
\beq
    B_\ell(k_1,k_2) = \frac{2\ell+1}{2}\int_{-1}^1d\mu_{12}\,B(k_1,k_2,\mu_{12})L_\ell(\mu_{12}) \approx \frac{2\ell+1}{2}\sum_{i=0}^N B(k_1,k_2,\mu_i)L_\ell(\mu_i)w_i.
\eeq
Here, we have approximated the integral via Gauss-Legendre quadrature, which defines a set of $N+1$ values of $\mu_i\in[-1,1]$ and corresponding weights $\{w_i\}$ from properties of Legendre polynomials and their derivatives. Here, we use tenth order quadrature which gives highly accurate results, requiring only $11$ $\mu$-bins to be computed for each $k_1,k_2$ pair. (Note that each $k_1,k_2$ pair can be computed independently, allowing for efficient parallelization.) As with the power spectrum, the FFT-based bispectrum estimator naturally picks up a significant shot-noise term, which, in periodic simulations, has the expected form
\beq
    B(k_1,k_2,k_3) = \frac{1}{n}\left[P(k_1)+P(k_2)+P(k_3)\right] + \frac{1}{n^2},
\eeq
for a simulation with number density $n$ \citep[e.g][]{2017PhRvD..96b3528C}. Here, $P(k)$ are the one-dimensional power spectra which are computed alongside the bispectra in the \texttt{Pylians} code. \resub{As discussed above, the \texttt{HIPSTER} algorithm does not include this shot-noise term, since it arises from self-counts (sets of particles at the same location, \textit{i.e.} $i=j$ in the sums over particle pairs $\{i,j\}$) that can be fully excluded in configuration-space as the data is not painted to a grid.} 
Due to the large number of FFTs that must be performed to compute the bispectrum in all combinations of bins, the process requires significant computation time; $\sim 200$ core-hours using $N_\mathrm{grid} = 512$ FFTs. For a given $k_1,k_2$ pair, computation of the $B_\ell(k_1,k_2)$ multipoles will involve a sum over triangles with $k_3\in[|k_1-k_2|,k_1+k_2]$, thus to avoid the Nyquist limit, we require $k_\mathrm{max} < k_\mathrm{Nyq}/2$, justifying our choice of $k_\mathrm{max}$.

Given that \texttt{HIPSTER} contains different assumptions to a standard FFT-based approach due to the window function (which leads to $k$-space convolutions and smoother results), it is useful to compare the results to an independent method that also implements a window function. To this end, we first compute the 3PCF multipoles $\zeta_\ell(r_1,r_2)$ via the $\mathcal{O}(N^2)$ algorithm of \citet{2015MNRAS.454.4142S}, then estimate the bispectrum via a numerical Fourier-space integral with the window function inserted. This gives the matrix product;
\beq\label{eq: bk-se-3pcf}
    B^\mathrm{3PCF}_\ell(k_1,k_2) \equiv (4\pi)^2(-1)^\ell \sum_i \left[x_i^2\Delta x_i W(x_i;R_0) j_\ell(k_1x_i)\right] \sum_j \left[x_j^2\Delta x_j W(x_j;R_0) j_\ell(k_2x_j)\right]\zeta^{ij}_\ell,
\eeq
where $\zeta_\ell^{ij}$ is the 3PCF estimate in bins $i,j$ centered at $x_i,x_j$ and the summations are over all bins with $x_i\leq R_0$. For simplicity we do not apply $k$-space binning (assuming $B_\ell^{ab} \approx B_\ell(k_a,k_b)$ where $k_x$ is the central value of bin $x$). Practically, we compute the 3PCF $RRR$ term analytically and mask the $i=j$ diagonal of $\zeta_\ell^{ij}$, since this has additional shot-noise contributions in the algorithm. Via this method, the bispectrum is computed in $\sim 10$ core-hours. Since the estimator contains a 3PCF discretized into $1\Mpch$ bins, we expect that the resulting spectrum will be inaccurate for $k_i\gtrsim 1\Mpch$ due to aliasing. This may be thought of as a discretized limit of \texttt{HIPSTER}; in Eq.\,\ref{eq: bk-se-3pcf} we compute the bispectrum by counting pairs of particles in given bins, before performing the Fourier integral, whilst in \texttt{HIPSTER} we perform the Fourier integral directly, effectively using infinite configuration-space bins. Whilst the 3PCF estimator asymptotically equals the \texttt{HIPSTER} algorithm in the limit of infinite 3PCF bins, this would require infinite memory, thus it is advantageous to perform the Fourier integral directly, as done by \texttt{HIPSTER}. 

\begin{figure}
    \centering
    \includegraphics[width=\textwidth]{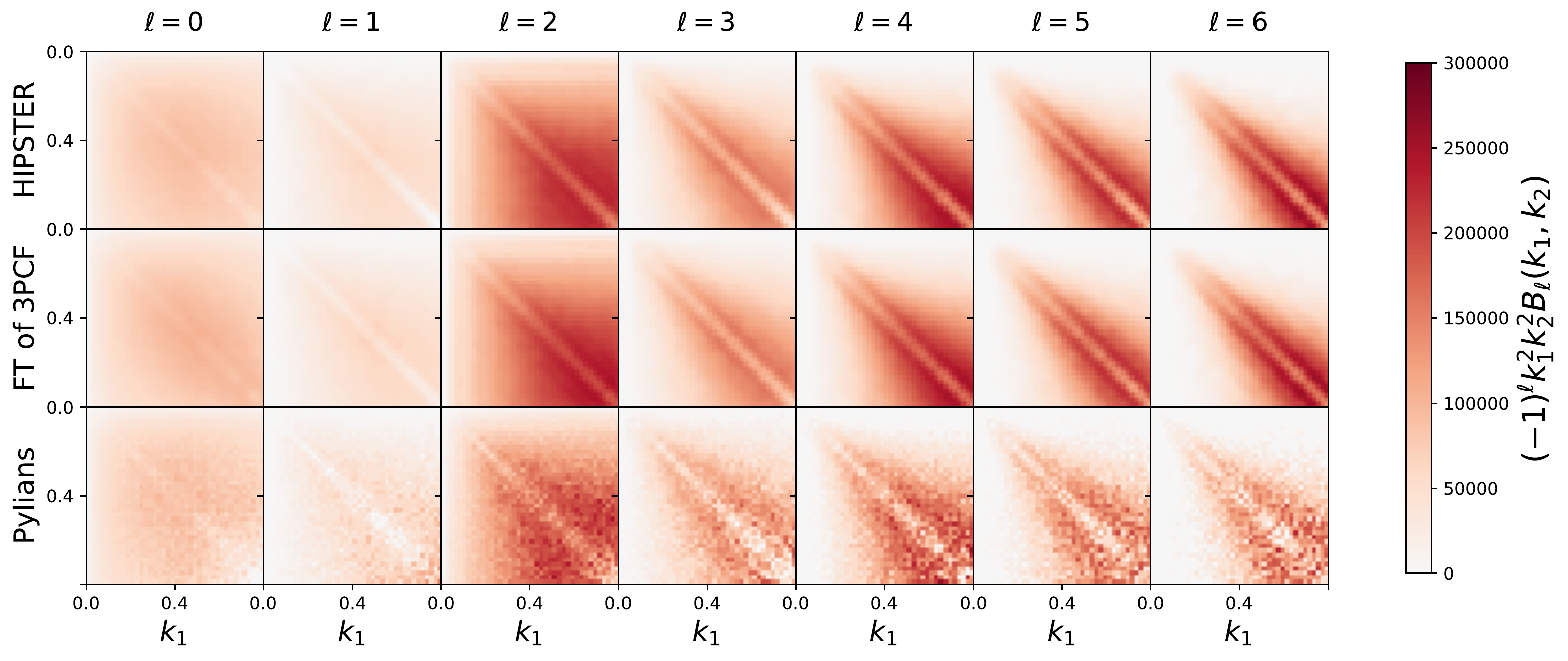}
    \caption{Bispectrum multipoles computed for a halo catalog using three approaches; the configuration-space estimators introduced in this work (top), numerical integration over a measured three-point correlation function (middle) and the canonical method using FFTs (bottom). The configuration-space \texttt{HIPSTER} routine is the fastest, and seen to produce comparable results with other routines at low-$k$. The first two routines are computed by counting pairs of particles up to $R_0 = 100\Mpch$ whilst \texttt{Pylians} uses repeated FFTs with a grid-size of $N_\mathrm{grid} = 512$. All $B_\ell(k_1,k_2)$ spectra are computed across 40 $k_1$ and $k_2$ bins (horizontal and vertical axes respectively) and we normalize by $(-1)^\ell k_1^2k_2^2$ to remove the leading scaling. Note that configuration-space estimates are naturally less noisy than those from FFTs, due to the implicit window function convolution. An analogous plot of the \texttt{HIPSTER} bispectrum up to high-$k$ is shown in Fig.\,\ref{fig: huge-bk}.}
    \label{fig: bispectrum-comparison}
\end{figure}

Fig.\,\ref{fig: bispectrum-comparison} compares the bispectra computed from the three methods. Firstly, we note good agreement between the spectra computed with \texttt{HIPSTER} and via numerical integration of the 3PCF across the range of $k$-bins and multipoles $\ell$ tested, implying that the configuration-space algorithm has been implemented successfully. Both methods include the same double convolution by the pair-separation window function (of width $R_0 = 100\Mpch$), as in Eq.\,\ref{eq: Bkk-convolution}. A more meaningful comparison is thus between the top and bottom plots of Fig.\,\ref{fig: bispectrum-comparison}, comparing the configuration-space and FFT-based approaches. We make two main observations: (1) the $k$ and $\ell$ dependencies of the bispectrum are comparable between the two approaches, especially at small $k$; (2) the \texttt{Pylians} FFT-based bispectra appear far more stochastic than those of \texttt{HIPSTER}, especially at large $k$ and $\ell$. The first indicates that \texttt{HIPSTER} is measuring the bispectrum correctly, though at small $k$, we expect small deviations from the convolution window (which are explored further in the subsequent section) and at large $k$ (roughly $k_1+k_2\gtrsim k_\mathrm{Nyq}/2 = 0.8\hMpc$, FFT aliasing starts to have an impact. \resub{To explain the different noise properties, we first note that there is an implicit convolution of the bispectrum with a smoothing window (primarily on scales $\Delta k \approx 3/R_0$ but with broad tails) that acts to `smear out' the underlying noise across bins. Secondly, whilst bispectrum estimates from FFT-based algorithms are fundamentally a sum of stochastic Fourier-space amplitudes that are binned to form a spectrum, the configuration-space estimators are sums over (many) smooth Bessel-like functions with random arguments. The smoothness in the basis functions leads to additional smoothness in the bispectra. Practically, these effects will be automatically taken into account in the sample covariances in any real analyses.}
\resub{The noise is further enhanced in Fig.\,\ref{fig: bispectrum-comparison} since we include only a single realization}: averaging over a number of simulations would significantly reduce it, though this is computationally expensive for FFT-type estimators. Based on the above, we conclude that the \texttt{HIPSTER} bispectrum estimator gives unbiased estimates of the true power spectra, surpassing conventional methods at moderate $k$-scales and above.

\subsubsection{Measuring the Bispectrum up to large $k$}
\begin{figure}
    \centering
    \includegraphics[width=\textwidth]{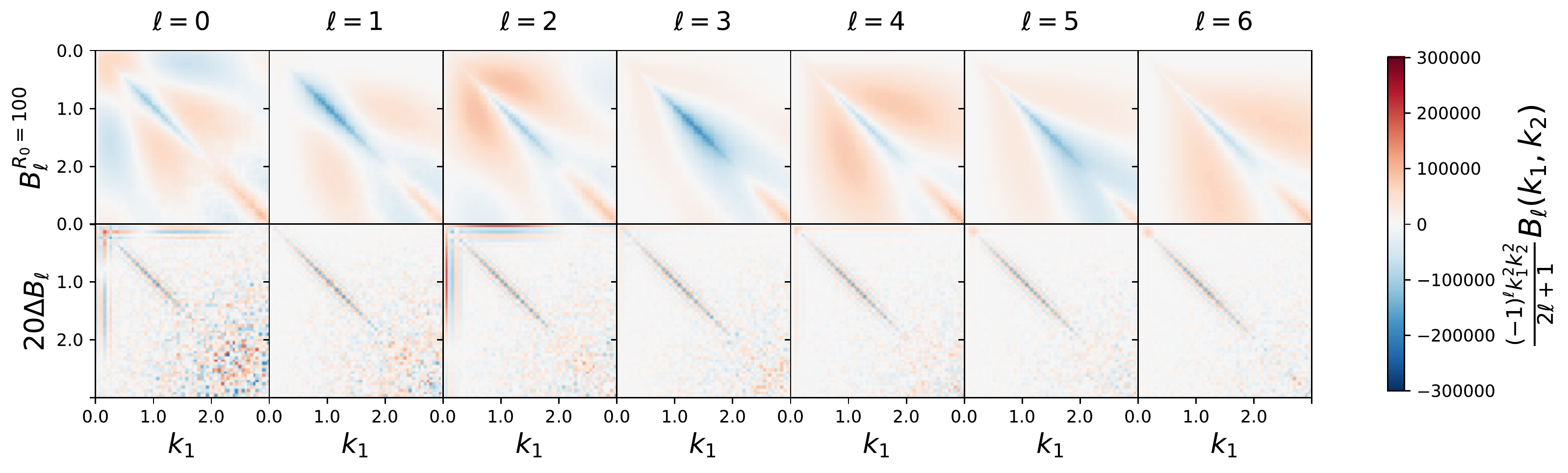}
    \caption{Bispectrum multipoles measured with the  configuration-space estimators introduced in this work up to large $k$. We display the multipoles computed by counting all particles up to $R_0 = 100\Mpch$ (top) and the difference between the bispectra at $R_0 = 100\Mpch$ and $R_0 = 50\Mpch$ (bottom, multiplied by 20 for visibility). This uses the average bispectrum across 100 halo catalogs in real-space, each containing $\sim 10^5$ halos at $z = 0$, subject to the selection function $M_\mathrm{halo} > 3.1\times 10^{13}h^{-1}M_\odot$. All spectra are computed using the \texttt{HIPSTER} code on a 20-core machine, with a run-time of $\sim 1$ and $5$ core-hours for $R_0 = 50\hMpc$ and $100\Mpch$ respectively. Note that we normalize by an additional factor of $(2\ell+1)$ compared to Fig.\,\ref{fig: bispectrum-comparison} \resub{and extend to $k_\mathrm{max}=3\hMpc$ rather than $0.8\hMpc$ (at which point regions with negative amplitudes become apparent).} The differences between the bispectra at the two truncation radii are well-understood, and discussed in the text.}
    \label{fig: huge-bk}
\end{figure}

With this in hand, we proceed to evaluate the bispectrum up to large $k$ using our configuration-space algorithm, implemented in \texttt{HIPSTER}. For this, we use 100 halo catalogs from the same simulations as before, this time using a mass cut of $M_\mathrm{min} = 3.1\times 10^{13}h^{-1}M_\odot$ to represent a more realistic galaxy sample. Since the isotropic bispectrum considered in this paper is insensitive to redshift-space distortions, we work in real-space for this test, though this will not affect the results. In the top panel of Fig.\,\ref{fig: huge-bk}, the bispectrum is plotted using 60 $k$-bins up to $k = 3\hMpc$ adopting a truncation radius of $R_0 = 100\Mpch$. This computation takes $\sim 5$ core-hours on a modern machine and we note that an analogous computation using FFTs would require $N_\mathrm{grid}\gtrsim 3800$ (setting $|k_1+k_2|<k_\mathrm{Nyq}/2$ to avoid aliasing) and $\sim 5000$ core-hours, which is computationally infeasible to run on a large number of simulations.

In the lower panel of Fig.\,\ref{fig: huge-bk}, we plot the difference in the bispectra obtained from $R_0 = 50\Mpch$ and $100\Mpch$. We first note that this is everywhere small compared to the bispectrum, indicating that accurate (and $\sim 8$ times faster) measurements can be obtained using the smaller radius. The first clear difference between the spectra is seen in the horizontal and vertical `stripes' at low $k$, which can be attributed to the window function convolution causing oscillations in the bispectra around the $R_0\rightarrow\infty$ limit. (Due to the sharply-peaked nature of the function in $k$-space, this is negligible for if all wavenumbers are greater than $\sim 0.3\hMpc$). Further, we observe a clear diagonal bias in the multipoles; this arises from the same effects, since the diagonal $k_1\approx k_2$ terms necessarily have contributions from $\hat{\vec k}_1\cdot\hat{\vec k}_2\approx 0$, i.e. modes with small $k_3$. If the bispectra were plotted in the full $(k_1,k_2,k_3)$-plane we would not observe this behavior away from the origin. Furthermore, we note increasing stochastic variations between the bispectra at large $k$. This can be explained by noting that (a) the two bispectra `see' a slightly different density field due to the different choices of $R_0$, and (b), the fractional error in the bispectrum scales as $(2\ell+1)/(k_1k_2)$. In the normalization factor used in Fig.\,\ref{fig: huge-bk}, we thus expect increasing statistical error with $k$. 

\section{Summary and Outlook}\label{sec: conclusion}
In this work, we have developed and implemented $\mathcal{O}(N^2)$ algorithms to compute the multipoles of the power spectrum and isotropic bispectrum that are optimized for cosmological simulations and highly efficient on small scales. This builds upon the work of \citet{2020MNRAS.492.1214P}, which introduced the technique of computing spectra via configuration-space weighted particle counts up to some maximum radius $R_0$. Such truncation allows for fast computation of the spectra and has negligible impact on the measurements at small-scales. By specializing to the case of simulations with periodic boundary conditions, our algorithms have been remarkably simplified, obviating the need for random particles in most aspects of the estimators. The bispectrum estimator has been developed in detail, following the realization that this can be written as a pair-count making use of spherical harmonic decompositions. Additionally, the covariances of the estimators have been discussed, and the key aspects verified with cosmological simulations.

The main benefits of our estimators, which are publicly available in the \texttt{HIPSTER}\footnote{\href{http://HIPSTER.readthedocs.io}{HIPSTER.readthedocs.io}} package are as follows:
\begin{itemize}
    \item \textbf{Speed}: Unlike Fourier-transform methods, the runtime of the configuration-space estimators decreases with the modulus of the wavenumber considered, since a smaller truncation scale $R_0$ can be used. Furthermore, a bispectrum measurement requires only two $\mathcal{O}(N^2)$ pair counts, rather than repeated Fourier transforms. Accurate galaxy power spectrum and bispectrum measurements are possible down to small scales ($k>1\hMpc$) in minutes using \texttt{HIPSTER}, and the runtime can be easily reduced by subsampling the data. 
    \item \textbf{Memory}: On small scales, Fourier-transform methods require a fine grid, whose memory scales as $N_\mathrm{grid}^3$, thus small-scale spectral computation must be performed on high-memory nodes. (As an example, consider $N_\mathrm{grid} = 1024$. At single precision, this requires almost $9\,\mathrm{GB}$ of memory per Fourier-space grid.) Since \texttt{HIPSTER} carries only the initial particles and arrays containing the output spectra, its memory usage is significantly less, and does not depend on scale.
    \item \textbf{Aliasing}: Since the estimators does not require gridding, the corresponding spectra do not suffer from aliasing. There is hence no notion of a Nyquist frequency, and the spectra can be computed up to arbitrarily large $k$.  
    \item \textbf{Accuracy}: \texttt{HIPSTER} has been rigorously tested and found to be in sub-percent agreement with FFT-based methods. At low $k$, we observe effects from the truncation of the pair-counts at $R_0$, but these are negligible for $kR_0\gtrsim 3$.
    \item \textbf{Shot-noise}: Since we work in configuration-space, we can easily exclude self-counts from the estimators. This avoids having to remove shot-noise in post-processing which is non-trivial for the bispectrum.
\end{itemize}

Whilst \texttt{HIPSTER} is excellent for measuring small-scale spectra, it is suboptimal on the largest scales since a great number of particle pairs must be counted. By combining FFT-based estimators with \texttt{HIPSTER} we can measure spectra efficiently on all scales, using both a small $N_\mathrm{grid}$ and low truncation radius $R_0$. This combined method requires orders of magnitude less computational resources than using either method alone. An additional application is to scale-free simulations, since \texttt{HIPSTER} is not limited by the fundamental frequency of the box, thus we are free to choose any desired $k$-space binning.

A number of extensions may be possible. An easily achievable one is the computation of the \textit{anisotropic} bispectrum, which includes the effects of redshift-space distortions \citep{2015PhRvD..92h3532S,2018MNRAS.478.1468S,2019MNRAS.484..364S}. Whilst this increases the dimensionality of the space by two (as we must parametrize the orientation of the redshift-space triangle with respect to the line-of-sight), it can be computed using a very similar algorithm requiring negligible additional computation time. In essence, this corresponds to replacing the sum over spherical harmonics in the bispectrum estimators with a product of two harmonics of different order, following \citet{2018MNRAS.478.1468S}. Furthermore, whilst we have focused on periodic simulations in this work, it is possible to compute the bispectrum in $\mathcal{O}(N^2)$ time for arbitrary survey geometries, which simply requires abandonment of the analytic random integrals discussed above and inclusion of sets of random particles, as done for the power spectrum in \citet{2020MNRAS.492.1214P}. We ought also to consider the bispectrum covariance in greater detail (including redshift-space effects, non-Gaussianity and the pair-separation window function), which, whilst algebraically difficult, is not intractable. We defer such considerations to future work.

\section*{Data Availability}
\resub{The data underlying this article will be shared on reasonable request to the corresponding author. The \texttt{HIPSTER} code is freely available, with documentation located at \href{HIPSTER.readthedocs.io}{HIPSTER.readthedocs.io}.}

\section*{Acknowledgements}
We thank Daniel Eisenstein, Lehman Garrison, David Spergel, Francisco Villaescusa-Navarro and, in particular, Zachary Slepian for insightful conversations and feedback on the manuscript. \resub{We additionally thank the anonymous referee for insightful comments that helped to improve the clarity of the manuscript.} OHEP acknowledges funding from the Herchel-Smith foundation and the Roman Telescope High Latitude Survey science team. 





\appendix
\section{An Analytic Form for the $\mathcal{E}^{II}_\ell$ kernel}\label{appen: E-II}
Here we consider the $\mathcal{E}^{II}$ kernel defined in Eq.\,\ref{eq: EII-intro}, which allows the bispectrum estimator (in particular the $\widetilde{DDR}^{II}$ term) to be written without any use of random particles. First, we write
\beq
    \mathcal{E}^{II}_\ell(\vec x_i,\vec x_j,k_1,k_2;R_0) &=& \int d\vec x_3\,j_\ell(k_1|\vec x_{i3}|)j_\ell(k_2|\vec x_{j3}|)W(|\vec x_{i3}|,R_0)W(|\vec x_{j3}|,R_0)L_\ell(\hat{\vec{x}}_{i3}\cdot\hat{\vec x}_{j3})\\\nonumber
    &=& \frac{4\pi}{2\ell+1}\int d\vec x_3\,\sum_{m=-\ell}^\ell j_\ell(k_1|\vec x_{i3}|)j_\ell(k_2|\vec x_{j3}|)W(|\vec x_{i3}|;R_0)W(|\vec x_{j3}|;R_0)Y^{}_{\ell m}(\hat{\vec x}_{i3})Y^{*}_{\ell m}(\hat{\vec x}_{j3})\\\nonumber
    &=& \frac{4\pi}{2\ell+1}\sum_{m=-\ell}^\ell\int d\vec x_3\,f^{}_{\ell m}(\vec x_i-\vec x_3;k_1)f_{\ell m}^{}(\vec x_3-\vec x_j;k_2),
\eeq
where we have used the spherical harmonic addition theorem in the second line \citep[Eq.\,14.30.9]{nist_dlmf} and defined the functions $f_{\ell m}$. We can express $f_{\ell m}$ in terms of its Fourier counterpart;
\beq
    \mathcal{E}^{II}_\ell(\vec x_i,\vec x_j,k_1,k_2;R_0) &=& \frac{4\pi}{2\ell+1}\sum_{m=-\ell}^\ell \int d\vec x_3\frac{d\vec p_1}{(2\pi)^3}\frac{d\vec p_2}{(2\pi)^3}\widetilde{f}^{}_{\ell m}(\vec p_1;k_1)\widetilde{f}^{*}_{\ell m}(\vec p_2;k_2)e^{i\vec p_1\cdot\vec x_{i3}}e^{-i\vec p_2\cdot\vec x_{j3}}\\\nonumber
    &=& \frac{4\pi}{2\ell+1}\sum_{m=-\ell}^\ell \int \frac{d\vec p}{(2\pi)^3}\widetilde{f}^{}_{\ell m}(\vec p;k_1)\widetilde{f}^{*}_{\ell m}(-\vec p;k_2)e^{i\vec p\cdot\vec x_{ij}},
\eeq
where we have integrated over $\vec x_3$ and the resulting Dirac delta function. Note that this result may be similarly obtained using the convolution theorem. To proceed we require expressions for $\widetilde{f}_{\ell m}$;
\beq
    \widetilde{f}_{\ell m}(\vec p;k) &=& \int d\vec r\,e^{-i\vec p\cdot\vec r}j_\ell(k|\vec r|)W(|\vec r|;R_0)Y_{\ell m}(\hat{\vec r})\\\nonumber
    &=& \sum_{L=0}^\infty\sum_{M=-L}^L 4\pi(-i)^L\int d\vec r\,j_L(p|\vec r|)Y^{}_{LM}(\hat{\vec p})Y^{*}_{LM}(\hat{\vec r})j_\ell(k|\vec r|)W(|\vec r|;R_0)Y_{\ell m}(\hat{\vec r})\\\nonumber
    &=& (-i)^\ell\left[4\pi \int_0^{R_0} r^2dr\,j_\ell(pr)j_\ell(kr)W(r;R_0)\right]Y_{\ell m}(\hat{\vec p}) = (-i)^\ell \omega_\ell(p;k,R_0)Y_{\ell m}(\hat{\vec p}),
\eeq
where $\omega_\ell$ was defined in Eq.\,\ref{eq: omega-ell-def}. This uses the plane-wave expansion of $e^{-i\vec p\cdot\vec r} = \sum_L(-i)^L(2L+1)j_L(pr)L_L(\hat{\vec p}\cdot\hat{\vec r})$ \citep[Eq.\,16.63]{arfken2013mathematical} and spherical harmonic orthonormality; $\int d\hat{\vec x}\,Y^{}_{\ell m}(\hat{\vec x})Y^{*}_{LM}(\hat{\vec x}) = \delta_{\ell L}^K\delta_{mM}^K$ \citep[Eq.\,14.30.8]{nist_dlmf}. We additionally restrict the domain of $r$ in the penultimate line, since $W(r)$ is zero for $r\geq R_0$. Given that the window function $W(\vec x;R_0)$ (Eq.\,\ref{eq: window_defn}) is a piecewise sum of polynomials, $\omega_\ell(p;k,R_0)$, and hence $\widetilde{f}_{\ell m}(\vec p;k)$, is expected to be analytic, using the techniques of \citet{2017arXiv170306428B}. Such an expression is highly complex however and not considered here. Using the expression above, we obtain the simplified kernel;
\beq\label{eq: DDRII-kernel}
    \mathcal{E}^{II}_\ell(\vec x_i,\vec x_j,k_1,k_2;R_0) &=& (-i)^{\ell-\ell}\frac{4\pi}{2\ell+1}\sum_{m=-\ell}^\ell \int \frac{d\vec p}{(2\pi)^3}\omega_\ell(p;k_1,R_0)\omega_\ell(p;k_2,R_0)Y_{\ell m}^{}(\hat{\vec p})Y_{\ell m}^{*}(\hat{\vec p})e^{i\vec p\cdot\vec x_{ij}}\\\nonumber
    &=& \int \frac{d\vec p}{(2\pi)^3}\omega_\ell(p;k_1,R_0)\omega_\ell(p;k_2,R_0)e^{i\vec p\cdot\vec x_{ij}} = \int \frac{p^2dp}{2\pi^2}j_0(p|\vec x_{ij}|)\omega_\ell(p;k_1,R_0)\omega_\ell(p;k_2,R_0),
\eeq
noting that $\sum_mY^{}_{\ell m}(\hat{\vec p})Y^{*}_{\ell m}(\hat{\vec p}) = (2\ell+1)/(4\pi)$ and that this is simply an inverse Fourier transform of a spherical function. The resulting kernel depends only on the distance $|\vec x_{ij}|$, and, if $\omega_{\ell m}(p;k,R_0)$ is known, reduces to a one-dimensional integral of (a large number of) sets of Bessel functions and polynomials. Numerical integration is not too difficult, since $\omega_\ell$ depend only on the $k$-bin and multipole used, of which there are a finite number. For speed, the kernel should be pre-computed for each set of $k$-bins and $\ell$ for an array of $|\vec x_i-\vec x_j|$ values and interpolated when needed. This is helped by the finite domain; $\mathcal{E}^{II}(\vec x_i,\vec x_j)=0$ for all $|\vec x_i-\vec x_j|\geq2R_0$. An alternative (but equivalent) expression for $\mathcal{E}^{II}$ is given in Appendix\,\ref{appen: alternative-cal-E-II}.

It is instructive to consider the special case of $R_0\rightarrow\infty$, practically corresponding to a pair-count over all pairs in the survey. Here $W(r)$ is unity for all $r$ and $\omega_\ell$ simplifies to
\beq\label{eq: omega-ell-lim}
    \lim_{R_0\rightarrow\infty} \omega_\ell(p;k,R_0) &=& 4\pi\int_0^\infty r^2dr j_\ell(pr)j_\ell(kr) = \frac{2\pi^2}{kp}\delta_D(k-p),
\eeq
by the closure relation for spherical Bessel functions.
\footnote{Note that $\delta_D(k-k')/k^2$ is simply the radial part of $\delta_D(\vec k+\vec k')$.} Inserting this into the expression for $\mathcal{E}_\ell^{II}$ gives
\beq
    \lim_{R_0\rightarrow\infty}\mathcal{E}_\ell^{II}(\vec x_i-\vec x_j,k_1,k_2) &=& \int \frac{p^2dp}{2\pi^2}j_0(p|\vec x_{ij}|)\frac{2\pi^2}{k_1^2}\frac{2\pi^2}{k_2^2}\delta_D(k_1-p)\delta_D(k_2-p) = \frac{2\pi^2}{k_1k_2}\delta_D(k_1-k_2)j_0(k_1|\vec x_{ij}|).
\eeq
Note that this (a) only contributes when $k_1=k_2$ and (b) has no dependence on $\ell$. For finite $R_0$, we thus expect the kernel to have little off-diagonal power and similar forms for each multipole moment (although we expect deviations as $k\rightarrow0$, since this is where the pair-count truncation is most important). 

When $k$-space binning is included, we obtain
\beq
    \mathcal{E}^{II,ab}_\ell(r;R_0) &=& \int \frac{p^2dp}{2\pi^2}j_0(pr)\omega_\ell^a(p;R_0)\omega_\ell^b(p;R_0),
\eeq
(where $\omega_\ell^a$ is defined in Eq.\,\ref{eq: omega-ell-a-def}), leading to
\beq
    \widetilde{DDR}_\ell^{II,ab}(R_0) &=& n(-1)^\ell(2\ell+1)\sum_{i\neq j}\mathcal{E}^{II,ab}_\ell(|\vec x_i-\vec x_j|;R_0).
\eeq
Using this approach, we remove all dependencies on a random catalog, though practically computation time is slow due to (a) compute a highly non-trivial analytic form for $\mathcal{E}^{II}$ or (b) store and call a large number of interpolators to evaluate $\mathcal{E}^{II}$.

\section{Alternative form for $\mathcal{E}^{II}_\ell$}\label{appen: alternative-cal-E-II}
We here derive an alternative, but equivalent form for the second $DDR$ kernel, $\mathcal{E}^{II}_\ell$. Starting from Eq.\,\ref{eq: DDRII-kernel} and inserting the definition of $\omega_\ell$ (Eq.\,\ref{eq: omega-ell-def}) we obtain
\beq
    \mathcal{E}^{II}_\ell(\vec x_i-\vec x_j,k_1,k_2;R_0) = 8\int_0^\infty p^2dp\,j_0(p|\vec x_{ij}|)\left[\int_0^{R_0} r_1^2dr_1\,j_\ell(pr_1)j_\ell(k_1r_1)W(r_1;R_0)\right]\left[\int_0^{R_0} r_2^2dr_2\,j_\ell(pr_2)j_\ell(k_2r_2)W(r_2;R_0)\right].
\eeq
Note that this contains an integral over all $p$ of the product of three spherical Bessel functions, which can be rewritten using the relation
\beq
    \int_0^\infty p^2dp\,j_0(p|\vec x_{ij}|)j_\ell(pr_1)j_\ell(pr_2) = \frac{\pi\beta(\Delta)}{4|\vec x_{ij}|r_1r_2}L_\ell(\Delta),
\eeq
for $\Delta = (r_1^2+r_2^2-|\vec x_{ij}|^2)/(2r_1r_2)$, Legendre polynomial $L_\ell$ and
\beq
    \beta(\Delta) = \begin{cases} \frac{1}{2}& \Delta=\pm 1\\ 1 & -1<\Delta<1\\0&\text{else}\end{cases}
\eeq
(\citealt{2009arXiv0909.0494M,2010JPhA...43S5204M,2017JCAP...11..039F}, a special case of \citealt[Eq.\,6.578.8]{2007tisp.book.....G}). Inserting this relation gives
\beq
    \mathcal{E}^{II}_\ell(\vec x_i-\vec x_j,k_1,k_2;R_0) &=& 2\pi \int_0^{R_0}\int_0^{R_0}dr_1dr_2\,\frac{r_1r_2}{|\vec x_{ij}|}\beta(\Delta)j_\ell(k_1r_1)j_\ell(k_2r_2)W(r_1;R_0)W(r_2;R_0)L_\ell(\Delta),
\eeq
simplifying the kernel to a two-dimensional integral, which must be evaluated numerically. Including $k$-space binning in bins $a$ and $b$, this becomes
\beq
    \mathcal{E}_\ell^{II,ab}(|\vec x_i-\vec x_j|;R_0) = 2\pi\int_0^{R_0}\int_0^{R_0}dr_1dr_2\,\frac{r_1r_2}{|\vec x_{ij}|}\beta(\Delta)j_\ell^a(r_1)j_\ell^b(r_2)W(r_1;R_0)W(r_2;R_0)L_\ell(\Delta),
\eeq
with $j_\ell^a$ defined in Eq.\,\ref{eq: jla-def}. This can be simplified somewhat further by consideration of the function $\beta(\Delta)$, which is non-zero only for $|r_1-r_2|\leq |\vec x_{ij}| \leq r_1+r_2$, giving;
\beq
    \mathcal{E}_\ell^{II,ab}(|\vec x_i-\vec x_j|;R_0) = \begin{cases} 2\pi\int_0^{R_0}\int_{| x_{ij}-r_1|}^{\operatorname{min}(R_0,r_1+ x_{ij})}dr_1dr_2\,\frac{r_1r_2}{|\vec x_{ij}|}j_\ell^a(r_1)j_\ell^b(r_2)W(r_1;R_0)W(r_2;R_0)L_\ell(\Delta)& x_{ij}\leq R_0\\
    2\pi\int_{x_{ij}-R_0}^{R_0}\int_{| x_{ij}-r_1|}^{R_0}dr_1dr_2\,\frac{r_1r_2}{|\vec x_{ij}|}j_\ell^a(r_1)j_\ell^b(r_2)W(r_1;R_0)W(r_2;R_0)L_\ell(\Delta)& \text{else.}
    \end{cases}
\eeq

\section{Recursive form for $D_\ell(u)$}\label{appen: Dellfn}
Below, we derive an analytic form for the $D_\ell(u)$ functions appearing in Eq.\,\ref{eq: jla-def}. Starting from the definition of $D_\ell(u)$ as an indefinite integral,
\beq
    D_\ell(u) \equiv \int u^2du\,j_\ell(u)
\eeq
we note that $j_\ell(u)$ satisfies the Sturm-Liouville equation for integer $\ell$;
\beq
    -\frac{d}{du}\left(u^2j_\ell'(u)\right) + \ell(\ell+1)j_\ell(u) = u^2j_\ell(u),
\eeq
\citep[Eq.\,10.47.1]{nist_dlmf} and hence
\beq
    D_\ell(u) = -u^2j'_\ell(u) + \ell(\ell+1)\int du\,j_\ell(u) = u^2j_{\ell+1}(u) - \ell u j_\ell(u) + \ell(\ell+1)\int du\,j_\ell(u),
\eeq
neglecting an arbitrary constant of integration and inserting a recursion relation for the Bessel function derivative \citep[Eq.\,10.51.2]{nist_dlmf}. Next, we define
\beq
    I_\ell(u) &\equiv& \int du\,j_\ell(u) = \frac{1}{\ell}\int du\,\left[(\ell-1)j_{\ell-2}(u) - (2\ell-1)j'_{\ell-1}(u)\right],
\eeq
using an additional recursion relation \citep[Eq.\,10.51.1][]{nist_dlmf} to substitute for $j_\ell(u)$. This yields the recursive definition
\beq
    D_\ell(u) &=& u^2j_{\ell+1}(u) - \ell u j_\ell(u) + \ell (\ell+1) I_{\ell}(u)\\\nonumber
    \ell I_\ell(u) &=& (\ell-1)I_{\ell-2}(u) - (2\ell-1)j_{\ell-1}(u)\\\nonumber
\eeq
for $\ell \geq 2$ with standard results $I_0(u) = \operatorname{Si}(u)$, $I_1(u) = -j_0(u)$, where $\operatorname{Si}(u)$ is the Sine integral.

\bibliographystyle{mnras}
\bibliography{adslib,other,oldlib} 

\begin{thebibliography}{}
\makeatletter
\relax
\def\mn@urlcharsother{\let\do\@makeother \do\$\do\&\do\#\do\^\do\_\do\%\do\~}
\def\mn@doi{\begingroup\mn@urlcharsother \@ifnextchar [ {\mn@doi@}
  {\mn@doi@[]}}
\def\mn@doi@[#1]#2{\def\@tempa{#1}\ifx\@tempa\@empty \href
  {http://dx.doi.org/#2} {doi:#2}\else \href {http://dx.doi.org/#2} {#1}\fi
  \endgroup}
\def\mn@eprint#1#2{\mn@eprint@#1:#2::\@nil}
\def\mn@eprint@arXiv#1{\href {http://arxiv.org/abs/#1} {{\tt arXiv:#1}}}
\def\mn@eprint@dblp#1{\href {http://dblp.uni-trier.de/rec/bibtex/#1.xml}
  {dblp:#1}}
\def\mn@eprint@#1:#2:#3:#4\@nil{\def\@tempa {#1}\def\@tempb {#2}\def\@tempc
  {#3}\ifx \@tempc \@empty \let \@tempc \@tempb \let \@tempb \@tempa \fi \ifx
  \@tempb \@empty \def\@tempb {arXiv}\fi \@ifundefined
  {mn@eprint@\@tempb}{\@tempb:\@tempc}{\expandafter \expandafter \csname
  mn@eprint@\@tempb\endcsname \expandafter{\@tempc}}}

\bibitem[\protect\citeauthoryear{{Alcock} \& {Paczynski}}{{Alcock} \&
  {Paczynski}}{1979}]{1979Natur.281..358A}
{Alcock} C.,  {Paczynski} B.,  1979, \mn@doi [\nat] {10.1038/281358a0}, \href
  {https://ui.adsabs.harvard.edu/abs/1979Natur.281..358A} {281, 358}

\bibitem[\protect\citeauthoryear{{Anderson} et~al.,}{{Anderson}
  et~al.}{2014}]{2014MNRAS.441...24A}
{Anderson} L.,  et~al., 2014, \mn@doi [\mnras] {10.1093/mnras/stu523}, \href
  {https://ui.adsabs.harvard.edu/abs/2014MNRAS.441...24A} {441, 24}

\bibitem[\protect\citeauthoryear{{Arfken}, {Weber}  \& {Harris}}{{Arfken}
  et~al.}{2013}]{arfken2013mathematical}
{Arfken} G.,  {Weber} H.,   {Harris} F.,  2013, Mathematical Methods for
  Physicists: A Comprehensive Guide.
Elsevier Science, \url {https://books.google.com/books?id=qLFo\_Z-PoGIC}

\bibitem[\protect\citeauthoryear{{Assassi}, {Simonovi{\'c}}  \&
  {Zaldarriaga}}{{Assassi} et~al.}{2017}]{2017JCAP...11..054A}
{Assassi} V.,  {Simonovi{\'c}} M.,   {Zaldarriaga} M.,  2017, \mn@doi [\jcap]
  {10.1088/1475-7516/2017/11/054}, \href
  {https://ui.adsabs.harvard.edu/abs/2017JCAP...11..054A} {2017, 054}

\bibitem[\protect\citeauthoryear{{Beutler} et~al.,}{{Beutler}
  et~al.}{2017}]{2017MNRAS.464.3409B}
{Beutler} F.,  et~al., 2017, \mn@doi [\mnras] {10.1093/mnras/stw2373}, \href
  {https://ui.adsabs.harvard.edu/abs/2017MNRAS.464.3409B} {464, 3409}

\bibitem[\protect\citeauthoryear{{Bianchi}, {Gil-Mar{\'\i}n}, {Ruggeri}  \&
  {Percival}}{{Bianchi} et~al.}{2015}]{2015MNRAS.453L..11B}
{Bianchi} D.,  {Gil-Mar{\'\i}n} H.,  {Ruggeri} R.,   {Percival} W.~J.,  2015,
  \mn@doi [\mnras] {10.1093/mnrasl/slv090}, \href
  {https://ui.adsabs.harvard.edu/abs/2015MNRAS.453L..11B} {453, L11}

\bibitem[\protect\citeauthoryear{{Blake} et~al.,}{{Blake}
  et~al.}{2011}]{2011MNRAS.415.2876B}
{Blake} C.,  et~al., 2011, \mn@doi [\mnras] {10.1111/j.1365-2966.2011.18903.x},
  \href {https://ui.adsabs.harvard.edu/abs/2011MNRAS.415.2876B} {415, 2876}

\bibitem[\protect\citeauthoryear{{Bloomfield}, {Face}  \& {Moss}}{{Bloomfield}
  et~al.}{2017}]{2017arXiv170306428B}
{Bloomfield} J.~K.,  {Face} S. H.~P.,   {Moss} Z.,  2017, arXiv e-prints, \href
  {https://ui.adsabs.harvard.edu/abs/2017arXiv170306428B} {p. arXiv:1703.06428}

\bibitem[\protect\citeauthoryear{{Chan} \& {Blot}}{{Chan} \&
  {Blot}}{2017}]{2017PhRvD..96b3528C}
{Chan} K.~C.,  {Blot} L.,  2017, \mn@doi [\prd] {10.1103/PhysRevD.96.023528},
  \href {https://ui.adsabs.harvard.edu/abs/2017PhRvD..96b3528C} {96, 023528}

\bibitem[\protect\citeauthoryear{{D'Amico}, {Gleyzes}, {Kokron}, {Markovic},
  {Senatore}, {Zhang}, {Beutler}  \& {Gil-Mar{\'\i}n}}{{D'Amico}
  et~al.}{2019}]{2019arXiv190905271D}
{D'Amico} G.,  {Gleyzes} J.,  {Kokron} N.,  {Markovic} D.,  {Senatore} L.,
  {Zhang} P.,  {Beutler} F.,   {Gil-Mar{\'\i}n} H.,  2019, arXiv e-prints,
  \href {https://ui.adsabs.harvard.edu/abs/2019arXiv190905271D} {p.
  arXiv:1909.05271}

\bibitem[\protect\citeauthoryear{{Davis}, {Efstathiou}, {Frenk}  \&
  {White}}{{Davis} et~al.}{1985}]{1985ApJ...292..371D}
{Davis} M.,  {Efstathiou} G.,  {Frenk} C.~S.,   {White} S.~D.~M.,  1985,
  \mn@doi [\apj] {10.1086/163168}, \href
  {https://ui.adsabs.harvard.edu/abs/1985ApJ...292..371D} {292, 371}

\bibitem[\protect\citeauthoryear{{DeRose} et~al.,}{{DeRose}
  et~al.}{2019}]{2019ApJ...875...69D}
{DeRose} J.,  et~al., 2019, \mn@doi [\apj] {10.3847/1538-4357/ab1085}, \href
  {https://ui.adsabs.harvard.edu/abs/2019ApJ...875...69D} {875, 69}

\bibitem[\protect\citeauthoryear{{Eisenstein} et~al.,}{{Eisenstein}
  et~al.}{2005}]{2005ApJ...633..560E}
{Eisenstein} D.~J.,  et~al., 2005, \mn@doi [\apj] {10.1086/466512}, \href
  {https://ui.adsabs.harvard.edu/abs/2005ApJ...633..560E} {633, 560}

\bibitem[\protect\citeauthoryear{Fabrikant}{Fabrikant}{2013}]{fabrikant}
Fabrikant V.,  2013, \mn@doi [Quarterly of Applied Mathematics]
  {10.1090/S0033-569X-2012-01300-8}, 71

\bibitem[\protect\citeauthoryear{{Feldman}, {Kaiser}  \& {Peacock}}{{Feldman}
  et~al.}{1994}]{1994ApJ...426...23F}
{Feldman} H.~A.,  {Kaiser} N.,   {Peacock} J.~A.,  1994, \mn@doi [\apj]
  {10.1086/174036}, \href
  {https://ui.adsabs.harvard.edu/abs/1994ApJ...426...23F} {426, 23}

\bibitem[\protect\citeauthoryear{{Fergusson}, {Regan}  \&
  {Shellard}}{{Fergusson} et~al.}{2012}]{2012PhRvD..86f3511F}
{Fergusson} J.~R.,  {Regan} D.~M.,   {Shellard} E.~P.~S.,  2012, \mn@doi [\prd]
  {10.1103/PhysRevD.86.063511}, \href
  {https://ui.adsabs.harvard.edu/abs/2012PhRvD..86f3511F} {86, 063511}

\bibitem[\protect\citeauthoryear{{Fonseca de la Bella}, {Regan}, {Seery}  \&
  {Hotchkiss}}{{Fonseca de la Bella} et~al.}{2017}]{2017JCAP...11..039F}
{Fonseca de la Bella} L.,  {Regan} D.,  {Seery} D.,   {Hotchkiss} S.,  2017,
  \mn@doi [\jcap] {10.1088/1475-7516/2017/11/039}, \href
  {https://ui.adsabs.harvard.edu/abs/2017JCAP...11..039F} {2017, 039}

\bibitem[\protect\citeauthoryear{{Fonseca de la Bella}, {Regan}, {Seery}  \&
  {Parkinson}}{{Fonseca de la Bella} et~al.}{2018}]{2018arXiv180512394F}
{Fonseca de la Bella} L.,  {Regan} D.,  {Seery} D.,   {Parkinson} D.,  2018,
  arXiv e-prints, \href {https://ui.adsabs.harvard.edu/abs/2018arXiv180512394F}
  {p. arXiv:1805.12394}

\bibitem[\protect\citeauthoryear{{Garrison}, {Eisenstein}, {Ferrer}, {Tinker},
  {Pinto}  \& {Weinberg}}{{Garrison} et~al.}{2018}]{2018ApJS..236...43G}
{Garrison} L.~H.,  {Eisenstein} D.~J.,  {Ferrer} D.,  {Tinker} J.~L.,  {Pinto}
  P.~A.,   {Weinberg} D.~H.,  2018, \mn@doi [\apjs] {10.3847/1538-4365/aabfd3},
  \href {https://ui.adsabs.harvard.edu/abs/2018ApJS..236...43G} {236, 43}

\bibitem[\protect\citeauthoryear{{Gil-Mar{\'\i}n}, {Nore{\~n}a}, {Verde},
  {Percival}, {Wagner}, {Manera}  \& {Schneider}}{{Gil-Mar{\'\i}n}
  et~al.}{2015a}]{2015MNRAS.451..539G}
{Gil-Mar{\'\i}n} H.,  {Nore{\~n}a} J.,  {Verde} L.,  {Percival} W.~J.,
  {Wagner} C.,  {Manera} M.,   {Schneider} D.~P.,  2015a, \mn@doi [\mnras]
  {10.1093/mnras/stv961}, \href
  {https://ui.adsabs.harvard.edu/abs/2015MNRAS.451..539G} {451, 539}

\bibitem[\protect\citeauthoryear{{Gil-Mar{\'\i}n} et~al.,}{{Gil-Mar{\'\i}n}
  et~al.}{2015b}]{2015MNRAS.452.1914G}
{Gil-Mar{\'\i}n} H.,  et~al., 2015b, \mn@doi [\mnras] {10.1093/mnras/stv1359},
  \href {https://ui.adsabs.harvard.edu/abs/2015MNRAS.452.1914G} {452, 1914}

\bibitem[\protect\citeauthoryear{{Gil-Mar{\'\i}n} et~al.,}{{Gil-Mar{\'\i}n}
  et~al.}{2016}]{2016MNRAS.460.4210G}
{Gil-Mar{\'\i}n} H.,  et~al., 2016, \mn@doi [\mnras] {10.1093/mnras/stw1264},
  \href {https://ui.adsabs.harvard.edu/abs/2016MNRAS.460.4210G} {460, 4210}

\bibitem[\protect\citeauthoryear{{Gradshteyn}, {Ryzhik}, {Jeffrey}  \&
  {Zwillinger}}{{Gradshteyn} et~al.}{2007}]{2007tisp.book.....G}
{Gradshteyn} I.~S.,  {Ryzhik} I.~M.,  {Jeffrey} A.,   {Zwillinger} D.,  2007,
  {Table of Integrals, Series, and Products}

\bibitem[\protect\citeauthoryear{{Hamana} et~al.,}{{Hamana}
  et~al.}{2020}]{2020PASJ...72...16H}
{Hamana} T.,  et~al., 2020, \mn@doi [\pasj] {10.1093/pasj/psz138}, \href
  {https://ui.adsabs.harvard.edu/abs/2020PASJ...72...16H} {72, 16}

\bibitem[\protect\citeauthoryear{{Hand}, {Li}, {Slepian}  \& {Seljak}}{{Hand}
  et~al.}{2017}]{2017JCAP...07..002H}
{Hand} N.,  {Li} Y.,  {Slepian} Z.,   {Seljak} U.,  2017, \mn@doi [\jcap]
  {10.1088/1475-7516/2017/07/002}, \href
  {https://ui.adsabs.harvard.edu/abs/2017JCAP...07..002H} {2017, 002}

\bibitem[\protect\citeauthoryear{{Hand}, {Feng}, {Beutler}, {Li}, {Modi},
  {Seljak}  \& {Slepian}}{{Hand} et~al.}{2018}]{2018AJ....156..160H}
{Hand} N.,  {Feng} Y.,  {Beutler} F.,  {Li} Y.,  {Modi} C.,  {Seljak} U.,
  {Slepian} Z.,  2018, \mn@doi [\aj] {10.3847/1538-3881/aadae0}, \href
  {https://ui.adsabs.harvard.edu/abs/2018AJ....156..160H} {156, 160}

\bibitem[\protect\citeauthoryear{{Hikage} et~al.,}{{Hikage}
  et~al.}{2019}]{2019PASJ...71...43H}
{Hikage} C.,  et~al., 2019, \mn@doi [\pasj] {10.1093/pasj/psz010}, \href
  {https://ui.adsabs.harvard.edu/abs/2019PASJ...71...43H} {71, 43}

\bibitem[\protect\citeauthoryear{{Hung}, {Fergusson}  \& {Shellard}}{{Hung}
  et~al.}{2019}]{2019arXiv190201830H}
{Hung} J.,  {Fergusson} J.~R.,   {Shellard} E.~P.~S.,  2019, arXiv e-prints,
  \href {https://ui.adsabs.harvard.edu/abs/2019arXiv190201830H} {p.
  arXiv:1902.01830}

\bibitem[\protect\citeauthoryear{{Ivanov}, {Simonovi{\'c}}  \&
  {Zaldarriaga}}{{Ivanov} et~al.}{2019a}]{2019arXiv190905277I}
{Ivanov} M.~M.,  {Simonovi{\'c}} M.,   {Zaldarriaga} M.,  2019a, arXiv
  e-prints, \href {https://ui.adsabs.harvard.edu/abs/2019arXiv190905277I} {p.
  arXiv:1909.05277}

\bibitem[\protect\citeauthoryear{{Ivanov}, {Simonovi{\'c}}  \&
  {Zaldarriaga}}{{Ivanov} et~al.}{2019b}]{2019arXiv191208208I}
{Ivanov} M.~M.,  {Simonovi{\'c}} M.,   {Zaldarriaga} M.,  2019b, arXiv
  e-prints, \href {https://ui.adsabs.harvard.edu/abs/2019arXiv191208208I} {p.
  arXiv:1912.08208}

\bibitem[\protect\citeauthoryear{{Jeong}}{{Jeong}}{2010}]{2010PhDT.........4J}
{Jeong} D.,  2010, PhD thesis, University of Texas at Austin

\bibitem[\protect\citeauthoryear{{Jing} \& {B{\"o}rner}}{{Jing} \&
  {B{\"o}rner}}{2001}]{2001MNRAS.325.1389J}
{Jing} Y.~P.,  {B{\"o}rner} G.,  2001, \mn@doi [\mnras]
  {10.1046/j.1365-8711.2001.04521.x}, \href
  {https://ui.adsabs.harvard.edu/abs/2001MNRAS.325.1389J} {325, 1389}

\bibitem[\protect\citeauthoryear{{Kaiser}}{{Kaiser}}{1987}]{1987MNRAS.227....1K}
{Kaiser} N.,  1987, \mn@doi [\mnras] {10.1093/mnras/227.1.1}, \href
  {https://ui.adsabs.harvard.edu/abs/1987MNRAS.227....1K} {227, 1}

\bibitem[\protect\citeauthoryear{{Landy} \& {Szalay}}{{Landy} \&
  {Szalay}}{1993}]{1993ApJ...412...64L}
{Landy} S.~D.,  {Szalay} A.~S.,  1993, \mn@doi [\apj] {10.1086/172900}, \href
  {https://ui.adsabs.harvard.edu/abs/1993ApJ...412...64L} {412, 64}

\bibitem[\protect\citeauthoryear{{Li}, {Hu}  \& {Takada}}{{Li}
  et~al.}{2014}]{2014PhRvD..89h3519L}
{Li} Y.,  {Hu} W.,   {Takada} M.,  2014, \mn@doi [\prd]
  {10.1103/PhysRevD.89.083519}, \href
  {https://ui.adsabs.harvard.edu/abs/2014PhRvD..89h3519L} {89, 083519}

\bibitem[\protect\citeauthoryear{{Li}, {Jing}, {Zhang}  \& {Cheng}}{{Li}
  et~al.}{2016}]{2016ApJ...833..287L}
{Li} Z.,  {Jing} Y.~P.,  {Zhang} P.,   {Cheng} D.,  2016, \mn@doi [\apj]
  {10.3847/1538-4357/833/2/287}, \href
  {https://ui.adsabs.harvard.edu/abs/2016ApJ...833..287L} {833, 287}

\bibitem[\protect\citeauthoryear{{Li}, {Schmittfull}  \& {Seljak}}{{Li}
  et~al.}{2018}]{2018JCAP...02..022L}
{Li} Y.,  {Schmittfull} M.,   {Seljak} U.,  2018, \mn@doi [\jcap]
  {10.1088/1475-7516/2018/02/022}, \href
  {https://ui.adsabs.harvard.edu/abs/2018JCAP...02..022L} {2018, 022}

\bibitem[\protect\citeauthoryear{{Li}, {Singh}, {Yu}, {Feng}  \& {Seljak}}{{Li}
  et~al.}{2019}]{2019JCAP...01..016L}
{Li} Y.,  {Singh} S.,  {Yu} B.,  {Feng} Y.,   {Seljak} U.,  2019, \mn@doi
  [\jcap] {10.1088/1475-7516/2019/01/016}, \href
  {https://ui.adsabs.harvard.edu/abs/2019JCAP...01..016L} {2019, 016}

\bibitem[\protect\citeauthoryear{{McAlpine} et~al.,}{{McAlpine}
  et~al.}{2016}]{2016A&C....15...72M}
{McAlpine} S.,  et~al., 2016, \mn@doi [Astronomy and Computing]
  {10.1016/j.ascom.2016.02.004}, \href
  {https://ui.adsabs.harvard.edu/abs/2016A&C....15...72M} {15, 72}

\bibitem[\protect\citeauthoryear{{McCarthy}, {Schaye}, {Bird}  \& {Le
  Brun}}{{McCarthy} et~al.}{2017}]{2017MNRAS.465.2936M}
{McCarthy} I.~G.,  {Schaye} J.,  {Bird} S.,   {Le Brun} A. M.~C.,  2017,
  \mn@doi [\mnras] {10.1093/mnras/stw2792}, \href
  {https://ui.adsabs.harvard.edu/abs/2017MNRAS.465.2936M} {465, 2936}

\bibitem[\protect\citeauthoryear{{Mehrem}}{{Mehrem}}{2009}]{2009arXiv0909.0494M}
{Mehrem} R.,  2009, arXiv e-prints, \href
  {https://ui.adsabs.harvard.edu/abs/2009arXiv0909.0494M} {p. arXiv:0909.0494}

\bibitem[\protect\citeauthoryear{{Mehrem} \& {Hohenegger}}{{Mehrem} \&
  {Hohenegger}}{2010}]{2010JPhA...43S5204M}
{Mehrem} R.,  {Hohenegger} A.,  2010, \mn@doi [Journal of Physics A
  Mathematical General] {10.1088/1751-8113/43/45/455204}, \href
  {https://ui.adsabs.harvard.edu/abs/2010JPhA...43S5204M} {43, 455204}

\bibitem[\protect\citeauthoryear{{Miyatake} et~al.,}{{Miyatake}
  et~al.}{2015}]{2015ApJ...806....1M}
{Miyatake} H.,  et~al., 2015, \mn@doi [\apj] {10.1088/0004-637X/806/1/1}, \href
  {https://ui.adsabs.harvard.edu/abs/2015ApJ...806....1M} {806, 1}

\bibitem[\protect\citeauthoryear{{More}, {Miyatake}, {Mandelbaum}, {Takada},
  {Spergel}, {Brownstein}  \& {Schneider}}{{More}
  et~al.}{2015}]{2015ApJ...806....2M}
{More} S.,  {Miyatake} H.,  {Mandelbaum} R.,  {Takada} M.,  {Spergel} D.~N.,
  {Brownstein} J.~R.,   {Schneider} D.~P.,  2015, \mn@doi [\apj]
  {10.1088/0004-637X/806/1/2}, \href
  {https://ui.adsabs.harvard.edu/abs/2015ApJ...806....2M} {806, 2}

\bibitem[\protect\citeauthoryear{{NIST}}{{NIST}}{DLMF}]{nist_dlmf}
{NIST} DLMF, NIST Digital Library of Mathematical Functions.
\url {http://dlmf.nist.gov/}

\bibitem[\protect\citeauthoryear{{Nishimichi} \& {Oka}}{{Nishimichi} \&
  {Oka}}{2014}]{2014MNRAS.444.1400N}
{Nishimichi} T.,  {Oka} A.,  2014, \mn@doi [\mnras] {10.1093/mnras/stu1528},
  \href {https://ui.adsabs.harvard.edu/abs/2014MNRAS.444.1400N} {444, 1400}

\bibitem[\protect\citeauthoryear{{O'Connell} \& {Eisenstein}}{{O'Connell} \&
  {Eisenstein}}{2019}]{2019MNRAS.487.2701O}
{O'Connell} R.,  {Eisenstein} D.~J.,  2019, \mn@doi [\mnras]
  {10.1093/mnras/stz1359}, \href
  {https://ui.adsabs.harvard.edu/abs/2019MNRAS.487.2701O} {487, 2701}

\bibitem[\protect\citeauthoryear{{O'Connell}, {Eisenstein}, {Vargas}, {Ho}  \&
  {Padmanabhan}}{{O'Connell} et~al.}{2016}]{2016MNRAS.462.2681O}
{O'Connell} R.,  {Eisenstein} D.,  {Vargas} M.,  {Ho} S.,   {Padmanabhan} N.,
  2016, \mn@doi [\mnras] {10.1093/mnras/stw1821}, \href
  {https://ui.adsabs.harvard.edu/abs/2016MNRAS.462.2681O} {462, 2681}

\bibitem[\protect\citeauthoryear{{Pearson} \& {Samushia}}{{Pearson} \&
  {Samushia}}{2018}]{2018MNRAS.478.4500P}
{Pearson} D.~W.,  {Samushia} L.,  2018, \mn@doi [\mnras]
  {10.1093/mnras/sty1266}, \href
  {https://ui.adsabs.harvard.edu/abs/2018MNRAS.478.4500P} {478, 4500}

\bibitem[\protect\citeauthoryear{{Pearson} \& {Samushia}}{{Pearson} \&
  {Samushia}}{2019}]{2019MNRAS.486L.105P}
{Pearson} D.~W.,  {Samushia} L.,  2019, \mn@doi [\mnras]
  {10.1093/mnrasl/slz067}, \href
  {https://ui.adsabs.harvard.edu/abs/2019MNRAS.486L.105P} {486, L105}

\bibitem[\protect\citeauthoryear{{Philcox} \& {Eisenstein}}{{Philcox} \&
  {Eisenstein}}{2019}]{2019MNRAS.490.5931P}
{Philcox} O. H.~E.,  {Eisenstein} D.~J.,  2019, \mn@doi [\mnras]
  {10.1093/mnras/stz2896}, \href
  {https://ui.adsabs.harvard.edu/abs/2019MNRAS.490.5931P} {490, 5931}

\bibitem[\protect\citeauthoryear{{Philcox} \& {Eisenstein}}{{Philcox} \&
  {Eisenstein}}{2020}]{2020MNRAS.492.1214P}
{Philcox} O. H.~E.,  {Eisenstein} D.~J.,  2020, \mn@doi [\mnras]
  {10.1093/mnras/stz3335}, \href
  {https://ui.adsabs.harvard.edu/abs/2020MNRAS.492.1214P} {492, 1214}

\bibitem[\protect\citeauthoryear{{Philcox}, {Ivanov}, {Simonovi{\'c}}  \&
  {Zaldarriaga}}{{Philcox} et~al.}{2020a}]{2020arXiv200204035P}
{Philcox} O. H.~E.,  {Ivanov} M.~M.,  {Simonovi{\'c}} M.,   {Zaldarriaga} M.,
  2020a, arXiv e-prints, \href
  {https://ui.adsabs.harvard.edu/abs/2020arXiv200204035P} {p. arXiv:2002.04035}

\bibitem[\protect\citeauthoryear{{Philcox}, {Eisenstein}, {O'Connell}  \&
  {Wiegand}}{{Philcox} et~al.}{2020b}]{2020MNRAS.491.3290P}
{Philcox} O. H.~E.,  {Eisenstein} D.~J.,  {O'Connell} R.,   {Wiegand} A.,
  2020b, \mn@doi [\mnras] {10.1093/mnras/stz3218}, \href
  {https://ui.adsabs.harvard.edu/abs/2020MNRAS.491.3290P} {491, 3290}

\bibitem[\protect\citeauthoryear{{Portillo}, {Slepian}, {Burkhart}, {Kahraman}
  \& {Finkbeiner}}{{Portillo} et~al.}{2018}]{2018ApJ...862..119P}
{Portillo} S. K.~N.,  {Slepian} Z.,  {Burkhart} B.,  {Kahraman} S.,
  {Finkbeiner} D.~P.,  2018, \mn@doi [\apj] {10.3847/1538-4357/aacb80}, \href
  {https://ui.adsabs.harvard.edu/abs/2018ApJ...862..119P} {862, 119}

\bibitem[\protect\citeauthoryear{{Schmittfull}, {Regan}  \&
  {Shellard}}{{Schmittfull} et~al.}{2013}]{2013PhRvD..88f3512S}
{Schmittfull} M.~M.,  {Regan} D.~M.,   {Shellard} E.~P.~S.,  2013, \mn@doi
  [\prd] {10.1103/PhysRevD.88.063512}, \href
  {https://ui.adsabs.harvard.edu/abs/2013PhRvD..88f3512S} {88, 063512}

\bibitem[\protect\citeauthoryear{{Scoccimarro}}{{Scoccimarro}}{2015}]{2015PhRvD..92h3532S}
{Scoccimarro} R.,  2015, \mn@doi [\prd] {10.1103/PhysRevD.92.083532}, \href
  {https://ui.adsabs.harvard.edu/abs/2015PhRvD..92h3532S} {92, 083532}

\bibitem[\protect\citeauthoryear{{Scoccimarro}, {Couchman}  \&
  {Frieman}}{{Scoccimarro} et~al.}{1999a}]{1999ApJ...517..531S}
{Scoccimarro} R.,  {Couchman} H.~M.~P.,   {Frieman} J.~A.,  1999a, \mn@doi
  [\apj] {10.1086/307220}, \href
  {https://ui.adsabs.harvard.edu/abs/1999ApJ...517..531S} {517, 531}

\bibitem[\protect\citeauthoryear{{Scoccimarro}, {Zaldarriaga}  \&
  {Hui}}{{Scoccimarro} et~al.}{1999b}]{1999ApJ...527....1S}
{Scoccimarro} R.,  {Zaldarriaga} M.,   {Hui} L.,  1999b, \mn@doi [\apj]
  {10.1086/308059}, \href
  {https://ui.adsabs.harvard.edu/abs/1999ApJ...527....1S} {527, 1}

\bibitem[\protect\citeauthoryear{{Scoccimarro}, {Feldman}, {Fry}  \&
  {Frieman}}{{Scoccimarro} et~al.}{2001}]{2001ApJ...546..652S}
{Scoccimarro} R.,  {Feldman} H.~A.,  {Fry} J.~N.,   {Frieman} J.~A.,  2001,
  \mn@doi [\apj] {10.1086/318284}, \href
  {https://ui.adsabs.harvard.edu/abs/2001ApJ...546..652S} {546, 652}

\bibitem[\protect\citeauthoryear{{Sefusatti}}{{Sefusatti}}{2005}]{2005PhDT........23S}
{Sefusatti} E.,  2005, PhD thesis, New York University, New York, USA

\bibitem[\protect\citeauthoryear{{Sefusatti}, {Crocce}, {Pueblas}  \&
  {Scoccimarro}}{{Sefusatti} et~al.}{2006}]{2006PhRvD..74b3522S}
{Sefusatti} E.,  {Crocce} M.,  {Pueblas} S.,   {Scoccimarro} R.,  2006, \mn@doi
  [\prd] {10.1103/PhysRevD.74.023522}, \href
  {https://ui.adsabs.harvard.edu/abs/2006PhRvD..74b3522S} {74, 023522}

\bibitem[\protect\citeauthoryear{{Sefusatti}, {Crocce}, {Scoccimarro}  \&
  {Couchman}}{{Sefusatti} et~al.}{2016}]{2016MNRAS.460.3624S}
{Sefusatti} E.,  {Crocce} M.,  {Scoccimarro} R.,   {Couchman} H.~M.~P.,  2016,
  \mn@doi [\mnras] {10.1093/mnras/stw1229}, \href
  {https://ui.adsabs.harvard.edu/abs/2016MNRAS.460.3624S} {460, 3624}

\bibitem[\protect\citeauthoryear{{Slepian} \& {Eisenstein}}{{Slepian} \&
  {Eisenstein}}{2015}]{2015MNRAS.454.4142S}
{Slepian} Z.,  {Eisenstein} D.~J.,  2015, \mn@doi [\mnras]
  {10.1093/mnras/stv2119}, \href
  {https://ui.adsabs.harvard.edu/abs/2015MNRAS.454.4142S} {454, 4142}

\bibitem[\protect\citeauthoryear{{Slepian} \& {Eisenstein}}{{Slepian} \&
  {Eisenstein}}{2016}]{2016MNRAS.455L..31S}
{Slepian} Z.,  {Eisenstein} D.~J.,  2016, \mn@doi [\mnras]
  {10.1093/mnrasl/slv133}, \href
  {https://ui.adsabs.harvard.edu/abs/2016MNRAS.455L..31S} {455, L31}

\bibitem[\protect\citeauthoryear{{Slepian} \& {Eisenstein}}{{Slepian} \&
  {Eisenstein}}{2017}]{2017MNRAS.469.2059S}
{Slepian} Z.,  {Eisenstein} D.~J.,  2017, \mn@doi [\mnras]
  {10.1093/mnras/stx490}, \href
  {https://ui.adsabs.harvard.edu/abs/2017MNRAS.469.2059S} {469, 2059}

\bibitem[\protect\citeauthoryear{{Slepian} \& {Eisenstein}}{{Slepian} \&
  {Eisenstein}}{2018}]{2018MNRAS.478.1468S}
{Slepian} Z.,  {Eisenstein} D.~J.,  2018, \mn@doi [\mnras]
  {10.1093/mnras/sty1063}, \href
  {https://ui.adsabs.harvard.edu/abs/2018MNRAS.478.1468S} {478, 1468}

\bibitem[\protect\citeauthoryear{{Slepian} et~al.,}{{Slepian}
  et~al.}{2017}]{2017MNRAS.469.1738S}
{Slepian} Z.,  et~al., 2017, \mn@doi [\mnras] {10.1093/mnras/stx488}, \href
  {https://ui.adsabs.harvard.edu/abs/2017MNRAS.469.1738S} {469, 1738}

\bibitem[\protect\citeauthoryear{{Slepian}, {Li}, {Schmittfull}  \&
  {Vlah}}{{Slepian} et~al.}{2019}]{2019arXiv191200065S}
{Slepian} Z.,  {Li} Y.,  {Schmittfull} M.,   {Vlah} Z.,  2019, arXiv e-prints,
  \href {https://ui.adsabs.harvard.edu/abs/2019arXiv191200065S} {p.
  arXiv:1912.00065}

\bibitem[\protect\citeauthoryear{{Springel}}{{Springel}}{2005}]{2005MNRAS.364.1105S}
{Springel} V.,  2005, \mn@doi [\mnras] {10.1111/j.1365-2966.2005.09655.x},
  \href {https://ui.adsabs.harvard.edu/abs/2005MNRAS.364.1105S} {364, 1105}

\bibitem[\protect\citeauthoryear{{Sugiyama}, {Saito}, {Beutler}  \&
  {Seo}}{{Sugiyama} et~al.}{2019a}]{2019arXiv190806234S}
{Sugiyama} N.~S.,  {Saito} S.,  {Beutler} F.,   {Seo} H.-J.,  2019a, arXiv
  e-prints, \href {https://ui.adsabs.harvard.edu/abs/2019arXiv190806234S} {p.
  arXiv:1908.06234}

\bibitem[\protect\citeauthoryear{{Sugiyama}, {Saito}, {Beutler}  \&
  {Seo}}{{Sugiyama} et~al.}{2019b}]{2019MNRAS.484..364S}
{Sugiyama} N.~S.,  {Saito} S.,  {Beutler} F.,   {Seo} H.-J.,  2019b, \mn@doi
  [\mnras] {10.1093/mnras/sty3249}, \href
  {https://ui.adsabs.harvard.edu/abs/2019MNRAS.484..364S} {484, 364}

\bibitem[\protect\citeauthoryear{{Szapudi}}{{Szapudi}}{2004}]{2004ApJ...605L..89S}
{Szapudi} I.,  2004, \mn@doi [\apjl] {10.1086/420894}, \href
  {https://ui.adsabs.harvard.edu/abs/2004ApJ...605L..89S} {605, L89}

\bibitem[\protect\citeauthoryear{{Szapudi} \& {Szalay}}{{Szapudi} \&
  {Szalay}}{1998}]{1998ApJ...494L..41S}
{Szapudi} I.,  {Szalay} A.~S.,  1998, \mn@doi [\apjl] {10.1086/311146}, \href
  {https://ui.adsabs.harvard.edu/abs/1998ApJ...494L..41S} {494, L41}

\bibitem[\protect\citeauthoryear{{Takada} \& {Hu}}{{Takada} \&
  {Hu}}{2013}]{2013PhRvD..87l3504T}
{Takada} M.,  {Hu} W.,  2013, \mn@doi [\prd] {10.1103/PhysRevD.87.123504},
  \href {https://ui.adsabs.harvard.edu/abs/2013PhRvD..87l3504T} {87, 123504}

\bibitem[\protect\citeauthoryear{{Tegmark} et~al.,}{{Tegmark}
  et~al.}{2006}]{2006PhRvD..74l3507T}
{Tegmark} M.,  et~al., 2006, \mn@doi [\prd] {10.1103/PhysRevD.74.123507}, \href
  {https://ui.adsabs.harvard.edu/abs/2006PhRvD..74l3507T} {74, 123507}

\bibitem[\protect\citeauthoryear{{Verde} et~al.,}{{Verde}
  et~al.}{2002}]{2002MNRAS.335..432V}
{Verde} L.,  et~al., 2002, \mn@doi [\mnras] {10.1046/j.1365-8711.2002.05620.x},
  \href {https://ui.adsabs.harvard.edu/abs/2002MNRAS.335..432V} {335, 432}

\bibitem[\protect\citeauthoryear{{Villaescusa-Navarro}}{{Villaescusa-Navarro}}{2018}]{2018ascl.soft11008V}
{Villaescusa-Navarro} F.,  2018, {Pylians: Python libraries for the analysis of
  numerical simulations} (\mn@eprint {ascl} {1811.008})

\bibitem[\protect\citeauthoryear{{Villaescusa-Navarro}
  et~al.,}{{Villaescusa-Navarro} et~al.}{2019}]{2019arXiv190905273V}
{Villaescusa-Navarro} F.,  et~al., 2019, arXiv e-prints, \href
  {https://ui.adsabs.harvard.edu/abs/2019arXiv190905273V} {p. arXiv:1909.05273}

\bibitem[\protect\citeauthoryear{{Wadekar} \& {Scoccimarro}}{{Wadekar} \&
  {Scoccimarro}}{2019}]{2019arXiv191002914W}
{Wadekar} D.,  {Scoccimarro} R.,  2019, arXiv e-prints, \href
  {https://ui.adsabs.harvard.edu/abs/2019arXiv191002914W} {p. arXiv:1910.02914}

\bibitem[\protect\citeauthoryear{{Watkinson}, {Majumdar}, {Pritchard}  \&
  {Mondal}}{{Watkinson} et~al.}{2017}]{2017MNRAS.472.2436W}
{Watkinson} C.~A.,  {Majumdar} S.,  {Pritchard} J.~R.,   {Mondal} R.,  2017,
  \mn@doi [\mnras] {10.1093/mnras/stx2130}, \href
  {https://ui.adsabs.harvard.edu/abs/2017MNRAS.472.2436W} {472, 2436}

\bibitem[\protect\citeauthoryear{{Yamamoto}, {Nakamichi}, {Kamino}, {Bassett}
  \& {Nishioka}}{{Yamamoto} et~al.}{2006}]{2006PASJ...58...93Y}
{Yamamoto} K.,  {Nakamichi} M.,  {Kamino} A.,  {Bassett} B.~A.,   {Nishioka}
  H.,  2006, \mn@doi [\pasj] {10.1093/pasj/58.1.93}, \href
  {https://ui.adsabs.harvard.edu/abs/2006PASJ...58...93Y} {58, 93}

\bibitem[\protect\citeauthoryear{{Yamamoto}, {Sato}  \& {H{\"u}tsi}}{{Yamamoto}
  et~al.}{2008}]{2008PThPh.120..609Y}
{Yamamoto} K.,  {Sato} T.,   {H{\"u}tsi} G.,  2008, \mn@doi [Progress of
  Theoretical Physics] {10.1143/PTP.120.609}, \href
  {https://ui.adsabs.harvard.edu/abs/2008PThPh.120..609Y} {120, 609}

\makeatother
\end{thebibliography}
.

\bsp	
\label{lastpage}
\end{document}